\documentclass[12pt]{article}
\usepackage{amsmath}
\usepackage{amssymb}
\usepackage{amsthm}
\usepackage{graphicx}
\usepackage{graphics}
\numberwithin{equation}{section}
\usepackage[left=0.925in,right=0.925in,bottom=1.4in]{geometry}
\usepackage{setspace}
\usepackage [linktocpage]{hyperref}
\date{(Last Revised: March 19, 2014)}
\begin{document}
\title{\bf Nucleon-Nucleon Interaction: \\
    A Typical/Concise Review\ \\ \ }
\author{{\bf M. Naghdi \footnote{E-Mail: m.naghdi@mail.ilam.ac.ir} } \\
\textit{Department of Physics, Faculty of Basic Sciences}, \\
\textit{University of Ilam, Ilam, West of Iran.}}
 \setlength{\topmargin}{0.1in}
 \setlength{\textheight}{9.2in}
  \maketitle
  \vspace{0.4in}
    \thispagestyle{empty}
    \begin{center}
\textbf{Abstract}
\end{center}

Nearly a recent century of work is divided to Nucleon-Nucleon (NN) interaction issue. We review some overall perspectives of NN interaction with a brief discussion about deuteron, general structure and symmetries of NN Lagrangian as well as equations of motion and solutions. Meanwhile, the main NN interaction models, as frameworks to build NN potentials, are reviewed concisely. We try to include and study almost all well-known potentials in a similar way, discuss more on various commonly used plain forms for two-nucleon interaction with an emphasis on the phenomenological and meson-exchange potentials as well as the constituent-quark potentials and new ones based on chiral effective field theory and working in coordinate-space mostly. The potentials are constructed in a way that fit NN scattering data, phase shifts, and are also compared in this way usually. An extra goal of this study is to start comparing various potentials forms in a unified manner. So, we also comment on the advantages and disadvantages of the models and potentials partly with reference to some relevant works and probable future studies.

  \newpage
  \setlength{\topmargin}{-0.7in}
  \pagenumbering{arabic} 
  \setcounter{page}{2} 

\tableofcontents
\section{Introduction}
In 1953, Bethe stated \cite{Bethe1} that in the quarter of the current century, many experiments, labor and mental works is allocated to the Nucleon-Nucleon (NN) problem; probably more than any other question in the history of humankind. NN interaction is the most fundamental problem in nuclear physics yet. In fact, since the discovery of neutron by Chadwick in 1932, the subject has been in the focus of attention; as, at the first, "nuclear physics" was often equal to "nuclear force". The reasons for this outstanding role are clear. The main reason is that describing the atomic-nuclei properties in terms of the interactions between the nucleon pairs is indeed the main goal of nuclear physics. \\
In nuclear structure studies, "nucleons" are always considered as "fundamental" objects, which is of course reasonable in the scale of nuclear physics with MeV energies. Although by the coming of Quantum Chromo Dynamics (QCD), it is established that nucleons are not fundamental, but by comparing the results from this traditional approach with the more fundamental ones, one may still understand better the advantages and disadvantages of the approaches. NN interaction is nowadays known more than any other parts of strong interaction both because of long-term researches (more than 80 years) and many experimental data as well as improved theoretical understanding of its various aspects. \\
The oldest theory of nuclear forces was presented by Yukawa \cite{Yukawa2} based on which the mesons mediate the NN ($pp, pn, nn$) interactions. Again, although the meson theory is not fundamental in the view of QCD, the meson-exchange approach has improved our understanding of nuclear forces besides giving some good qualitative results. Still, the mesons need in today's standard NN models/potentials, with the quarks and gluons, is avoidable to describe well many nuclear interactions and to build better models/potentials with more satisfactory results. In fact, by the advent of Effective Field Theory (EFT) and applying it to the low-energy QCD, we are somehow coming back to the meson-exchange theories with the aid of Chiral Perturbation Theory (CHPT). \\
Most basic questions were settled in the 1960's - 1990's. In recent years, the focuses are on the subtleties and various extensions of the idea for this special force leading to setting up more sophisticated two- and few-nucleon potentials. As a result, various high-quality models and forms for NN interaction are present nowadays. According to this, we can absolutely not address all on this rich and long-lived subject here but some basic facts and important issues of our favorite of course. By the way, we will discuss various potentials in more details in that one may intend to study and compare them in future studies--For some general and up-to-date views to the subject, look, for instance, at \cite{0811.1338}, \cite{1110.3761} and \cite{1210.0992}.

This note is organized as follows. In Section 2, we briefly discuss some basics of NN interaction, deuteron as the unique bound-state of two-nucleon systems, the symmetries of two-nucleon Lagrangian, general forms of NN potentials in configuration/coordinate-space (r-space from now on), equations of motion and partial-wave analysis. There, we also present a brief view to the scattering-length, effective-range and momentum space (p-space from now on) formalisms as well as relativistic NN scattering. In Section 3, we review the four main NN interaction models qualitatively. There are the Phenomenological models with many free parameters to be fitted to experimental NN data, the Boson-Exchange models based on the field-theoretical and dispersion-relations methods, the QCD-inspired models based on the fundamental quarks and gluons degrees of freedom, and the models based on EFT by using the chiral symmetry of QCD. As there are many NN interaction models and potentials forms and detailed studies need more times and places so, in Section 4, we try to review almost all-important two-nucleon potentials together with addressing the original papers for technical studies. There, we also mention the road of modeling and improving exact NN potentials. In addition, we study some high-precession potentials in more details as samples of the various existing potentials to do further studies and comparisons in an almost common scheme. Next, in Section 5, we mention few other models and potentials not mentioned in Section 4, which are the Mean Field Theory (MFT) methods and the Renormalization Group (RG) approaches as well as the Lattice QCD techniques. Finally, in Section 6, we make few comments about the current status and problems as well as the probable futures tries to be made on the rich way of nuclear force studies.

\section{A Brief of Nucleon-Nucleon Interaction}
One can estimate, with an introductory evaluation (e.g. by uncertainty principle) that two-nucleon interaction has the greatest contribution to nuclear force and four- and few-body interactions have almost negligible roles in most nuclear calculations. \\
In this section, we discuss some basics about NN interaction mainly in r-space and nonrelativistic theory. The aim is to introduce the beginners with the subject by referring the interested readers to the relevant textbooks and lecture notes for various technical and advanced studies.	

\subsection{Three Interaction Parts in Two-Nucleon Systems}
\begin{figure}[htp]
\centering
  \includegraphics[height=3.4in, width=3.65in, scale=1]{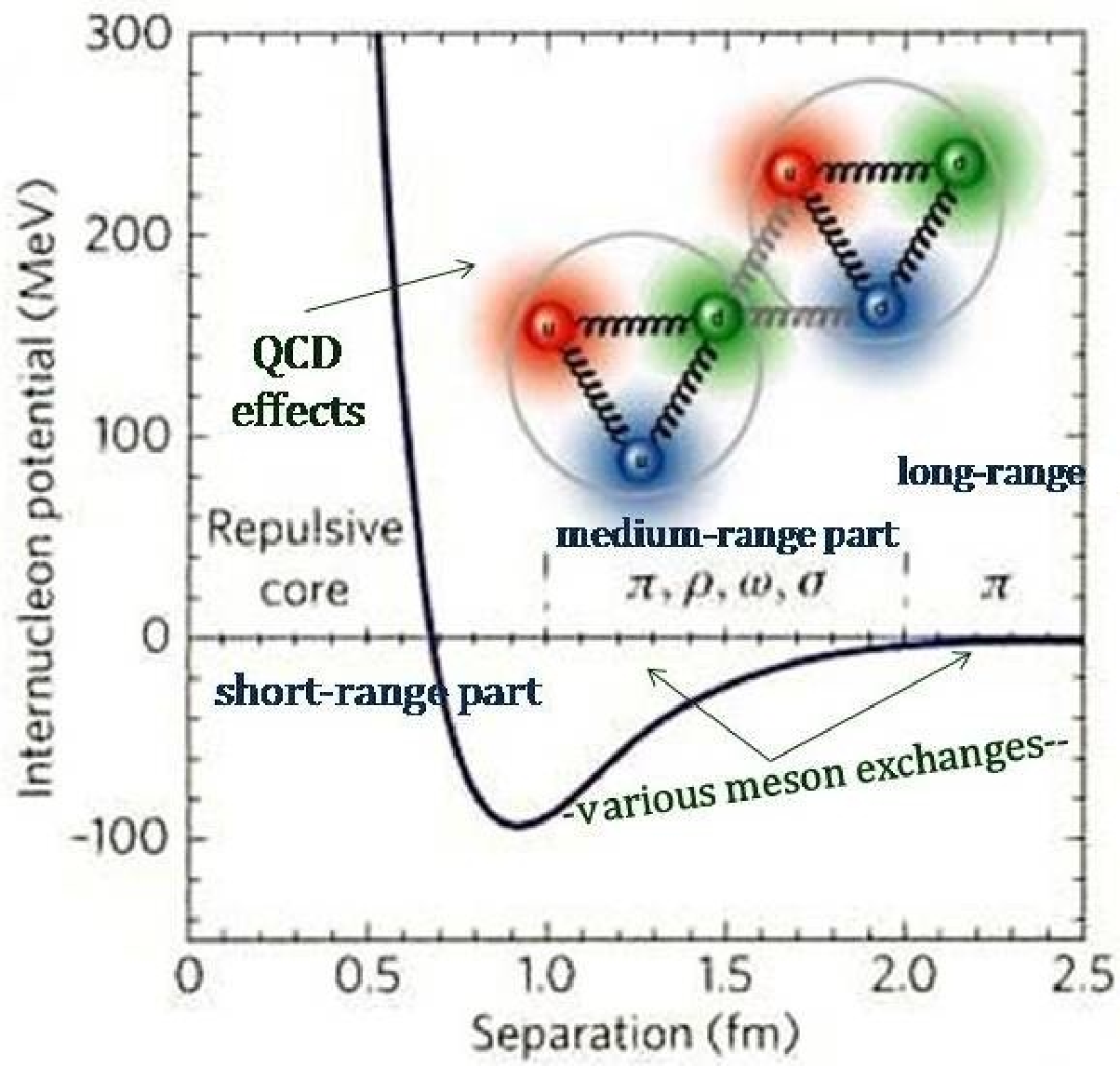} \\
  \caption{\textit{A general scheme for nucleon-nucleon potential.}} \label{Fig1.}
\end{figure}

Nucleon-Nucleon interaction is always divided into three parts, first in \cite{Tokyo1}, as follows:\\
a) The long-range (LR from now on) part $(r\gtrsim 2fm)$: In most models, it is considered as One-Pion-Exchange Potential (OPEP) and is added to the other parts of the potential as a tail. In a simple form in r-space, it reads
\begin{equation} \label{GrindEQ__1_}
 V_{OPEP}^{(1)}(r)=\frac{g_{pi}^2}{3} (\vec{\tau}_{1}.\vec{\tau}_{2}) \left[\frac{e^{-\mu r}}{r} (\vec{\sigma}_{1}.\vec{\sigma}_{2}) + \left(1+\frac{3}{\mu r}+\frac{3}{(\mu r)^{2}} \right) \frac{e^{-\mu r}}{r} S_{12} \right],
\end{equation}
where $\mu=\frac{1}{r_0}, \, r_0=\frac{\hbar}{m_{pi} c}$ and $S_{12}= 3(\vec{\sigma}_{1}.\hat{r}) (\vec{\sigma}_{2} .\hat{r}) - (\vec{\sigma}_{1} .\vec{\sigma}_{2})$ is the usual tensor operator; and $g_{pi}$ is the coupling constant, which is obtained from the experiments with mesons (meson-nucleon scattering). This potential has earned some improvements such as considering the difference between the neutral and charged pions and that it is different for \emph{pp}, \emph{nn}, \emph{np} interactions besides the clear forms raised from some new models of NN interaction.\\
b) The intermediate/medium (MR from now on)-range part $(1 fm\lesssim  r\lesssim 2fm)$: It comes from the various single-meson exchanges and mainly from the scalar-meson exchanges (two pions and heavier mesons). \\
c) The short-range (SR from now on) part $(r\lesssim 1fm)$: It is always given by exchanges of the vector bosons (heavier mesons and multi-pion exchanges) as well as the QCD effects.\\
In some of the potential forms, various Feynman diagrams, depended on the considered exchanges, in each of the three mentioned parts, are used. A general scheme for NN potential is shown in Figure \ref{Fig1.}.

\subsection{Deuteron: The Sole Bound-State of Two-Nucleon Systems}
One way to study the nuclear two-body interactions is using a two-nucleon system such as deuteron (the ${}^{2} H$ nuclei). Detailed studies need a general system of two-nucleon, which is, in turn, framed through scattering a nucleon from another nucleon. Nevertheless, deuteron is still fundamental to understand some basic properties of NN interaction. Deuteron is the exclusive loosely bound-state of two-nucleon system. From the symmetry considerations, ${}^{3} S_{1}$ and ${}^{3} D_{1}$ are its states. Its nonzero electric quadrupole-moment \cite{3},
\begin{equation} \label{GrindEQ__2_}
Q = \frac{\sqrt{2}}{10} \int_{0}^{\infty} u w r^{2} dr - \frac{1}{20} \int_{0}^{\infty} w^{2} r^{2} dr,
\end{equation}
confirms the presence of D-state and leads to introduce the tensor force. As a partial way to measure the quality of a potential, one may insert the wave functions for the S-state (u(r)) and the D-state (w(r)), gained from a special potential, into the last equation and then compares the results with the experimental values. For a technical study of deuteron, look at \cite{Garcon}.

\subsection{General Symmetries of Two-Nucleon Hamiltonian}
In general, the invariance of NN interaction under both rotation of coordinate system (the isotropic property of space) and translation of the origin of coordinate system (the homogeneous property of space) as well as time reversal, charge-independent (CI) and charge-symmetry (CS) are considered commonly. There are already some witnesses for symmetry breaking of the interaction, such as violating CI and CS \cite{4}--Look also at section 4 for more references. Nowadays, almost all accurate and high-precession NN potentials include these violations--We remember that charge dependence (CD) of NN interaction means that the interactions of \emph{pp} ($T_z=1$), \emph{np} ($T_z=0$), and \emph{nn} ($T_z=-1$) are different, whereas CS of NN interaction means that just the interactions of \emph{pp} and \emph{nn} are different. \\
From symmetry considerations, one can find out the various two-nucleon states under the condition that
\begin{equation} \label{GrindEQ__3_}
P_{r} P_{\sigma} P_{\tau} \psi (\vec{r}, \vec{\sigma}_{1} , \vec{\sigma}_{2} , \vec{\tau}_{1} , \vec{\tau}_{2}) = -\psi (\vec{r}, \vec{\sigma}_{1} , \vec{\sigma}_{2} , \vec{\tau}_{1} , \vec{\tau}_{2}),
\end{equation}
where $P_{r}, P_{\sigma}, P_{\tau}$ are the space-exchange (Majorana), spin-exchange (Bartlett) and isospin-exchange operators, respectively. For instance, in \emph{np} system, some states read
\begin{equation} \label{GrindEQ__4_}
\left\{\begin{array}{l} {S=0\, :\, {}^{1} P_{1} \, ,\, {}^{1} F_{3} \, ,\, {}^{1} H_{5} \, ,{}^{1} K_{7} \, ,\, {}^{1} M_{9} \, ,\, ...} \\
 {S=1\, :\, ({}^{3} S_{1} \, -{}^{3} D_{1} )\, ,\, {}^{3} D_{2} \, ,\, ({}^{3} D_{3} \, -{}^{3} G_{3} )\, ,\, {}^{3} G_{4} \, ,\, ({}^{3} G_{5} \, -{}^{3} I_{5} )\, ,\, {}^{3} I_{6}, \, ({}^{3} I_{7} \, -{}^{3} L_{7} ), ...\ .} \end{array}\right.
\end{equation}
The references \cite{Cohen8, Krane7, Wong6, Pal5, Nutshell1}, and \cite{Philips10} may be useful to earn more basic and general information about NN interaction.

\subsection{More About NN Interaction}
\subsubsection{Potential Forms, Equations of Motion, and Wave Functions in r-Space}
\begin{figure}[htp]
\centering
 \includegraphics[height=3in, width=4in, scale=1]{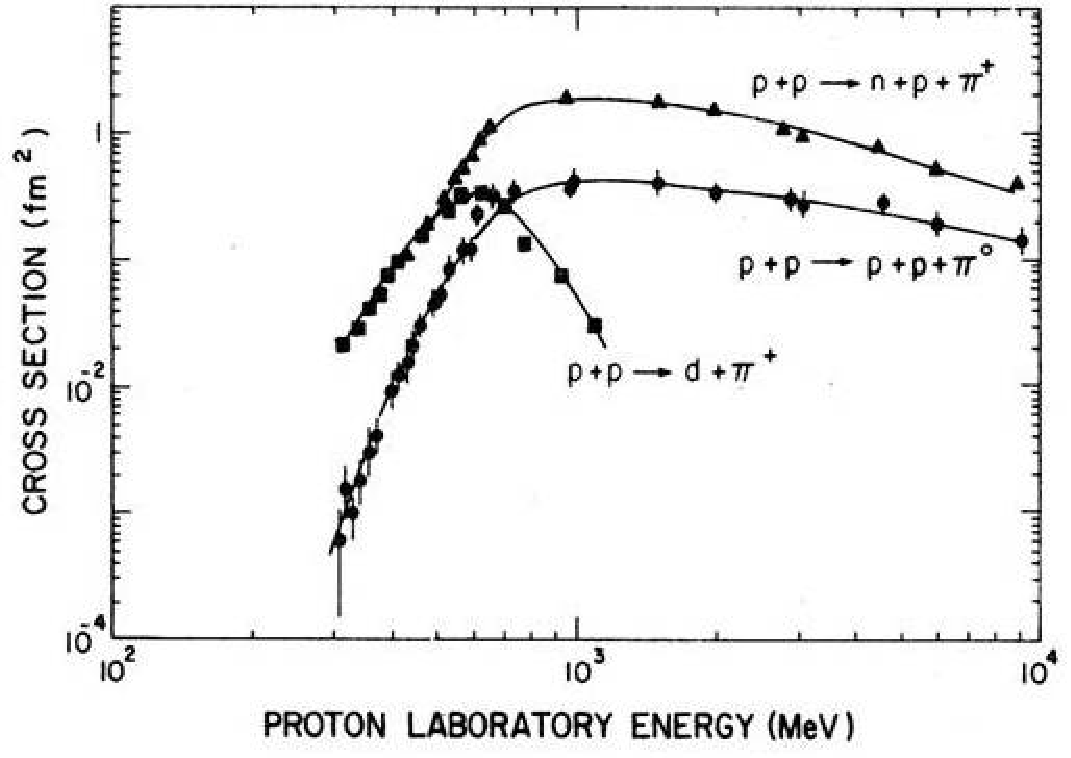} \\
  \caption{\textit{The energy dependence of the cross-section for pion production in the np scattering through the reactions of $p+p \rightarrow d + \pi^{+}$, $p+p \rightarrow p+p+\pi^{+}$ and $p+p\to p+p+\pi^{0}$}; \cite{11}.} \label{Fig2.}
\end{figure}

Generally, one can construct the following combinations
\begin{equation} \label{GrindEQ__5_}
\left\{\begin{array}{l} {\vec{A}\, .\, \vec{B}\, \, \, \, \, \, \, \, \, \, \, \, \, \, \, \, \, \, \, \, \, \, \, \, \, \, \, \, \, \, \, \, \, \, \, \, \, \, \, \, \, \, \, \, \, \, \, \, \, \, \, \, \, \, \, \, \, \, \, \, \, \, \, \ \ \ \ \ \ \ \ \ \ \ \ (scalar)} \\
{\vec{A}\, \times \, \vec{B},\, \, \vec{A}\pm \, \vec{B}\, \, \, \, \, \, \, \, \, \, \, \, \, \, \, \, \, \, \, \, \, \, \, \, \, \, \, \, \, \, \, \, \, \, \, \, \, \, \, \, \, \, \, \, \, \, \, \ \ \ \ \ \ (vector)} \\
{S_{ij} \, =\, \frac{1}{2} \, (A_{i} \, B_{j} \, +\, A_{j} B_{i} )\, -\frac{1}{3} \, \delta _{ij\, } \vec{A}\, .\, \vec{B}\, \, \, \, \, \, (rank-2\, spherical\, tensor)} \end{array}\right.
\end{equation}
from two vectors of $\vec{A}, \vec{B}$. For spin, isospin, space, and momentum vectors and also their combinations, one can consider many cases that obey the symmetry conditions as well. General form for the central potential is a linear combination of $I$, $\vec{\sigma}_{1} . \vec{\sigma}_{2}$, $\vec{\tau}_{1} . \vec{\tau}_{2}$ by multiplying each operator in a suitable radial function such as $V(r/a)$, where the range parameter $a$ is different for various operators. In general, these spin-isospin operators make the potential state-dependent.\\
The generic forms for the central and noncentral terms always read
\begin{equation} \label{GrindEQ__6a_}
V_{central}= V_{c}(r) + V_{\sigma}(r) (\vec{\sigma}_{1}.\vec{\sigma}_{2}) + V_{\tau}(r) (\vec{\tau}_{1}.\vec{\tau}_{2})+V_{\sigma \tau}(r)({\vec{\sigma}}_{1}.{\vec{\sigma}}_{2}) ({\vec{\tau}}_{1}.{\vec{\tau}}_{2}),
\end{equation}
\begin{equation} \label{GrindEQ__6b_}
 \begin{split}
V_{non-central}= V_{ls}(r)\vec{L}.\vec{S} + V_{t}(r) S_{12} & + V_{ls \tau}(r) (\vec{L}.\vec{S}) (\vec{\tau}_{1}.\vec{\tau}_{2}),
              + V_{ls \sigma}(r)\ (\vec{L}.\vec{S}) (\vec{\sigma}_{1}.\vec{\sigma}_{2}) \\
              & +V_{ls \sigma \tau}(r) (\vec{L}.\vec{S}) (\vec{\sigma}_{1}.\vec{\sigma}_{2}) (\vec{\tau}_{1}.\vec{\tau}_{2})+..... \ .
 \end{split}
\end{equation}
On the other hand, the matrix elements for some of the operators
\begin{equation} \label{GrindEQ__7_}
\left. \begin{array}{l} {I} \\ {\vec{\tau}_{1} . \vec{\tau}_{2} } \end{array}\right\}\, \times \, V \left(\frac{r}{a} \right)\, \times \, \left\{\begin{array}{l} {I} \\ {\vec{\sigma}_{1} . \vec{\sigma}_{2} } \\ {\left(\vec{r} \times \vec{p}\right) . \left(\vec{\sigma}_{1} . \vec{\sigma}_{2} \right) \, \, \, (spin-orbit)} \\
{S_{12} = 3(\vec{\sigma}_{1} . \hat{r})(\vec{\sigma}_{2} . \hat{r}) - \vec{\sigma}_{1} . \vec{\sigma}_{2} } \end{array}\right.
\end{equation}
are as follows:
\begin{equation} \label{GrindEQ__8_}
<\vec{\sigma}_{1}.\vec{\sigma}_{2}>= \left\{\begin{array}{cc} {1} & {\begin{array}{cc} {} & {} \end{array}; S=1 \, \, \, \, \, \, \, (spin-triplet state)} \\
                                    {-3} & {\begin{array}{cc} {} & {} \end{array}; S=0\, \, \, \, \, \, \, (spin-singlet state)} \end{array}\right. ,
\end{equation}
\begin{equation} \label{GrindEQ__9_}
\left\langle \ell ' S j m, TM_{T} \right. \left|\vec{L}.\vec{S}\right| \left. \ell S j m, TM_{T} \right\rangle =\frac{1}{2} \delta _{\ell \ell '} \left[j (j+ 1) - \ell  (\ell + 1) -S (S + 1)\right],
\end{equation}

\begin{equation} \label{GrindEQ__10_}
\begin{split}
  & \left\langle \ell, S =1, jm\right. \left|S_{12} \right| \left. \ell ', S= 1, j m\right\rangle= \\
  &  \ \ \ \ \ \ \ \ \ \ \ \ \ \ \ \ \ \ \ \begin{array}{ccc}
     \begin{tabular}{c|c}
       $\ell$ $\backslash$ ${\ell'}$ & $j$ \\
       \hline
       $j$ & 2 \\
     \end{tabular},
      &  \ \ \ \  &  \begin{tabular}{c|c c}
       $\ell$ $\backslash$ ${\ell'}$ & $j-1$ & $j+1$ \\
       \hline
       $j-1$ & $\frac{-2(j-1)}{2j+1}$ & $\frac{6\sqrt{j(j+1)}}{2j+1}$ \\
       $j+1$ & $\frac{6\sqrt{j(j+1)}}{2j+1}$ & $\frac{-2(j+2)}{2j+1}$ \\
     \end{tabular}.
   \end{array}
\end{split}
\end{equation}
The uncoupled radial Schrodinger equation (without the Coulomb force) reads
\begin{equation} \label{GrindEQ__11_}
\frac{d^{2} u}{dr^{2}} - \frac{j(j+1)}{r^{2}} u - {\left\langle j S j m,TM_{T} \right|} \left. \upsilon \right| \left. j S j m,TM_{T} \right\rangle u +  k^{2} u= 0,
\end{equation}
in which $\upsilon=-\frac{M}{\hbar^2} V$ and $k^2=\frac{M}{\hbar^2} E$, where $M$ and $E$ are the nucleon mass and  center of mass (c.m. from now on) energy, respectively. The potential of $V$ is indeed from the already mentioned forms of (\ref{GrindEQ__6a_}) and (\ref{GrindEQ__6b_}). Actually, that is composed of a form-function as $V(r/a)$, a linear combination of the various exchange operators and noncentral operators such as $(\vec{L}.\vec{S})$, $(\vec{L}.\vec{S})^{2}$, $S_{12}$ and so on. \\
Then, the asymptotic solutions to the equations read
\begin{equation} \label{GrindEQ__12_}
r \to 0 \Rightarrow u (r)=\left\{\begin{array}{l} {r^{-\ell}} \\ {r^{\ell+1}} \end{array}\right.; \quad r\to \infty \Rightarrow  u=\frac{1}{k} A_{\ell} \sin \left(kr - \frac{1}{2} \ell \pi -\frac{1}{2} + \delta _{\ell} \right).
\end{equation}
As one could find an asymptotic solution, the solution for all $r$'s is obtained by numerical integration. Next, with phase shifts, one can earn a potential from Schrodinger equation. \\
For the coupled states (without the Coulomb force), in turn, we also have
\begin{equation} \label{GrindEQ__13_}
\begin{split}
&\frac{d^{2}u}{dr^{2}}-\frac{j(j-1)}{r^{2}} u + k^{2} u + F(r) u + H(r) w= 0, \\
&\frac{d^{2}w}{dr^{2}}-\frac{(j+1)(j+2)}{r^{2}} w + k^{2} w + G(r) w + H(r) u = 0.
\end{split}
\end{equation}
It is notable that the ground-state of deuteron is a special case of the last equation with $k^{2}=-\gamma^{2}$ and $j=1$, where $\gamma^2=\frac{M}{\hbar^2} E_B$ with $E_B$ for deuteron binding-energy, and $u$ and $w$ stand now for the radial functions of ${}^{3} S_{1}$- and ${}^{3} D_{1}$- states, respectively. Because two partial-wave channels are coupled, an incoming wave in either $\ell=j-1$ or $\ell=j+1$ channel is scattered into either $\ell=j-1$ or $\ell=j+1$ channel. Therefore, we have two phase shifts (proper phases) $\delta_{j}^{\alpha}$, $\delta_{j}^{\beta}$, and a mixing parameter of $\varepsilon$. In presence of the Coulomb potential, a Coulomb phase-shift is also added and the problem becomes a little more complicated \cite{Pal5}.\\
On the other hand, as the c.m. kinetic energy of two-nucleon system is larger than the necessary amount to produce a meson, inelastic reactions becomes possible (see Figure \ref{Fig2.}). Since the lightest meson ($\pi$) mass is about 140 MeV, we expect, when the bombarding energy is upper than the threshold, some kinetic energy in the system transfers into the pion. By increasing the energy, the excitations because of the nucleon internal degrees of freedom, and the probable production of other particles, become more and more important. The inelastic scattering shows losing the flow from the incident channel and so, the probability amplitude is no longer conserved. Such a condition may be described by a complex scattering potential while some other relativistic effects come into account; therefore the Schrodinger equation for two-nucleon system is no longer enough. When discussing various direct potential forms, we return to the issue partly.

\subsubsection{Scattering-Length and Effective-Range}
One can simply show, by semiclassical reasoning's, that for the low-energy scatterings, only the S-state is important. By increasing the energy, the high-momentum states come into play because of the short-range properties of NN interaction. If we show the range by $a$ and the momentum by $\vec{p}$, the maximum angular-momentum, which is affected by the scattering potential, is $p a$ trivially. By squaring the last quantity and equating it with $\ell(\ell+1) \hbar^2$, one can easily get the energy in which a given $\ell$ comes into play. A rough estimate says that, for $\ell=1$ state, that energy is nearly 10 MeV.

For \emph{np} scattering below this energy, we have the following expression (see, for instance, Sec. II.C of \cite{3}, or Sec. 9.a of \cite{Pal5}):
\begin{equation} \label{GrindEQ__14_}
k\cot\delta =-\frac{1}{a} + \frac{1}{2} r_{e} k^{2} - P r_{e}^{3} k^{4}.
\end{equation}
As the term in $k^{4}$ becomes important (for $E \gtrsim$ 10 MeV), the P- and D-waves come into the play and so, it is not easy to break the $k^{4}$ dependence of $k\cot \delta$ for the S-wave phase-shift. Therefore, the useful energy region for the legality of (\ref{GrindEQ__14_}) is where the first two terms answer. In the case where the Coulomb potential is present (\emph{pp} scattering), a near effective-range expansion reads (see, for instance, Sec. IV.C of \cite{3}, or Sec. 9.b of \cite{Pal5}, and or see \cite{12}):
\begin{equation} \label{GrindEQ__15_}
C^{2} k \cot \delta + 2 k \eta h(\eta) = -\frac{1}{a} + \frac{1}{2} r_{e} k^{2} - P r_{e}^{3} k^{4},
\end{equation}
where
\begin{equation} \label{GrindEQ__15a_}
\begin{split}
& C^{2} = \frac{2\pi \eta}{\left(e^{2\pi \eta} -1\right)}, \quad \eta=\frac{Me^{2}}{\left(2\hbar^{2} k\right)}, \\
& h(\eta) = -\gamma - \log \eta  + \eta^{2} \sum_{M = 1}^{\infty}\left[M \left(M^{2} + \eta^{2} \right)\right]^{-1}, \quad  \gamma= 0.577.
\end{split}
\end{equation}
By using these relations (or similar ones) and three low-energy $^{1}S_{0}$ phase shifts, one can hold the parameters of $a$ (scattering-length), $r_{e}$ (effective-range) and $P$ (e.g. for a special potential).

\subsubsection{P-Space, Relativistic Scattering and so on}
For many calculations associated with NN interaction, it is suitable to express Schrodinger equation as an integral equation in momentum-space; look, for instance, at \cite{Goto, Hoshizaki} for some typical studies. The nonrelativistic scattering theory that leads to Lippmann-Schwinger (LS) equation and T-matrix are useful in this approach. Bethe-Salpeter (BS) equation in relativistic scattering theory, as relativistic counterpart of LS equation, studying separable potentials, separable expansions for arbitrary potentials, inverse scattering problem (see, for instance, \cite{nucl-th/0103010}, \cite{Shirokov} for some particular studies with references therein), are among the topics covered in this theory. The interested readers may also refer to \cite{3, Pal5, Nutshell1} for more fundamental discussions--As a side, we should note that many modern NN potentials (mainly meson-exchange and chiral EFT potentials) are always written in p-space originally and then Fourier transform into r-space. So, the p-space formalism is important and commonly used in the standard and relativistic approaches to NN interaction.

By the way, let us discuss a little more on the need for relativistic approaches to the problem. In fact, one may adjust the LS equation by including relativistic considerations. At the first look, one may suppose that these corrections are not so important below the first inelastic threshold. That is because the c.m. energy of the system is an almost small fraction of the nucleon rest mass. But, for the high momenta, it does not seem that describing the interaction through just nonrelativistic equations is satisfactory. In other words, the short-range repulsion of NN interaction, known from the various models based on at least phenomenological investigations, rapidly reduces the S-wave functions at the distances less than almost 0.5 $fm$. Therefore, it brings the high-momentum components into the wave functions for all energies. Meanwhile, one should note that to regularize the potentials at the origin, various parameterizations or form-factors with cutoffs are used; although in the potentials based on chiral EFT, more standard approaches are employed.  \\
Nevertheless, as one uses the phenomenological approach to describe NN interaction, the shortages in nonrelativistic approaches are not so important. That is because the parameterizations of the phenomenological models/potentials have enough flexibility to describe NN scattering in terms of the mesons with various coupling constants, masses and other free parameters. These phenomenological approaches are valid until they provide at least good quantitatively descriptions of experimental scattering data, and then they could be good alternatives for the complete relativistic descriptions.\\
On the other hand, while there is not a comprehensive theory for strong interactions, looking for a relativistic equation is somehow notional. Indeed, one may start from adapting the LS equation to satisfy the least needs of every relativistic equation. The basic want is that the scattering amplitude must satisfy the relativistic unitary along the elastic cutoff. The resultant equation is not unique ever; though, it is a relativistic version of the LS equation. For some studies on the relativistic NN scattering, look, for instance, at \cite{Faassen}, \cite{Wallace} and references therein.

\section{Nucleon-Nucleon Interaction Models}
There are some substantial models to build NN interaction potentials. In this section, we specify some qualitative features and, for more technical and quantitative studies, refer the interested readers to other relevant studies. 

\subsection{Almost Full Phenomenological Models}
These models always use the general form of a potential allowed by the symmetries like rotation, translation, isospin, and so on. In general, the phenomenological potentials often have the following features: \\
a) They are somewhat in a similar spirit as EFT, as we describe below, but much older and restricted to the space-time, spin, and isospin symmetries. b) Four important terms in the potentials are the central ($I$), spin-spin (${\vec{\sigma}}_{1}.{\vec{\sigma}}_{2}$), spin-orbit ($\vec{L}.\vec{S}$), and tensor ($S_{12}$) interactions. c) Each term occurs twice; one time without isospin-dependence and one time with the dependence ($\vec{\tau}_{1} . \vec{\tau}_{2}$), which in turn measures total isospin of NN system. d) The potential terms are responsible to describe various phenomena remarked in NN interactions. For example, the tensor term is important for the LR part of potential and arises naturally from pion-exchange.\\
In these potentials, the MR and SR parts are usually determined in a fully phenomenological way while for the LR part, an OPEP is often used. Examples for thepotentials are Hamada-Johnston potential \cite{HamadaJohnston}, Yale-group potential \cite{Yale}, Reid potentials (Reid68 \cite{Reid68}, Reid68-Day \cite{Reid68-Day}, Reid93 \cite{Nijm93}), Urbana-group potentials (e.g., UrbanaV14 \cite{UV14}), Argonne-group potentials (e.g., ArgonneV14 \cite{AV14}, ArgonneV18 \cite{AV18}), etc. Look at \cite{Gintautas} for a new study of phenomenological NN potentials. In the next section, we concentrate more on some samples of these phenomenological potentials. \\
The phenomenological potentials have almost many free parameters to be fitted to experimental scattering data and phase shifts. Less physics one may earn from them rather than the physics one may earn from the other potentials with tight theoretical grounds. Nevertheless, their ability to describe the practical facts of NN (\emph{pn}, \emph{nn}, \emph{pp}) interactions, their flexibility and convenience for using in nuclear structure calculations, are notable. These properties have still kept them in work more than the other potentials nowadays.

\subsection{Boson Exchange Models}
The potential acting between a pair of particles, because of a meson exchange, has the range of the meson's Compton-wavelength; which is in turn proportional to the meson-mass inversely. Because the pion $\pi$ is the lightest meson exchanged between nucleon pairs, it contributes the LR part of NN interaction, beyond its Compton's wavelength. So, the resultant potential is called OPEP. Similarly, to describe the MR and SR parts of the interaction, one should consider the exchanges of the heavier mesons than the pion as well as two, three and more pion exchanges. However, because considering these exchanges exactly is rather difficult, most people consider them phenomenologically while, in meson-exchange potentials, they are included clearly. Look at \cite{Taketani, BruecknerWatson} for the first meson and multi-pion exchange NN potentials.\\
On the other hand, it is already known that multi-pion systems have some strongly correlated resonances that behave often as a single meson. So, it is supposed that multi-pion resonances, when exchange between two nucleons, may contribute to the MR and SR parts of interactions. The potentials built in this way, by including single meson exchanges, are called One-Boson-Exchange Potentials (OBEP). Besides the traditional one $\pi(138)$ exchange, various meson exchanges are considered in OBEP's. There are the exchanges of $\rho(769)$ meson (as a $2\pi$ resonance), $\omega(783)$ meson (as a $3\pi$ resonance), $\eta(549)$ meson (the same quantum numbers with $\pi$ but its isospin that is $T=0$), $\acute{\eta}(958)$ meson (the same quantum numbers with $\eta$ but heavier and a resonance of $\eta \pi \pi$), $\delta(983)$ (as a $4\pi$ resonance), $\phi(1020)$ meson (the same quantum numbers as $\omega$ but a resonance of $K^+ K^-$ system), and $S^*(975)$ meson (as $2\pi, K\bar{K}$ resonances). In addition, one always considers the experimentally undetermined scalar-isoscalar boson of $\sigma (500-700)$, which is usually considered as a good parameterization of $2\pi$ system in S-state. Still, there are two other mesons with the mass above 1 GeV that may act as $2\pi$ resonances. They are $\epsilon (1300)$ (or $f_0$) meson (with the same quantum numbers as $S^*(975)$ but just as a $2\pi$ resonance) and $f(1274)$ meson. Some other two-boson resonances may come from the mesons of $A_1(1275), A_2(1318)$ (as $\rho \pi$ resonances), $B(1234)$ (as a $\omega \pi$ resonance) and $D(1283)$ (as $\eta \pi \pi, 4\pi$ resonances). But, because of the importance of the hadrons' structure in the energy region of 1 GeV and with respect to the energies involved in the common NN interactions, the roles of the heavier mesons may not be so important.\\
In general, $\pi$-meson (and also $\phi$-meson) exchange provides the most LR (tensor) force, whereas $\omega$-meson exchange provides the most SR repulsive force and SR spin-orbit force. The intermediate attractive force is often explained by $2\pi$ (as $\rho$- and/or fictitious $\sigma$-meson) exchanges, whereas the potential contribution by $\eta$-meson is weak and always ignorable. Therefore, these few mesons describe the main features of NN interaction; but to describe well experimental data and other subtle properties, depended on the case, the exchanges of the other mentioned mesons are also included.

From the differences among various NN potentials, based on meson theory, are their methods to deal with the $2\pi$ exchange. In one approach, its effect is simulated through one or two scalar and isoscalar mesons. Considering the $2\pi$ environmental effects as well as employing scattering-length and effective-range formalism for the S-state of the system are the efforts in the line. By the way, each group uses its own methods and details to evaluate the potential. In general, the settled "field theoretical techniques" and the methods based on "dispersion relations" are two main ways for handling the problem. Look at \cite{PartoviLomon} for a senior field-theoretical NN model, \cite{Erkelenz} for an old review on OBEP, \cite{Bonn2} for a comprehensive review, and \cite{Gross} for another useful typical study. See also \cite{nucl-th/9609054, nucl-th/9611056, Lebed1, 1201.0443} for some other relevant studies.

In other words, by discovering the vector mesons of $\rho$ and $\omega$, with the massed in the ranges of 770-780 MeV, more progresses in understanding NN interaction were archived and led to expand OPEP's. In OBEP's, mostly, the unrelated contributions of the single-meson exchanges of the pseudoscalar mesons $\pi$, $\eta$ and the vector mesons $\rho$, $\omega$ as well as the scalar meson $\delta(983)$ are considered and iterated into scattering equation. There are also $2\pi$ exchanges, which are always parameterized by the artificial $\sigma$ meson with the masses in the range of 400-800 MeV. The core (SR) region of the potentials is always parameterized through phenomenological parameters and the form factors related to the meson-nucleon vertices. The form factors in turn hold on fundamental relations to QCD. Then, such OBEP's provide good (at least quantitative) description of scattering data. Many types of these potentials, each with its own characterizes and features already exist. Nowadays, it is almost clear that the meson-exchange potentials (MEP's) are almost the standard NN potentials. Some examples are the Partovi-Lomon model \cite{PartoviLomon}, and Stony Brook-group \cite{Jackson}, Paris-group \cite{Cottingham}, Bonn-group \cite{Machleidt}, Padua-group \cite{Minelli}, Nijmegen-group \cite{Nijm93} and Hamburg-group \cite{Jaede} potentials.\\
Boson-exchange methods are nowadays extended, besides NN systems, to many Baryon-Baryon (BB) interactions such as pion-nucleon, pion-pion, Hyperon-Nucleon (YN) and Hyperon-Hyperon (YY) interactions as well. Although these models do not refer to QCD deeply, but the baryon and meson fields are already considered as the asymptotic states that absorb all effects from the quark and gluon dynamics. It is also notable that not only phenomenological models but also the advanced models of NN interactions, such as QCD-inspired and chiral EFT models, which we describe below, use boson exchanges in some parts of studies.

To summary, we note that in the quark-antiquark pair (= meson) exchange model, there are the following features: a) It is similar to the quark exchange but the reverse direction of one quark. b) It gives a good description of many aspects of NN interaction. c) It is preferred because the meson states are colorless and have almost lower masses or larger ranges. d) It studies OPEP and generalizes it to other mesons that results in OBEP's and more. e) Next to full phenomenological potentials and chiral EFT potentials, BEP's are the best physical potentials that give perfect agreement with the data for the LR and MR parts especially.

\subsection{The Models Based on QCD} \label{3.3}
In these models, the aim is to connect hadronic processes to the underlying theory of strong interactions that is QCD. In other words, hadron-hadron interactions are described in terms of quark and gluon degrees of freedom. Look at \cite{FredMyhrer}, \cite{nucl-th/9809093}, \cite{hep-ph/0211443}, \cite{nucl-th/0212044}, \cite{hep-ph/0205138} and \cite{Wu0}, \cite{nucl-th/0404004} for some reviews and typical studies of NN interaction in QCD and quark models. \\
In low energies, relevant to NN interaction, QCD is nonperturbative and could not solved exactly. Chiral Perturbation Theory (CHPT) (see, e.g., \cite{1110.3022}), Skyrme Model (see, e.g., \cite{nucl-th/0007051}, \cite{Wambach}), and Nambu-Jona-Lasinio (NJL) models (see, e.g., \cite{Rashdan}) are examples of this approach. The models describe the characteristic phenomena observed in nucleon-nucleon, pion-nucleon, and pion-pion scattering well qualitatively but they fail quantitatively. Common features of the "QCD-inspired" models, that reduce the demand for them, are cumbersome mathematics, large numbers of parameters and limits in applying especially to very low energies. Therefore, if one wants a good quantitative description of experimental data, phenomenological approaches such as boson-exchange and phenomenological models are preferred. Nevertheless, in some models just for the short distances, the QCD approach is used whereas for the remaining parts of interaction, the two former approaches are used with satisfactory results. We deal with the issue more when discussing plain potentials.

In summary, there are two subsets of QCD-inspired models, with basics features and main characteristics, as follows:\\
1) The gluon and quark exchange among nucleons plus the Pauli-repulsion between similar quarks in overlapping nucleons, with the following features: a) The gluon exchanges based on "constituent quark model"(CQM) besides one-gluon-exchange-potential (OGEP). It does not give a good description for reasonable distances because of confining the colorless singlets. b) The Pauli-repulsion is related to a minimum energy to excite a nucleon (that is to move a quark into a different state) of 300 MeV. c) The quark exchanges between two nucleons and may change nucleon charges (i.e., $n\to p$ and at the same time $p\to n$). d) It gives a reasonable and semi-quantitative description of the SR repulsive part and maybe the MR part of NN interaction. Look at, for instance, \cite{hep-ph/0211443, nucl-th/0212044, hep-ph/0205138, Wu0, nucl-th/0404004} for some general studies. Among the exact potentials of this type are the Moscow-group \cite{Kukulin} and Oxford-group \cite{Oxford1} potentials.

2) Chiral symmetry and CHPT can also be considered as a subset of QCD methods, with the following features: a) That is based on chiral symmetry of QCD Lagrangian. That symmetry means that the quarks with opposite helicity are indistinguishable and do not couple to each other except for their masses. b) Chiral symmetry is spontaneously broken because QCD prefers the quark-antiquark pairs with negative parity to the quark-quark pairs with positive parity. Thus, the low-mass modes (zero-mass theoretically) of the "quark condensation" are called "Goldstone bosons" (pions, kaons, etc.). This, in turn, limits the Lagrangian to the processes involving nucleons and pseudoscalar mesons. In other words, for the energies around $\Lambda_{QCD} \approx$ 1GeV, there is a "phase transition" from 'fundamental" theory to an "effective" theory through spontaneous breaking of the chiral symmetry of QCD Lagrangian. During this procedure, pseudoscalar "Goldstone" bosons are produced. As a result, in low energies ($E<\Lambda_{QCD}$), the proper degrees of freedom are the pseudoscalar mesons and other similar hadrons, and not the quarks and gluons of the original theory. The standard effective theory to describe this process is called CHPT. c) Chiral symmetry is also violated by the (small) quark masses; so, the Goldstone bosons are not massless totally. Nevertheless, one can expand the interaction in small parameters to make definite predictions (as in CHPT). Look, for some related studies, at \cite{0802.2484, 0803.4190} and \cite{1110.3022} and also references therein. In the following subsection, we discuss this issue further.\\
Still, we should note to some other studies, on NN interaction, in language of "lattice QCD", for instance in \cite{Beane, nucl-th/0611096, hep-lat/0601006, 1005.1908} and \cite{nucl-th/9807024}.

\subsection{Effective Field Theory Approach} \label{3.4}
Effective-field-theories (EFT's) are the low-energy descendants to the high-energy parent theories. Some of the features are as follows: a) In general, one notes that there are different/separate energy scales in the nature each with its own degrees of freedom. In each energy level, just some degrees of freedom are relevant and as the energy decreases, some others are frozen and become irrelevant. An example of this is the chiral symmetry. b) About NN interaction, as first hinted by Weinberg \cite{Weinberg}, EFT means applying all symmetries including the chiral symmetry of QCD Lagrangian but not directly considering the underlying degrees of freedom like pions or quarks. This gives the most general Lagrangian that contains many parameters to be constrained with data. In other words, the Lagrangian must include all possible terms to guarantee that the "effective" theory is indeed the low-energy limit of the "fundamental" high-energy theory. So, no presumptions about, for example, renormalizability or simplifying the Lagrangian are permissible. This, in turn, means that we probably have an infinite set of interactions. Therefore, to have a reasonable theory with well-defined results, one must organize the perturbative expansion up to some defined orders. Look, for instance, at \cite{nucl-th/9801034, nucl-th/9902015, nucl-th/9909011, 0704.0807} and references therein, for some reviews of EFT approach to NN interaction.

In general, a systematic improvement in the ability of the model to reproduce NN data is observed when the orders of chiral expansion increase. One of the first extended models (in Next-to-Next-to-Leading Order: NNLO) of CHPT described \emph{np} phase shifts well up to the energies about 100 MeV; but, for the higher energies, some inconsistencies occurred in some partial waves--See \cite{1303.4674} for a recent study of this approach and developments. Although NNLO and the most recent higher-order chiral NN potentials show significant progress towards the earlier ones and are almost perfect (in fact, the new NNNLO potentials describe data well up to 350 MeV with similar quality as the high-precession phenomenological and boson-exchange potentials), still to apply well the resultant potentials to all nuclear structure calculations, more quantitative and even qualitative improvements are necessary. We should, of course, note that chiral EFT models have more standards and great theoretical bases to be known as the most reasonable models to describe the strong nuclear interactions. \\
In summary, we can say that CHPT from EFT in low energies is as fundamental as QCD in high energies. In addition, because of the perturbative nature of CHPT, it can be evaluated order by order in chiral expansions. As long as we are looking for a substantial theory of nuclear forces, applicable to nuclear structure calculations as well, CHPT is likely able to overcome the discrepancies between experiment and theory. For some other typical studies on the subject of EFT and CHPT, look, for instance, at \cite{nucl-th/9801034, nucl-th/9910064, nucl-th/0608068}, and \cite{0811.1338, 1110.3022, 1302.3241} for some recent views. Meanwhile, among the high-quality potentials of this type are those by Texas-group \cite{Ordonez1}, Sao Paulo-group \cite{Robilotta1}, Munich-group \cite{Munich1}, Idaho-Group \cite{Idaho1}, and Bochum-Julich-Group \cite{BochumJulich2}.

\section{Nucleon-Nucleon Interaction Potentials}
\subsection{Basic Potentials and General Remarks}
In this subsection, we discuss on the main preliminary potentials and a brief on the methods of making them. As already mentioned, the range of nucleon-nucleon interaction is divided into three parts, which are the short-range (SR), the intermediate or medium-range (MR) and the long-range (LR). For the MR and LR parts, many workers have always taken the phenomenological and boson-exchange pictures. However, in most models, for the LR part, one-pion-exchange (OPE) is usually included. For the SR part, phenomenological parameterizations are often employed. In some models, form factors are included to regularize the potentials at the origin; whereas, in some other models, severe hard cores are included. The first major approach to describe the MR part was to include two-pion-exchange (TPE) contributions. The first samples of TPE potentials were given by Taketani-Machida-Ohnuma \cite{Taketani} and Brueckner-Watson \cite{BruecknerWatson}. However, those TPE potentials did not provide good descriptions of NN scattering data as one reason was the lack of a spin-orbit potential therein. Next, Gammel, Christian and Thaler \cite{Gammel1} discovered the need to include a spin-orbit potential when they tried to fit the NN scattering data at that time with a velocity-dependent local phenomenological NN potential as
\begin{equation} \label{GrindEQ__16_}
V=V_{c}(r) + V_{t}(r) S_{12},
\end{equation}
for each of the four combinations of the spin and isospin. Nevertheless, they failed! \\
In 1957, the efforts to build further phenomenological potentials, by including the phenomenological spin-orbit potentials as well, got started. The purely phenomenological potential of Gammel-Thaler \cite{Gammel2} provided a good description of the scattering data at that time below $T_{lab}=$310 MeV (note that we use throughout this note the laboratory energy unless otherwise be told). At the same time, the semiclassical potential of Singell-Marshak \cite{SingellMarshak}, which was consist of the TPE potential of Gartenhaus \cite{Gartenhaus}, next to the phenomenological spin-orbit potential, provided a satisfactory description of the data below 150 MeV. \\
Then, Okubo-Marshak \cite{OkuboMarshak} showed that the most general two-nucleon potential, by considering symmetry conditions, reads
\begin{equation} \label{GrindEQ__17_}
\begin{array}{l} {V(\vec{r},\vec{p},\vec{\sigma}_{1},\vec{\sigma}_{2},\vec{\tau}_{1},\vec{\tau}_{2}) = V_{c}(r) + V_{\sigma}(r) \left(\vec{\sigma}_{1}.\vec{\sigma}_{2} \right) + V_{\tau}(r) \left(\vec{\tau}_{1} . \vec{\tau}_{2} \right) + V_{\sigma \tau}(r) \left(\vec{\sigma}_{1} . \vec{\sigma}_{2} \right)\left(\vec{\tau}_{1} . \vec{\tau}_{2} \right)} \\ {\, \, \, \, \, \, \, \, \, \, \, \, \, \, \, \, \, \, \, \, \, \, \, \, \, \, \, \, \, \, \, \, \, \, \, \, \, \, \, \, \, \, \, \, \, \, \ \ +V_{ls}(r)( \vec{\tau}_{1} .\vec{\tau}_{2} )+ V_{ls \tau}(r) \left(\vec{L}. \vec{S}\right) \left(\vec{\tau}_{1} . \vec{\tau}_{2} \right)} \\ {\, \, \, \, \, \, \, \, \, \, \, \, \, \, \, \, \, \, \, \, \, \, \, \, \, \, \, \, \, \, \, \, \, \, \, \, \, \, \, \, \, \, \, \, \, \, \ \ +V_{t}(r)S_{12}+ V_{t\tau}(r)S_{12} \left(\vec{\tau}_{1} . \vec{\tau}_{2} \right)} \\ {\, \, \, \, \, \, \, \, \, \, \, \, \, \, \, \, \, \, \, \, \, \, \, \, \, \, \, \, \, \, \, \, \, \, \, \, \, \, \, \, \, \, \, \, \, \, \ \ + V_{q}(r)Q_{12} + V_{q\tau}(r) Q_{12} \left(\vec{\tau}_{1} . \vec{\tau}_{2} \right)} \\ {\, \, \, \, \, \, \, \, \, \, \, \, \, \, \, \, \, \, \, \, \, \, \, \, \, \, \, \, \, \, \, \, \, \, \, \, \, \, \, \, \, \, \, \, \, \, \ \ + V_{pp}(r) \left(\vec{\sigma}_{1} . \vec{p}\right) \left(\vec{\sigma}_{2} . \vec{p}\right) +  V_{pp\tau}(r) \left(\vec{\sigma}_{1} . \vec{p}\right) \left(\vec{\sigma}_{2} . \vec{p}\right) \left(\vec{\tau}_{1} .\vec{\tau}_{2} \right)}, \end{array}
\end{equation}
where $\vec{L}.\vec{S}$ is the usual spin-orbit operator and
\begin{equation} \label{GrindEQ__18_}
Q_{12}=\frac{1}{2} \left\{(\vec{\sigma}_{1} . \vec{L}) (\vec{\sigma}_{2} . \vec{L}) + (\vec{\sigma}_{2} . \vec{L}) (\vec{\sigma}_{1} . \vec{L})\right\}
\end{equation}
is the quadratic spin-orbit operator. The twelve terms in the potential are given by the twelve radial functions $V_{c}(r), ...$ . These functions can be obtained from our knowledge about the nature of nuclear forces. Information to find out $V(r)$'s could be from, for example, the exchanges of various mesons or phenomenological mechanisms in which some given radial functions, with maybe some arbitrary free parameters to be fixed to experimental data, exist. Once our understanding of underlying theories (such as QCD) improves further, we may be able to get these functions from the basics. The first four terms in (\ref{GrindEQ__17_}) stand for the complete central potential and, in the case, $L$ and $S$ are the good quantum numbers. By adding other terms, the good quantum number is $J$ as the two-nucleon system is now invariant under the combined space of $L$ and $S$. The main reason for two terms in the spin-orbit potential of
\begin{equation} \label{GrindEQ__19_}
V_{Spin-Orbit}(r) = V_{ls}(r)\vec{L}.\vec{S} + V_{ls \tau}(r) (\vec{L}.\vec{S}) (\vec{\tau}_{1} . \vec{\tau}_{2})
\end{equation}
is that the radial dependence of the potentials may be different for the isospin-independent and isospin-dependent parts, for examples, because of different meson exchanges. The seventh and eighth terms stand for the tensor forces while the ninth and tenth terms are for the quadratic spin-orbit forces. The latter two terms enter just when momentum dependence exists in the potential. The last, 11th and 12th, terms are always omitted because, at least for elastic scattering, they can be written as linear combinations of the other terms. So, their role cannot be determined from elastic scattering from which most of our information about NN interaction comes--For a useful study about NN interaction, including the potentials and ideas, before 1960, look also at \cite{Philips10}.

Soon after, better potentials were constructed. Among the 1960's meson-exchange and field-theoretical potentials, the NN potential by Sugawara and others \cite{Sugawara1} , \cite{Sugawara2} are also mentionable. Other important phenomenological potentials then were Hamada-Johnston \cite{HamadaJohnston} and Yale \cite{Yale} potentials and also various hard- and soft-core potentials by Reid \cite{Reid68}.

Before going more into discussing some other potentials, it is useful to mention that almost all experimental elastic phase shifts are derived from the differential cross sections of \emph{pp} and \emph{np} scattering's. For most potentials the data are often fitted in the energy range of 0-350 MeV. That is because, in the higher energies, inelastic processes (with the threshold of about 280 MeV), such as pion production and other relativistic effects, come into play and so, the two-body Schrodinger equation is no longer enough. It should be mentioned that the modern analysis with more improved relativistic equations (for example with BS equation) have tried to account for all effects at once.\\
To know the methods of making the early potentials, we note that, for instance, the Hamada-Johnston (HJ) \cite{HamadaJohnston} and Yale-group \cite{Yale} determined all two-nucleon scattering data and polarization parameters as a function of energy for the energies of a few hundred MeV. The Yale-group potential was initially framed to reproduce the phase shifts in various states as a smooth function of energy. As a first step, the phase parameters (that is the phase shifts and mixing parameters of the coupled states) were determined as functions of energies by fitting to all experimental scattering and polarization data. The procedure was performed by several groups mainly Yale-group \cite{Yale2} and Livermore-group \cite{Livermore} then--For a updated analysis of NN scattering data by the latter group look at \cite{0706.2195}. As a second step, the potentials, with their adjusted parameters, reproduced the phase parameters. The more standard procedure is to present  scattering amplitudes as a sum of all partial waves up to a maximum orbital angular-momentum, which is more or less $\ell_{\max}=5$. The contributions of the higher partial waves are always indicted by OPE contribution to the scattering amplitude. In the Yale-group potential, OPEP was used as a fixed part while the remaining parts of the potential were fixed by fitting the energy-dependent phase parameters up to $\ell_{\max}$. It is mentionable that, for the current up-to-date potentials, the basic analyses are drastically improved although the procedures are more or less similar. \\
By the way, in most NN potentials, for the LR part, OPEP is usually used while for the MR part the multi-pions and single mesons such as $\rho, \omega, \sigma$, ... are often used. Still for the SR repulsive part, various methods including neutral vector-meson exchanges, velocity-dependent potentials, phenomenological parameterizations and QCD substructure techniques are used.

\subsection{NN Potential's Road} \label{3.5}
The original try to find the fundamental theory of nuclear forces was started around 1935 by Yukawa. The Yukawa \cite{Yukawa2} meson-exchange model for nuclear force and the other old pion-exchange potentials, such as those by Taketani-Machida-Ohnuma  \cite{Taketani}, Brueckner-Watson \cite{BruecknerWatson}, Singell-Marshak \cite{SingellMarshak}, Gartenhaus \cite{Gartenhaus}, etc. were not so successful.  That was both because of the failing of their structures and the pion dynamics, which we now know is restricted by chiral symmetry. By discovering heavy mesons in the early 1960's, modeling better one-boson-exchange-potentials (OBEP's) was started in \cite{Hoshizaki00}, \cite{Wong00} and \cite{Bryan} as well, and was then developed more by framing some better potentials. Therefore, the field-theoretical and quantum-dispersion methods were involved with making the potentials such as Partovi-Lomon model \cite{PartoviLomon}, Stony Brook-group \cite{Jackson}, Paris-group \cite{Cottingham}, Nijmegen-group \cite{Nijm78} and Bonn-group \cite{Machleidt} potentials \footnote{Among the other  boson-exchange potentials are those in \cite{Virgina1} and \cite{Bochum1}, where the former is a relativistic OBE model and the latter is constructed from the meson-exchange and nucleon structure properties.}. But there were still some problems with the boson-exchange potentials. Among them was the $\sigma$-boson exchange for which experimental evidence was polemic. Nevertheless, because that equals a $2\pi$ resonance, there were many efforts to find two-pion contributions to the interactions. Anyhow, then, more high-precession potentials such as parameterized Paris potential \cite{Paris2}, the high-quality potentials of Nijm93, NijmI, NijmII \cite{Nijm93}, CD-Bonn \cite{Bonn4} and many other interesting potentials based on meson-exchange pictures were constructed. So, it seemed that the nuclear force problem was solved! But, no! \\
With the coming of QCD and its underside quarks and gluons degrees of freedom, the studies came into new phases. Still, the problem with QCD was its nonperturbative structure when applying to the MeV low-energy limit, where nuclear physics is valid. The QCD-inspired quark models were the first tries in the phase \cite{FredMyhrer}. Lattice QCD was/is also a way to deal with the problem; see for instance \cite{Beane}. Still, the QCD-inspired potentials were/are qualitatively successful but no quantitatively well as are the phenomenological meson-exchange potentials. Among these potentials, the potential set up by some members of the Paris-group in \cite{VinhMau1}, the Moscow-group potentials \cite{Kukulin} and the Oxford potential \cite{Oxford1} are mentionable. Nevertheless, some potentials, such as the high-quality Nijmegen-group ones \cite{Nijm93} (and even two former ones) use a mixture of the mesons and quarks in some parts of the interaction.\\
Meanwhile, many phenomenological potentials composed of meson-exchanges, operators and functions with adjusted parameters to fit experimental data, with wide applications in nuclear computations, were constructed. Among them are the Reid \cite{Reid68} and UrbanaV14 \cite{UV14} potentials, and the high-precession Nijmegen-group potentials \cite{Nijm93} and ArgonneV18 \cite{AV18} potential.

By coming EFT and applying it to the low-energy QCD, first by Weinberg \cite{Weinberg}, the new phase to set up NN potentials got started. In such models, one usually starts by writing the most general Lagrangian including the assumed symmetries and especially chiral symmetry of QCD. In low-energies, chiral symmetry breaks down and then the suitable degrees of freedom are not quarks and gluons but there are pions and nucleons, while heavy mesons and nucleon resonances are integrated out. So, it seems that we are going back to the meson theory! of course with much more experiences. \\
The chiral effective Lagrangian is composed of a set of the sentences increasing in derivative terms or nucleon fields. Indeed, one use a perturbative expansion in $(Q/\Lambda_{QCD})^\nu$, where $Q$ refers to the soft scale associated with external momenta or pion mass, $\Lambda_{QCD}\approx 1$ GeV is the chiral symmetry breaking scale and $\nu\geq0$. By applying the Lagrangian to NN scattering, there are the suiting Feynman diagrams whose importance becomes less as the order of the chiral perturbation theory ($\chi$PT) expansion increases. Besides describing the nuclear two-body problem, the model makes some good predictions for nuclear few-body forces as well. The first potential of this type was constructed by Texas-group (Ordonez, Ray and van Kolck) \cite{Ordonez1} and among the further developed ones are those by Idaho-group \cite{Idaho1} and Bochum-Julich-group \cite{BochumJulich2} up to NNNLO. These new CHPT potentials are quantitatively and qualitatively best so far candidates to describe two-nucleon as well as few-nucleon interactions.\\
It is also notable that there are some tries to construct NN potentials based on renormalization-group (RG) approach to NN interaction by another Stony-Brook group \cite{nucl-th/0108041}. As a result, they have earned many creditable and satisfactory results that we comment more in Section 4. Anyway, in what follows we continue studying some of the potentials which are of course more important with established results in nuclear structure calculations, briefly.

\subsection{Hamada-Johnston Potential}
The Hamada-Johnston (HJ) potential \cite{HamadaJohnston} is a leading phenomenological NN (\emph{pp}+\emph{np} here) energy-independent potential. It described well the scattering data below 350 MeV and deuteron properties as well as the effective-range parameters. The general form of HJ potential \cite{HamadaJohnston} reads
\begin{equation} \label{GrindEQ__19a_}
  V=V_{c}(r) + V_{t}(r) S_{12} + V_{ls}(r) \vec{L}.\vec{S} + V_{ll}(r) L_{12},
\end{equation}
where
\begin{equation} \label{GrindEQ__19b_}
  S_{12} = 3 (\vec{\sigma}_{1}. \hat{r}) (\vec{\sigma}_{2} . \hat{r}) - (\vec{\sigma}_{1} . \vec{\sigma}_{2}), \quad L_{12}=\left(\delta_{\ell j} + \vec{\sigma}_{1}.\vec{\sigma}_{2} \right) L^{2} - (\vec{L} . \vec{S})^{2},
\end{equation}
and
\begin{equation} \label{GrindEQ__19c_}
 \left\{\begin{array}{l} {V_{c}(r) = 0.08 \left(\frac{1}{3} m_{pi} \right) \left(\vec{\tau}_{1} . \vec{\tau}_{2} \right) \left(\vec{\sigma}_{1} . \vec{\sigma}_{2} \right) Y(x) \left[1+ a_{c}\ Y(x)+ b_{c}\ Y^{2}(x)\right]}, \\ {V_{t}(r) = 0.08 \left(\frac{1}{3} m_{pi} \right) \left(\vec{\tau}_{1} . \vec{\tau}_{2} \right) Z(x) \left[1+a_{t}\ Y(x) + b_{t}\ Y^{2}(x)\right]}, \\ {V_{ls}(r)=m_{pi} \ G_{ls}\ Y^{2}(x) \left[1+b_{ls} Y(x)\right]}, \\ {V_{ll}(r)=m_{pi} \ G_{ll} x^{-2} Z(x) \left[1+a_{ll}\ Y(x)+ b_{ll}\ Y^{2}(x)\right]}, \end{array}\right.
\end{equation}
in which $m_{pi}$, $x$ and $M$ are the pion mass (139.4 MeV), the internucleon distance measured in the units of the pion Compton's wavelength ($r_0=1.415 fm$), and the nucleon mass (taken to be $6.73\mu$), respectively. Note also that $x=\mu r$, $\mu=m_{pi} c/{\hbar}=r_0^{-1}$ with respect to Eq. (\ref{GrindEQ__1_}), and that
\begin{equation} \label{GrindEQ__19d_}
  Y(x)=\frac{e^{-x}}{x}, \quad  Z(x)=\left(1+\frac{3}{x}+\frac{3}{x^{2}} \right) Y(x).
\end{equation}
We should note that the quadratic spin-orbit potential was mainly introduced to describe \emph{np} data satisfactorily. For the $r$ large enough, $V_{c}(r)$ and $V_{t}(r)$ reduce to the well-known OPEP with the pseudovector coupling constant of 0.08. The coefficients $a_{c}$, $b_{c}$, $a_{t}$ and $b_{t}$ represent the potential diversion from OPEP at small $r$'s. $G_{ls}$ is the strength of the short ranged spin-orbit potential $V_{ls}(r)$ and is depended on the parity of state. $G_{ll}$, as the strength of $V_{ll}(r)$, originated from special evaluations, is determined phenomenologically. All the coefficients are determined from the detailed fit to scattering data and are given in the original paper \cite{HamadaJohnston}. The hard cores are considered for all states with their radius at $x_{c}=0.343$. The HJ potential, as originally proposed, included a strong long-range quadratic spin-orbit potential in triplet-even states, and also a strong short-range spin-orbit potential in triplet $(\ell =j)$-odd states, where it is known that the latter does not exist. So, the potential for triplet-odd states was modified as follows \cite{HamadaJohnstonM}: It was defined to be - 0.26744 $m_{pi}$ around $x_{c} < x \le 0.487$ and by above standard relations for $x > 0.487$. The values of the binding energy, electric quadratic-moment, effective-range, D-state probability and the asymptotic D-wave to S-wave ratio of deuteron were determined by the potential to be 2.226 MeV, 0.285 $fm^2$, 1.77 $fm$, 6.97 $\%$ and $A_D/A_S=0.02656$, respectively.

An improvement of HJ potential was made in \cite{Bressel} (we call it \textbf{Massachusetts-group} potential) to replace mainly the HJ hard cores (for $x\leq x_c$) by finite square-well cores. Outside the square-well radius (for $x> x_c$), the potential is the same as HJ except for a few changes in parameters such as considering the pion mass differences, and that the $a_c$ values of the singlet-even and triplet-odd states as well as the triplet-odd $b_c$ are changed slightly. The pion mass splitting leads to charge-independent breaking (CIB) while CS is still preserved. Now, $m_{pi}$ is replaced by the effective pion mass and $x_c=0.4852$, which in turn implies the larger core-radius of 0.7 \emph{fm}. Describing NN scattering data and deuteron properties with the potential were good. Indeed, the main aim to form the latter potential was to show that the hard cores were not necessary since all data could be described by the finite soft-core potentials.

\subsection{Yale-Group Potential}
The Yale-group potential \cite{Yale} is a \emph{pp}+\emph{np} phenomenological potential similar to HJ potential \cite{HamadaJohnston} that is fitted to its time phase parameters as well. There, an OPEP is included directly and the quadratic spin-orbit potential is considered in a somewhat different form than that of HJ. The whole NN potential reads
 \begin{equation} \label{GrindEQ__20e_}
V =V^{(2)}_{OPEP} + V_{c}(r) + V_{t}(r) S_{12} + V_{ls}(r) \vec{L}. \vec{S} + V_{ql}(r) \big[Q_{12} - (\vec{L} . \vec{S})^{2} \big],
\end{equation}
where
\begin{equation} \label{GrindEQ__20f_}
\left[Q_{12} - (\vec{L}. \vec{S})^{2} \right] = (\vec{L}. \vec{S})^{2} + \vec{L}. \vec{S} - \vec{L}^{2},
\end{equation}
and
\begin{equation} \label{GrindEQ__20g_}
V^{(2)}_{OPEP} = \left(\frac{g_{pi}^{2}}{12} \right) m_{pi}c^2 \left(\frac{m_{pi}}{M} \right)^{2} (\vec{\tau}_{1} . \vec{\tau}_{2}) \left[(\vec{\sigma}_{1} . \vec{\sigma}_{2}) + S_{12} \left(1+\frac{3}{x} + \frac{3}{x^{2}} \right)\right] \frac{e^{-x}}{x}.
\end{equation}
This OPEP is used for the distances larger than nearly 3$fm$, with the same parameter definitions as in HJ potential. For the coupling constant, $g_{pi}^{2}/14= 0.94$ is used in singlet-even states and 1 elsewhere. For singlet-even and triplet-odd states, the neutral-pion mass ($m_{pi}=m_{\pi_{0}}$) is used while for singlet-odd and triplet-even states, a mean of the charged- and neutral-pion masses [$m_{pi}=\left(m_{\pi_{0}}+2m_{\pi{\pm}}\right)/3$] is used. The hard-core radius is considered at $x_{c}=0.35$, and except in the OPEP part, all the radial functions $V_{c}$, $V_{t}$, $V_{ls}$ and $V_{ql}$ are taken as
\begin{equation} \label{GrindEQ__20h_}
V= \sum_{n=1}^{7} a_{n} \frac{e^{-2x}}{x^{n}}.
\end{equation}
The potential's parameters are determined by fitting to data for various states and involved potentials. It is also notable that HJ and Yale potentials are OPEP for $L>5$, and that the Yale potential sets $V_{ls}=0$ for $J>2$.

\subsection{Reid68 and Reid-Day Potentials}
\subsubsection{Reid68 Potential}
Among the failures of HJ \cite{HamadaJohnston} and Yale \cite{Yale} hard-core potentials were that they could not reproduce reasonable results when applying to many-body calculations. It appeared that the Reid soft-core potentials \cite{Reid68} were better. The Reid potentials are static and local phenomenological potentials similar to those of HJ and Yale. Reid determined the potential for each two-nucleon state independent of the other states. So, one may suppose that this approach is problematic in that, with many two-nucleon states each with its own potential, fitting the experimental data could be probably meaningless. But, because the highest energy in the analyses was about 350 MeV, just the two-nucleon states with $J\le2$, which are more important in nuclear calculations, were considered in practice.\\
Reid used only a central potential in the singlet- and uncoupled triplet-states while, for the coupled triplet-states, he used
\begin{equation} \label{GrindEQ__22_}
V = V_{c}(r) + V_{t}(r) S_{12} + V_{ls}(r) \vec{L}. \vec{S},
\end{equation}
which has the central, tensor and usual spin-orbit components. For the LR part, he used the OPEP of (\ref{GrindEQ__20g_}) as a tail attached to the potential, with $g_{pi}^{2}=14$, $m_{pi}=$ 138.13 MeV, $M=$ 938.903 MeV and $\mu=0.7 fm^{-1}$. On the other hand, to remove the $x^{-2}$ and $x^{-3}$ behaviors at small distances, an SR potential was subtracted from the tensor part of the potential. For the MR's, the potentials were expressed as the sums of the Yukawa's functions of $e^{-nx}/x$, where $n$ was an integer. The SR repulsions were also some combinations of the severe hard-core and the Yukawa soft-core potentials--It is mentionable that the criterion for a potential to be soft-core is that the wave functions do not vanish in nonzero radiuses. For the hard-core radius, when needed, the radiuses of $x_c\leq 0.1$ could be used there. One should, of course, note that because of fitting the potentials to the energies often below 350 MeV, finding a unique formalism for the SR part was almost difficult. Finally, it is notable that the Reid potentials did not describe well some of the scattering data and deuteron properties at that time. It was also hinted the need for velocity-dependence and nonlocality in NN potentials, imposed by experimental data.

\subsubsection{Reid-Day Potential} \label{4.5.2}
In 1980, B. D. Day \cite{Reid68-Day} expanded the Reid68 soft-core potentials up to the higher partial waves to solve three-body equation in nuclear matter calculations. In fact, he used three two-nucleon potentials in calculations. The first one (called $V_2$) was just the central part of the Reid68 potential in ${}^3S_1-{}^3D_1$ channel for all states. The second one (called $V_6$(Reid)) had four forms for the four ($S,T$) states. Indeed, in the latter case, for all $S=0$ states, just two central $V_c(r)$ potentials (Reid68 ${}^1S_0$ and ${}^1P_1$ for $T=1$ and $T=0$ respectively) were used; meanwhile for all $S=1$ states, just two central $V_c(r)$ and two tensor $V_t(r)$ potentials (Reid68 ${}^3P_2-{}^3F_2$ and ${}^3S_1-{}^3D_1$ for $T=1$ and $T=0$ respectively) were used. The third one (called Full-Reid potential that we call Reid-Day potential) used the original Reid68 potentials for all $J\le2$ states; meanwhile for the states with $J\geq3$, he set up the potentials based on the Reid68 ones almost roughly. Clearly, for the states up to $J=5$, the potential structures were similar to the original Reid68 ones. For example, in the coupled sate of ${}^3D_3-{}^3G_3$, he used
\begin{equation} \label{GrindEQ__22aa_}
 V_{c}(r)= -10.463\ Y(x)-103.4\ Y^2(x)-419.6\ Y^4(x)+9924.3\ Y^6(x),
\end{equation}
\begin{equation} \label{GrindEQ__22bb_}
 V_{t}(r)= -10.463\ \left[Z(x)-\left(\frac{12}{x}+\frac{3}{x^{2}} \right) Y^4(x) \right]+351.77\ Y^4(x)-1673.5\ Y^6(x),
\end{equation}
\begin{equation} \label{GrindEQ__22cc_}
 V_{ls}(r)= 650\ Y^4(x)-5506\ Y^6(x),
\end{equation}
where $x=0.7 r$, and $r$ is the internucleon distance measured in \emph{fm} as usual. For all other not clearly mentioned states, he used the $V_6$(Reid) potentials. Therefore, that new expansion was not based on any fundamental underlying argument on NN interaction, and was just to sake of applying the wanted potentials in some nuclear calculations.

\subsection{Partovi-Lomon Potential}
Partovi-Lomon potential \cite{PartoviLomon} is among the early NN potentials based on quantum field theory methods. The advantages of boson exchanges and multi-pion resonances especially in short distances were considered. They also considered some TPEP's and OPEP's to improve the quality of previous similar potentials. The resultant (Schrodinger-equation) potential was originated from reducing BS equation to an LS equation.
In fact, by starting from a relativistic tow-body equation, they arrived in a nonrelativistic LS equation and presented a potential as a solution of the integral equation. Then, they tried to build r-space potentials with momentum operators, resulting in a potential composed of the central, spin-orbit, tensor and spin-spin parts. Contributions of $\rho, \omega, \eta$ bosons were included and then, by using experimental masses and coupling parameters, the complete potential was calculated. The potential has some likeness to Hamada-Johnston potential \cite{HamadaJohnston}, and appears to dissolve some of the problems hinted in Reid68 potential \cite{Reid68}.

\subsection{Paris-Group Potentials}
Paris-group potentials are based on dispersion relations and field-theoretical techniques. In their first major potential, Paris72 \cite{Cottingham}, they included some TPE contributions for the potentials by considering pion-nucleon phase shifts and pion-pion interactions. They also included $\pi$- and $\rho$-meson exchanges. Indeed, for the LR and MR parts, the accurate potentials of $\pi+2\pi+\rho$ exchanges were used; while for the SR part of $r\leq 0.8 fm$, a constant soft-core potential was used. The Paris72 potential includes the central, spin-spin, spin-orbit, tensor and quadratic spin-orbit components for each isospin state. Fitting the potential to the \emph{pp}, \emph{np} scattering data of the Livermore-group \cite{Livermore1969} of 1969, needed 12 adjustable parameters. The potential described the data with similar qualities as the phenomenological potentials of HJ and Yale-group with more adjustable parameters. Describing the LR and MR parts by the potentials was more sensible. Nevertheless, describing short distances was not satisfactory besides the problems in its applications to many-body nuclear calculations.\\
The next improved version of the potential came in 1979, named as "parameterized Paris potential" or Paris79 potential. In that version, they employed a unique expression for the whole potential, which was a sum of the Yukawa's functions that had simple forms in both configuration and momentum spaces. Indeed, those 12 local Yukawa functions could provide a semi-phenomenological description of the Paris72 potential. Meanwhile, the older contributions for the LR and MR parts were used yet. The potentials for both values of isospin ($T=1,0$) have the following nonrelativistic forms in r-space:
\begin{equation} \label{GrindEQ__22aa_}
V(\vec{r},p^2) = V_0(r,p^2) SS_1 + V_1(r,p^2) SS_2 + V_{ls}(r) \vec{L}. \vec{S} + V_{t}(r) S_{12} + V_{q}(r) Q_{12},
\end{equation}
where
\begin{equation} \label{GrindEQ__22bb_}
SS_1= \left(\frac{1-\vec{\sigma}_{1}.\vec{\sigma}_{2}}{4}\right), \qquad  SS_2= \left(\frac{3+\vec{\sigma}_{1}.\vec{\sigma}_{2}}{4}\right).
\end{equation}
Clear forms for the velocity-dependent functions of $V_0$ and $V_1$, and especial forms for the Yukawa functions of $V_{ls}, V_{t}, V_{q}$, as well as coupling constants and  other parameters, under special conditions, are given in the original paper \cite{Paris2}. The potential in p-space, by Fourier transform of (\ref{GrindEQ__22aa_}), reads
\begin{equation} \label{GrindEQ__22cc_}
\tilde{V}(\vec{p}_i,\vec{p}_f) = \tilde{V}_0(\vec{p}_i,\vec{p}_f) SS_1 + \tilde{V}_1(\vec{p}_i,\vec{p}_f) SS_2 + \tilde{V}_{ls}(k^2) \tilde{LS}_1 + \tilde{V}_{t}(k^2) \tilde{S}_{12} + \tilde{V}_{q}(k^2) \tilde{Q}_{12},
\end{equation}
where
\begin{equation} \label{GrindEQ__22dd_}
 \tilde{LS}_1=i \vec{S}.\vec{n}, \quad \tilde{S}_{12}=\left[ k^2 (\vec{\sigma}_{1}.\vec{\sigma}_{2})-3(\vec{\sigma}_{1}.\vec{k})(\vec{\sigma}_{2}.\vec{k}) \right], \quad \tilde{Q}_{12}= (\vec{\sigma}_{1}.\vec{n}) (\vec{\sigma}_{2}.\vec{n}),
\end{equation}
with the definitions
\begin{equation} \label{GrindEQ__22dddd_}
\vec{k}=\vec{p}_{f}-\vec{p}_{i}, \quad  \vec{q}=\frac{1}{2} (\vec{p}_{f}+\vec{p}_{i}), \quad \vec{n}=\vec{p}_{i} \times \vec{p}_{f}=\vec{q} \times \vec{k},\quad \vec{S}=\frac{1}{2} (\vec{\sigma}_1+\vec{\sigma}_2),
\end{equation}
and especial Fourier transformations for the velocity-dependent central and noncentral components. Note that $\vec{p}_f$ and $\vec{p}_i$ are in- and outgoing two-nucleon momentum transfers, respectively. The results for fitting its time \emph{pp} and \emph{np} scattering data were good up to the energies about 350 MeV except for the low energies below about 13 MeV.\\
In the next related work \cite{Paris3A}, in 1984, a separable representation of the Paris79 potential, through using a special method, was presented. That representation offered a good approximation of the on-shell and off-shell properties of the potential. In 1985, another adjustment of the separable representation for the states of $^{1}S_{0}$ and $^{3}P_{0}$ was performed \cite{Paris3} to improve the previous problem in representations.

\subsection{Stony-Brook Potential}
Stony-Brook potential is also among the original NN potentials based on dispersion relations and field theory. The group included the contributions from $\pi$, $\omega$ and $\pi\pi$ exchanges. They tried to set up a local and energy-dependent regularized potential in p-space by using the field theoretical elastic NN scattering amplitudes. Indeed, by solving Blankenbecler-Sugar (BbS) equation, by using a proposed interaction potential, they estimated NN phase parameters. Some phenomenological parameters were adjusted to get satisfactory results compared with experimental data. The short-range repulsion was weaker than the phenomenological potentials such as HJ and Reid68 as well as its time OBE potentials, mainly because of $\omega$-boson exchange. It is also mentionable that describing experimental data and deuteron properties by the potential was not as good as the phenomenological potentials at that time. For detailed studies on the potential and the techniques used there, look at \cite{3} and \cite{Jackson}.

\subsection{dTRS Super-Soft-Core Potentials}
The earlier super-soft-core (SCC) potential, called dTS potential in \cite{deTourreil0}, described physical observables better than the harder-core potentials. But in dTS potential, only the OPEP was purely theoretical while in the next SCC potential \cite{deTourreil}, called dTRS B potential, by the same group, more theoretical components were added. In addition, dTRS B improved fitting NN scattering data besides giving better results for nuclear-matter and many-body calculations rather than dTS potential.\\
In dTRS B, the OBEP's, because of $\pi, \rho, \omega$ exchanges, were considered directly; and the remaining contributions for the MR part due to the other probable OBE's and TPE's were parameterized phenomenologically by special OBE functions that we mention below. In other words, the OBEP functions with 32 free ranges and amplitudes were used instead. In the SR part, below about 1 $fm$, the potential components were regularized, and the core region phenomenological potentials were chosen so that the previous results for the LR and MR parts could not be disrupted. The general form of dTRS B potential in ($S,T$) space, reads
 \begin{equation} \label{GrindEQ__22ee_}
V = V_{c}(r) + V_{t}(r) S_{12} + V_{ls}(r) \vec{L}. \vec{S} + V_{q}(r) Q_{12} + V_{ll}(r) L^2,
\end{equation}
in which the $L^2$ potential term is to account the difference between ${}^1S_0$ and ${}^1D_2$ potentials. In the LR part, the radial dependence of every component reduces to the OPE contribution of $V_c, V_t$. For the phenomenological OBEP's, they used the radial functions of $V_{c}(r),...$ as linear combinations of the following functions:
\begin{equation} \label{GrindEQ__22ff_}
\begin{split}
& \ \ \ \ \ \ \ \ \ \ \ Y_c(x)=\frac{e^{-x}}{x}=Y(x), \quad Y_{ls}(x)=(\frac{1}{x}+\frac{1}{x^2}) Y(x), \\
&  Y_q(x)=\left(\frac{1}{x^2}+\frac{2}{x^3}\right) Y(x), \quad Y_{t}(x)=\left(1+\frac{3}{x}+\frac{3}{x^{2}} \right) Y(x)=Z(x),
\end{split}
\end{equation}
and
\begin{equation} \label{GrindEQ__22ff22_}
F(r)=\frac{(1.2r)^{20}}{[1+(1.2r)^{20}]^{20}},
\end{equation}
for various states in subspaces of ($S,T$) independently. $x$ is the same as that we already used in HJ, Yale and Reid68 potentials except that we have to use $m_\lambda$ with $\lambda=pi, \rho, \omega$ here instead of $m_{pi}$ there. The constant coefficients in the linear combinations are, in turn, some functions of the involved masses and other parameters determined by fitting to experimental data and from other sources. In the last relation, $F(r)$ is a step-like function that is used as a cutoff to define the core region. Meanwhile, for \emph{pp} scattering, $M=M_p, m_{pi}=m_{\pi_{0}}$; and for \emph{np} scattering, $M=M_r=2M_p M_n/(M_p+M_n)$ and $m_{pi}=\left(m_{\pi_{0}}+2m_{\pi_{\pm}}\right)/3$ are used.\\
Describing the experimental data below 350 MeV and NN bound states were good as the other phenomenological potentials at that time. Meanwhile, although describing  nuclear matter and some many-body results by using TRS B potential, as seen further in \cite{deTourreil2}, were reasonable, more improvement were yet required. It is also mentionable that there are some likenesses between this and Paris72 potential \cite{Cottingham} framed first.

It is good here to mention another potential built in 1981, with a similar meson content and operators as dTRS potentials, which we call it \textbf{Melbourne potential} \cite{Petris9}. In fact, it is especially a \emph{np} potential that includes the OBE's of $\pi, \rho, \omega$ and TPE of $2\pi$ next to some phenomenological features to reproduce experimental elastic scattering data and neutron properties (mainly its binding energy) and low-energy parameters (mainly scattering-length). There, a special form function was used for each meson contributing to a special energy range. Reproducing the data and deuteron properties as well as the basic results from nuclear-structure and -matter calculations were satisfactory.

\subsection{Funabashi Potentials}
Funabashi potentials are among meson-exchange potentials based on field-theoretical methods. They included the OBEP of $\pi, \rho, \omega, \eta$ and the scalar mesons of $\delta, \sigma$ for LR and MR parts. For the core region, they included the hard cores, Gaussian soft cores and velocity-dependent cores. Indeed, the potentials are nonstatic OBEP's with retardation in r-space. The nonstationary, mainly because of recoiling, is considered by including the spin-orbit, quadratic spin-orbit and velocity-dependent terms; whereas the retardation of the meson propagations causes the off-energy shell effects that in turn contribute to two-nucleon processes, and are even more important in many-body systems.\\
The general form of the Funabashi OBEP's in r-space reads \cite{Funabashi00}
\begin{equation} \label{GrindEQ__22gg_}
V =V_{core}+ V_{c}(r) + V_{t}(r) S_{12} + V_{ls}(r) \vec{L}. \vec{S} + V_{qll}(r) \grave{Q}_{12} + V_{ll}(r) L^2-\frac{1}{M} \left[\nabla^2 V_p+V_p\nabla^2\right],
\end{equation}
in which
\begin{equation} \label{GrindEQ__22hh_}
\grave{Q}_{12}=Q_{12}-\frac{2}{3}L^2 \vec{S}, \quad V_i(r)=U_i(r)+R_i(r), \quad i=c,t,ls,qll,ll,p,
\end{equation}
where $U_i(r)$ and $R_i(r)$ stand for the usual Yukawa and Retarded potential functions, respectively. These functions are in turn expressed as combinations of the functions in (\ref{GrindEQ__22ff_})
where the involved masses, coupling constants and other parameters are used in combinations as coefficients in various two-nucleon states and for the various included mesons. The core potential also reads
\begin{equation} \label{GrindEQ__22jj_}
V_{core}= V_{core}^c(r) + V_{core}^{ls}(r) \vec{L}. \vec{S},
\end{equation}
where, depended on the case, three different cores are included. a) The hard core (step-like) potential (OBEH) plus a spin-orbit core as
\begin{equation} \label{GrindEQ__22kk_}
 V_{core}^c(r)=\left\{\begin{array}{l} 8, \ \ \ \  r\leq r_c, \\ 0, \ \ \ \  r> r_c, \end{array}\right., \quad V_{core}^{ls}(r)=-V_{(0)}^{ls} \exp{\left[-(r\diagup {r_{ls}})^2\right]}.
\end{equation}
b) The Gaussian soft core (OBEG) plus a special spin-orbit core as
\begin{equation} \label{GrindEQ__22ll_}
 V_{core}^c(r)= V_{(0)}^{G} \exp{\left[-(r\diagup {r_{G}})^2\right]}, \quad V_{core}^{ls}(r)=\frac{1}{M^2}\frac{1}{r}\frac{\partial V_{core}^c(r)}{\partial r}.
\end{equation}
c) The velocity-dependent core (OBEV) plus a spin-orbit core as
\begin{equation} \label{GrindEQ__22mm_}
\begin{split}
 & V_{core}^c(r)= \frac{p^2}{M}\phi(r)+\phi(r)\frac{p^2}{M}, \quad \phi(r)=\phi_0^p \exp{\left[-(r\diagup {r_{p}})^2\right]}, \\
 & V_{core}^{ls}(r)=-V_{(0)}^{ls} \exp{\left[-(r\diagup {r_{ls}})^2\right]}.
\end{split}
\end{equation}
In addition, to remove singularities and to make OBEP's in the core region, the $V_i(r)$'s in (\ref{GrindEQ__22hh_}) are multiplied by the following cutoff factor
\begin{equation} \label{GrindEQ__22nn_}
F_i(r)=1-\exp{\left[-(r\diagup {r_{cc}})^2\right]^n}, \quad n=\left\{\begin{array}{l} 1, \ when \ i=c,t,ls, \\ 6, \ when \ i=qll,ll, \end{array}\right. .
\end{equation}
The parameters of $r_c, r_{cc}, r_{ls}, r_{G}, r_p, \phi_0^p, V_{(0)}^{ls}, V_{(0)}^{G}, .... $ are the constants properly chosen for the potentials. It is also mentionable that, to have nonrelativistic potentials, the higher-order terms than $p^2/M^2$ are avoided, where $\vec{p}=\vec{p}_i$ here is the nucleon momentum.\\
Next, in \cite{Funabashi01}, the velocity-dependent tensor potentials were included to discuss better nonstatic effects. So, the improved potential reads
\begin{equation} \label{GrindEQ__22oo_}
V^{(2)} =V-\frac{1}{M}\left(\left[\nabla^2 V_{pt} S_{12}+V_{pt} S_{12}\nabla^2\right]\right).
\end{equation}
In addition, the core potentials were modified, rather than those in the first version of \cite{Funabashi00}, to improve the phase shifts of ${}^3P_3$ state by including the attractive spin-orbit cores in (\ref{GrindEQ__22jj_}), which were in turn set to zero in the first potentials. With these improvements, the properties of neutron- and nuclear-matter were evaluated. Describing experimental scattering data and low-energy parameters as well as deuteron properties with the latter potential was better than the first one. More improvements to give even better results in nuclear structure calculations were then done in \cite{Funabashi02}. \\
Later, a development of the potentials was given in \cite{Funabashi2}. In fact, it was shown that the radial dependence of the OBEP's, which was smooth and finite at origin, could be represented by a superposition of special Gaussian functions. In other words, the Yukawa functions in Funabashi's potentials were expanded as
\begin{equation} \label{GrindEQ__22pp_}
Y(\mu r)=\frac{e^{-\mu r}}{\mu r}=\sum_{n=1}^N a_n \ \exp{\left[-(r\diagup {r_{n}})^2\right]},
\end{equation}
where the coefficients of $a_n$ were determined by fitting the data; and for $N$ and $r_{n}$, some special finite values were chosen. As the authors have claimed, the new potentials give better fit of NN scattering data.

\subsection{Urbana-Group Potentials}
UrbanaV14 (Urb81) potential \cite{UV14} is a charge-independent fully phenomenological potential including the operators of central, spin-spin, tensor, spin-orbit, centrifugal, centrifugal spin-spin with general dependence on isospin. Besides an LR OPE part and a representation of MR part as TPE's with 14 parameters, the SR part is described by two Woods-Saxon potentials with free parameters fitted to experimental data. The whole potential reads
\begin{equation} \label{GrindEQ__23aa_}
V = \sum_{i=1}^{n} V^{i} O_i,
\end{equation}
in which the fourteen operators ($n=14$ here) read
 \begin{equation} \label{GrindEQ__23bb_}
 \begin{split}
O_{i= 1,...,14}=& 1, \vec{\sigma}_{1} .\vec{\sigma}_{2}, \vec{\tau}_{1} .\vec{\tau}_{2}, \big(\vec{\sigma}_{1} .\vec{\sigma}_{2} \big) \big(\vec{\tau}_{1} .\vec{\tau}_{2} \big), S_{12} , S_{12} \big(\vec{\tau}_{1} .\vec{\tau}_{2} \big), \big(\vec{L}.\vec{S}\big), \big(\vec{L}.\vec{S}\big) \big(\vec{\tau}_{1} .\vec{\tau}_{2} \big),\\
& L^{2}, L^{2} \big(\vec{\sigma}_{1} .\vec{\sigma}_{2} \big), L^{2} \big(\vec{\tau}_{1} .\vec{\tau}_{2} \big), L^{2} \big(\vec{\sigma}_{1} .\vec{\sigma}_{2} \big)\big(\vec{\tau}_{1} .\vec{\tau}_{2} \big), \big(\vec{L}.\vec{S}\big)^{2} , \big(\vec{L} .\vec{S}\big)^{2} \big(\vec{\tau}_{1} .\vec{\tau}_{2} \big),
\end{split}
\end{equation}
and the radial potentials are
\begin{equation} \label{GrindEQ__23cc_}
V^{i} = V_{\pi}^{i}(r) + V_{M}^{i}(r) + V_{S}^{i}(r),
\end{equation}
where $V_{\pi}, V_{I}, V_{S}$ stand for the pion-exchange potential, MR  and SR potentials, respectively. Further, we should note that the first eight operators of (\ref{GrindEQ__23bb_}) are obtained from fitting the phase shifts of $\ell <4$ up to the laboratory energies of 425 MeV, and deuteron properties. The next six "quadratic-$L$" operators are introduced to do many-body calculations with the potentials and have almost weak effects. We shorten the operators as $c, \sigma, \tau$, $\sigma \tau$, $t, t\tau$, $ls, ls \tau$, $ll, ll \sigma$, $ll \tau, ll \sigma\tau$, $ls2, ls2\tau$, to simplify their using, from now on.

The LR OPEP of $V_{\pi}^{i}(r)$ is nonzero just for $i=\sigma \tau, t\tau$ with
\begin{equation} \label{GrindEQ__44a_}
V_{\pi}^{\sigma \tau}(r)=3.488 \frac{e^{-0.7r} }{0.7r} \left(1-e^{-cr^{2}} \right),
\end{equation}
\begin{equation} \label{GrindEQ__44b_}
V_{\pi }^{t\tau}(r) = 3.488 \left(1 + \frac{3}{0.7r} + \frac{3}{(0.7r)^{2}} \right) \frac{e^{-0.7r}}{0.7r} \left(1-e^{-cr^{2}} \right)^{2} = 3.488 T_{\pi}(r),
\end{equation}
with a note that $x=\mu r$ with $\mu=0.7 fm^{-1}$ is considered. The cutoff parameter of $c$ is obtained by fitting the experimental phase shifts, and that the $1/r$ and $1/r^{3}$ singularities of OPEP's are removed. A remarkable point is that $(1-e^{-cr^{2}})^{2}$, as argued in \cite{Green}, simulates $\rho$-meson exchange effect. Another point is that because nucleon is not a point source, the two-nucleon interaction should not have the singular behavior of $1/r$ at small distances.\\
The MR potential of $V_{M}^{i}(r)$ is considered as
\begin{equation} \label{GrindEQ__45_}
V_{M}^{i}(r) = S^{i} T_{\pi}^{2}(r).
\end{equation}
With respect to the $T_{\pi}^{2}(r)$ included, this potential is usually owned to the second-order OPEP's. This form of $V_{M}^{i}(r)$ is suitable to include three-nucleon (3N) interactions as argued in \cite{Green2}. Besides, the strengths of $S^{i}$ are determined by fitting experimental phase shifts.\\
For the SR potential of $V_{S}^{i}(r)$, in contrast to the custom method where Yukawa functions are used, here a sum of two Woods-Saxon potentials is considered as
\begin{equation} \label{GrindEQ__46_}
  V_{S}^{i}(r) =S_1^{i} W_1(r) + S_2^{i} W_2(r),
\end{equation}
where
\begin{equation} \label{GrindEQ__46a_}
 W_1(r)=\left(1+\exp \left(\frac{r-R_1}{a_1} \right)\right)^{-1}, \quad W_2(r)=\left(1+\exp \left(\frac{r-R_2}{a_2} \right)\right)^{-1}.
\end{equation}
For all $i$'s, except for $ls$ and $ls \tau$, a good fit of data is achieved with $S_2^{i}=0$. \\
There are some likenesses between parameterizing the Urb81 potential with those used in Hamada-Johnston \cite{HamadaJohnston}, Yale-group \cite{Yale}, Reid68 \cite{Reid68} and also Bressel et al. \cite{Bressel} potentials. The values of free parameters are obtained mainly by fitting the \emph{np} phase shifts by Arndt et al. \cite{Arndt3} and its time analysis by Bugg et al. \cite{Bugg}, with some differences and adjustments. Describing scattering data and deuteron properties with Urb81 potential are satisfactory with similar results as the Reid68 and Paris79 \cite{Paris2} potentials. For more details see \cite{UV14}.

\subsection{Argonne-Group Potentials}
\subsubsection{ArgonneV14 and ArgonneV28 Potentials}
The basic potential of Argonne-group, ArgonneV14 (Arg84) potential \cite{AV14}, has a similar structure with UrbanaV14 (Urb81) \cite{UV14} potential with a few differences. The first difference is that the used $\pi NN$ coupling is larger than that used in Urb81 potential. Second, in Arg84, contrary to Urb81 potential, $S^i_2$ is nonzero for $i=t, t\tau$. Third, as a probable result of the former constraint, that is no need to insert a second SR Woods-Saxon function for $i=ls, ls \tau$ as is in Urb81 potential. As a result, in high-energies (low-distances), the phase shifts are fitted well and especially the D-state of deuteron takes more contribution than that in Urb81 potential. Further, the effects of the six quadratic-$L$ operators are confirmed in some nuclear structure calculations.
The potential was fitted to the phase-shift analyses of Arndt and Roper in 1981 (an update of the analyses in \cite{Arndt3}). Still, in the energy range of 25-350 MeV, the Arg84 potential provides an improvement over Urb81 potential.\\
It is good to mention another potential of Argonne-group, called ArgonneV28 \cite{AV14}, framed simultaneously with ArgonneV14 potential. It includes the $\Delta(1232)$ degrees of freedom, which play important roles in both TPE processes in the MR part of NN systems as well as TPE and repulsive parts of 3N systems. The effects because of including these degrees of freedom are shown by 14 extra operators next to the 14 operators used in Arg84 potential. From these 14 extra operators, 12 transition operators are for all $\pi N\Delta $ and $ \pi \Delta \Delta$ couplings while 2 central operators are for $N \Delta$ and $\Delta \Delta$ channels. The extra operators are so chosen that no other free parameters than those used in Arg84, but the coupling constants of $\pi N\Delta $ and $ \pi \Delta \Delta$, are required to fit the experimental scattering data. In general, ArgonneV28 potential has a more complicated structure and gives better results especially in many-body calculations.

\subsubsection{ArgonneV18 Potential} \label{4.12.2}
ArgonneV18 (Arg94) NN potential \cite{AV18} is an improved and updated version of Arg84 NN potential \cite{AV14}. It addition to 14 operators of Urb81 and Arg84 potentials, it includes three charge-dependent and one charge-asymmetry operators next to a complete electromagnetic interaction. Arg94 potential has the following general form:
\begin{equation} \label{GrindEQ__47_}
V=V_{EM}+V_{N}=V_{EM}+V_{\pi}+V_R,
\end{equation}
where $V_{\pi}$ is for an LR OPEP, $V_R$ is for  MR and SR parts (called the Remaining part), and $V_{EM}$ is for electromagnetic (EM) part.

The EM part, in turn, reads
\begin{equation} \label{GrindEQ__48_}
V_{EM}=V_{C1}(r)+V_{C2}(r)+V_{DF}(r)+V_{VP}(r)+V_{MM}(r),
\end{equation}
where the terms with the indices $C1, C2, DF, VP$ and $MM$ stand for one-photon, two-photon, Darwin-Foldy, vacuum-polarization and magnetic-moment interactions. In these interactions, some short-range terms and the effects due to finite size of nucleon are also included. The terms of $V_{C2}, V_{DF}, V_{VP}$ are used just for \emph{pp} scattering while the other terms have its own forms for each three scattering cases; and that for \emph{nn} scattering, just $V_{MM}$ is used. The various EM potentials are given through some combinations of the following functions, with masses, coupling constants, coefficients and other constants determined from other sources or from experimental data,
\begin{equation} \label{GrindEQ__50a_}
F_{c}(r)= 1-\left(1+ \frac{11}{6} x + \frac{3}{16} x^{2} +\frac{1}{48} x^{3} \right) Y^{(0)},
\end{equation}
\begin{equation} \label{GrindEQ__50b_}
F_{\delta}(r)= b^{3} \left(\frac{1}{16} + \frac{1}{16} x + \frac{1}{48} x^{2} \right) Y^{(0)},
\end{equation}
\begin{equation} \label{GrindEQ__50c_}
F_{t}(r)= 1- \left(1+ x +\frac{1}{2} x^{2} + \frac{1}{6} x^{3} + \frac{1}{24} x^{4} + \frac{1}{44} x^{5} \right) Y^{(0)},
\end{equation}
\begin{equation} \label{GrindEQ__50d_}
F_{\ell s}(r)= 1- \left(1+ x +\frac{1}{2} x^{2} + \frac{7}{48} x^{3} + \frac{1}{48} x^{4} \right) Y^{(0)},
\end{equation}
where $x=br$, $b = 4.27 fm^{-1}$. They are related as
\begin{equation} \label{GrindEQ__50e_}
F_{\delta } = -\nabla^{2} \left(\frac{F_{c}}{r} \right), \quad F_{t} = \left(\frac{F_{c}}{r} \right)^{''} - \left(\frac{F_{c}}{r} \right)^{'}/r, \quad F_{\delta}  = -\nabla^{2} \left(\frac{F_{c}}{r} \right),
\end{equation}
and for a point-like nucleon go to 1. In other words, these SR functions show the finite-size of the nucleon charge distribution with a dipole form-factor. It is also mentionable that for \emph{pp} case, $V_{C1}, V_{C2}, V_{VP}$ are in terms of $F_{c}$, with the mentioned adjusted parameters; $V_{DF}$ is in terms of $F_{\delta}$, while $V_{C1}$ is in terms of
\begin{equation} \label{GrindEQ__53_}
F_{np}(r) =\frac{b^{2}}{384} \left(15x + 15x^{2} + 6x^{3} + x^{4} \right) Y^{(0)}.
\end{equation}
$V_{MM}(pp)$ includes $F_{\delta}, F_{t}, F_{ls}$ together with spin-spin ($\vec{\sigma}_{1} .\vec{\sigma}_{2}$), spin-orbit ($\vec{L}.\vec{S}$) and tensor ($S_{12}$) operators, while $V_{MM}(np)$ includes the same functions and operators as in \emph{pp} case besides the CS operator of $\vec{L}.\vec{A}$ with $\vec{A}={\tfrac{1}{2}} \left(\vec{\sigma}_{1} - \vec{\sigma}_{2} \right)$ \cite{Henley}; and $V_{MM}(nn)$ includes $F_{\delta}, F_{t}$ with just spin-spin ($\vec{\sigma}_{1} .\vec{\sigma}_{2}$) and tensor ($S_{12}$) operators. One should note that the radial dependencies, various coefficients and combinations are different for all three cases. It is also notable that the vacuum-polarization and two-photon interactions are useful to fit the low-energy scattering data, and that $F_{c}, F_{c}^2$ used in $V_{VP}, V_{C2}$, are the estimated ways to remove the singularities of $1/r$, $1/r^{2}$, respectively.

The LR OPE part of Arg94 potential ($V_{\pi}$) is charge-dependent, because of the differences between the neutral- and charged-pion masses. It reads
\begin{equation} \label{GrindEQ__55a_}
    V_{\pi}(N_1N_2) = f_{N_1N_1}f_{N_2N_2} V_{OPEP}^{(3)} (m_{\pi_{0}}) + (-1)^{T+1} 2f_{c}^{2} V_{OPEP}^{(3)} (m_{\pi_{\pm}}),
\end{equation}
where the second term on the RHS is used just for \emph{np} system, i.e. with $N_1=n, N_2=p$ and that $f_{pp} =-f_{nn} =f_{c} \equiv f$ \cite{Stoks4} with  $f^{2} = 0.075$, and
\begin{equation} \label{GrindEQ__55b_}
V_{OPEP}^{(3)}(m_{pi})= \left(\frac{m_{pi}}{m_{\pi_{\pm}}} \right)^{2} \frac{1}{3} m_{pi}c^{2} \left[Y_{\mu}(r) \vec{\sigma}_{1} . \vec{\sigma}_{2} + T_{\mu}(r) S_{12} \right],
\end{equation}
in which
\begin{equation} \label{GrindEQ__56a_}
Y_{\mu}(r) = \frac{e^{-\mu r}}{\mu r} \left(1-e^{-cr^{2}} \right),
\end{equation}
\begin{equation} \label{GrindEQ__56b_}
T_{\mu}(r) =\left(1+ \frac{3}{\mu r} + \frac{3}{(\mu r)^{2} } \right) \frac{e^{-\mu r}}{\mu r} \left(1-e^{-cr^{2} } \right)^{2},
\end{equation}
where $Y_{\mu}(r)$ and $T_{\mu}(r)$ are the common Yukawa and tensor functions with exponential cutoffs similar to those in Urb81 and Arg84 potentials, $\mu = m_{pi}c/\hbar$ as before (with $m_{pi}=m_{\pi_{0}}, m_{\pi_{\pm}}$ in the formulas); and the scaling mass of $m_{\pi_{\pm}}$ in (\ref{GrindEQ__55b_}) makes the coupling-constant dimensionless--Look at the differences between (\ref{GrindEQ__55b_}) and (\ref{GrindEQ__20g_}), as well as among (\ref{GrindEQ__56a_}) and (\ref{GrindEQ__56b_}) with (\ref{GrindEQ__44a_}) and (\ref{GrindEQ__44b_}), respectively.

Similar to Urb81 and Arg84 potentials, the remaining (MR and SR) phenomenological parts could be written as a sum of all eighteen terms as
\begin{equation} \label{GrindEQ__57aa_}
V_R = \sum_{i=1}^{n=18} V^{i} O_i,
\end{equation}
where 14 out of the 18 operators are those in (\ref{GrindEQ__23bb_}) of the  Urb81 potential and the 4 remaining ones are
\begin{equation} \label{GrindEQ__57bb_}
O_{i= 15, 16, 17, 18}=T_{12},\ T_{12} (\vec{\sigma}_{1} .\vec{\sigma}_{2}),\ T_{12}\ S_{12},\ (\tau_{z1} +\tau_{z2}), \quad T_{12}=3\tau_{z1}.\tau_{z2}-\vec{\tau}_{1}.\vec{\tau}_{2},
\end{equation}
that in turn we mark them with the indices $i=T,\ \sigma T,\ t T,\ \tau z$, respectively.\\
Now, we note that in the operator form of (\ref{GrindEQ__57aa_}), the whole CD potential could be separated into a CI part and a CIB part, where the latter in turn could be separated into a CD part with three CD operators ($i=T,\ \sigma T,\ t T$) and a charge-asymmetry (CA) part with one operator ($i=\tau z$). In this procedure, the radial potentials of $V^{i}$ could be expressed in terms of the following potentials, with suitable weighting coefficients,
\begin{equation} \label{GrindEQ__57cc_}
V_{ST, NN}^{i} = S_{ST, NN}^{i} T_{\mu}^{2}(r) + \left[P_{ST, NN}^{i} + \mu r Q_{ST . NN}^{i} + (\mu r)^{2} R_{ST, NN}^{i} \right]W_3(r),
\end{equation}
where now $\mu ={\frac{1}{3}} (m_{\pi_{0}} + 2m_{\pi_{\pm}}) c/\hbar$ -- It is also mentionable that the potential is basically written in ($S,T,T_z$) space for various two-nucleon states \cite{AV18}. The same as in Urb81 potential, $T_{\mu}^{2}(r)$ is to simulate TPE force, and the only Woods-Saxon function is
\begin{equation} \label{GrindEQ__58_}
W_3(r)=\left(1+\exp \left(\frac{r-R_3}{a_3} \right)\right)^{-1}.
\end{equation}
All constant coefficients of $S_{ST, NN}^{i}, P_{ST, NN}^{i}, Q_{ST, NN}^{i}, R_{ST, NN}^{i}$ are obtained by fitting experimental data for each two-nucleon state. Further, they imposed the regularization conditions at the origin as
\begin{equation} \label{GrindEQ__59a_}
V_{ST, NN}^{i}(r=0) = 0, \quad \left. \frac{\partial V_{ST, NN}^{i\ne t} }{\partial r} \right|_{r=0} = 0,
\end{equation}
which reduces the number of free parameters for each $V_{ST, NN}^{i}$ by one.

It is also notable that the EM interaction (and also CD) of Arg94 potential is the same as that used in Nijmegen partial-wave-analysis (PWA93) \cite{NijmPWA} besides including the short-range terms and effects for the finite-size of nucleon. This potential is fitted to the Nijmegen \emph{pp} and \emph{pp} scattering database \cite{NijmPWA00}, \cite{NijmPWA}, low-energy \emph{nn} scattering data, and deuteron binding-energy. It has 40 adjustable parameters and gives a best description of the data in the energy range of 0-350 MeV as a high-quality NN potential. The effects of CD and CA are explicitly seen in finite nuclei systems and the results in many-body and nuclear-structure calculations are more satisfactory than the mentioned potentials so far.
Another extension of the Arg94 potential, called $ArgonneV18pq$ potential, is presented in \cite{AV18pq}, where various choices for the quadratic momentum-dependence in NN potentials, to fit the phase shifts of the high partial waves, are included. There is also a p-space proposal for Arg94 potential presented in \cite{1106.1934}.

\subsection{Bonn-Group Potentials}
\subsubsection{Full-Bonn Potential}
The Bonn-group has used the field-theoretical methods to deal with NN interaction problem. In the first version \cite{Machleidt}, in 1987, they presented a comprehensive NN potential by including various meson-exchanges that they thought were important below the pion-production threshold. To do so, the mesons of $\pi, \omega, \delta$ as OBE's and $\rho$, $2\pi$ (as the direct exchange and $\Delta(1232)$-isobar excitation) as TPE's as well as a special combination of $\pi \rho$ were considered. There were also $3\pi, 4\pi$ exchanges that did not have significant contributions. Indeed, the OBE contributions provided good descriptions of high $\ell$ phase shifts while the TPE's with $\pi \rho$ combination provided good descriptions of low $\ell$ phase shifts. So, in general, the exchanges of $\pi$ and $\omega$ together with $\rho$ and $2\pi$ provided good descriptions of the LR and MR (high $\ell$'s) parts while for good describing the SR part (low $\ell$'s), including the $\pi \rho$ combination next to $2\pi$ exchange was required. We should also mention that the $\delta$ meson was needed to provide a consistent description of S-wave phase shifts, and that including the crossed-box diagrams in the two-boson-exchanges (TBE's) made another fitting quality of the potential.

This Full-Bonn (Bonn87 or Bonn-A) potential is originally written in p-space and is energy-dependent that makes its applications in nuclear calculations problematic. To resolve some of the problems, a parameterization of the potential in terms of OBE's in both p-space and r-space is given, which is always called Bonn-B (or Bonn89) potential. As a first step, the retardation terms are neglected to suppress the energy dependence by applying the OBE's in the framework of reducing BS equation into BbS equation, where the latter equation is similar to nonrelativistic Schrodinger equation while it is a relativistic equation. The resultant energy-independent p-space OBE potentials are useful to apply in nuclear structure calculations. The details can be found in \cite{Machleidt}, \cite{Bonn2}. The latter expansion changes somewhat the original results because of some new adjustments. Anyhow, to do so, first the effects of $2\pi+\pi \rho$ exchanges are replaced by the scalar-isoscalar $\sigma$-meson exchange; and without the $\pi \rho$ contribution in the new OBE expansion, the $\eta$ meson is introduced to improve the ${}^3P_1$ phase shifts.\\
The general form of the expanded potential in r-space, coming from the Fourier transform of the agreeing p-space contributions, can be written as a sum on the six boson contributions as
\begin{equation} \label{GrindEQ__59aa_}
V = \sum_{\alpha=\pi, \rho, \eta, \omega, \delta, \sigma} V_{\alpha}^{OBE},
\end{equation}
which in turn divides into a local and a nonlocal part; or may be written as
\begin{equation} \label{GrindEQ__59bb_}
V = V_{c}(r) + V_{t}(r) S_{12} + V_{ls}(r) \vec{L}. \vec{S}-\frac{1}{M} \left[\nabla^2 V_p+V_p\nabla^2\right],
\end{equation}
in ($S,T$) space, where $V_{c}(r)$ includes the contributions from all six mesons of $\pi$ and $\eta$ (pseudoscalar mesons), $\delta$ and $\sigma$ (scalar mesons), $\rho$ and $\omega$ (vector mesons), and is written in terms of $c, \sigma, \tau, \sigma \tau$ operators together with $Y_c(x)$ ($x=m_{\alpha} r$) and the nucleon and included meson masses and couplings as well as some constants; and similarly for the other two functions $V_{t}(r), V_{ls}(r)$. But $V_{t}(r)$ includes the contributions from the four pseudoscalar and vector mesons together with $Y_t(x)$; and $V_{ls}(r)$ includes the contributions from the four scalar and vector mesons together with $Y_{ls}(x)$. Some of the scalar functions are defined in (\ref{GrindEQ__22ff_}) and that here
\begin{equation} \label{GrindEQ__59cc_}
 V_p=\sum_{\beta} C_0\ \frac{g_{\beta}^2}{4\pi} \frac{m_{\beta}}{4M} Y(m_{\beta}r), 
\end{equation}
where $\beta=\delta, \sigma,\rho, \omega$; $g_{\beta}$ is the suiting meson coupling, and $C_0=1$ for the scalar mesons and $C_0=3$ for the vector mesons. It is also mentionable that general structure of the potential for the pseudoscalar mesons is similar to the OPEP used, for example, in the Yale-group and Reid68 potentials (\ref{GrindEQ__20g_}), which is in turn related to the fact that the pion provides the main long-range part of the interaction here as well. It is also notable that the potentials are regularized at the origin by the dipole form factors, which are coming from the Fourier transformations of
\begin{equation} \label{GrindEQ__59dd_}
 F_{\alpha}(k^2)=\left(\frac{\Lambda_{\alpha}^2-m_{\alpha}^2}{\Lambda_{\alpha}^2+m_{\alpha}^2}\right)^{n_{\alpha}},
\end{equation}
where for each vertex $n_{\alpha}=1$ and $\Lambda_{\alpha}$ is the so-called cutoff mass. In other words, the SR part of the interaction is parameterized through the phenomenological form factors attached to the p-space Feynman diagrams, while the high-momentum part of the scattering amplitudes are then regularized with the cutoffs. The cutoff masses ($\Lambda_{\alpha}$) are adjusted to fit the data and are given in \cite{Machleidt} next to other potential parameters, coupling and constants.

By the way, the Bonn87 potential described very good its time experimental NN data up to $T_{lab}=$300 MeV, low-energy parameters and deuteron properties. Meanwhile, we note that, the weak tenor force there, because of the $\rho$-meson exchange and including a real $\pi NN$ form-factor as well as introducing the meson retardation, caused a smaller contribution of the deuteron D-state; and at the same time, the larger quadruple moment and the asymptotic $D/S$ state of deuteron were in full agreement with experimental results---However, in a work done in 1993 \cite{nucl-th/9211013} to compare some of the potential forms with \emph{pp} scattering data, it was shown that the adjusted r-space versions \cite{Bonn2}, i.e. Bonn-A and Bonn-B potentials, give a very poor description of the scattering data ($\chi^{2}/N_{data} >8$ in the energy range of 2-350 MeV). That was not strange of course, in that Full-Bonn potential was originally fitted to \emph{np} scattering data and not to \emph{pp} scattering data. \\
In addition, these potentials have many other special advantages to describe well NN interactions. The nucleons, isobars (nucleon resonances) and mesons are discussed on an equal footing. Because of relativistic approach, the meson retardation (recoil effects) and the off-shell behavior of the nuclear force were included besides that a consistent expansion to the regions above the pion-production threshold was possible. Further, the potential could discuss about three-body nuclear forces (at least because of an almost complete set of the diagrams contributing to the NN interaction and expandable to the 3N case), the meson-exchange currents contributing to the electromagnetic properties of nuclei, the medium effects of the NN interaction in many-body calculations and also CSB and CIB issues. It is also notable that the cutoff masses, used in the meson-nucleon vertex form functions, to explain the extended structure of hadrons, are obtained in a consistent way to be $\Lambda_{\alpha}=$1.2-1.5 GeV, where applying the meson-exchange picture is suppressed. For detailed studies of various aspects of Bonn-A and Bonn-B potentials and the already mentioned issues, look at the original study of \cite{Bonn2}, where the final version of the Full Bonn potential was presented in 1989.

To remind the main differences, we note that Bonn-A potential includes the correlated $2\pi$ and $\pi \rho$ contributions with an intermediate $\Delta$-isobar, while Bonn-B potential is a so-called OBE potential that uses a fictitious $\sigma$-meson (and also a $\eta$-meson) to simulate these two meson exchanges. In contrast to Bonn-A potential, Bonn-B potential is energy-independent that in turn simplifies its applications in nuclear structure and nucleon-nucleus scattering calculations. Despite its greater simplicity, Bonn-B potential gives a good description of its time data, and the other results almost identical with those found in Bonn-A potential. \\
Still, we note that the p-space Full-Bonn potential was fitted just to \emph{np} scattering data. In 1989, another development of the potential to apply it to \emph{pp} scattering data was presented in \cite{Haidenbauer}; see also \cite{nucl-th/9301019}. To do so, a Coulomb interaction (similar to $V_{C_1}$ in Arg94 potential \cite{AV18}) was introduced in the p-space calculations. Then, after a few minor adjustments (for example, the coupling constants of the scalar mesons changed) to face the potential with data, a good description of \emph{pp} data was found as well.

\subsubsection{CD-Bonn Potential} \label{4.13.2}
The Bonn Charge-Dependent (CD-Bonn) NN potential \cite{Bonn4} is an improved and updated version of the previous Bonn-A and Bonn-B potentials \cite{Machleidt, Bonn2}. It is based on the OBE contributions of $\pi, \rho, \omega$ mesons next to two scalar-isoscalar mesons of $\sigma_1, \sigma_2$, which the latter simulates the roles of $2\pi+\pi \omega$ exchanges. The resultant potential is energy-independent in the framework of nonrelativistic LS equation and produces the results of Full-Bonn potential. In addition, the predictions of the latter potential such as CSB and CIB (for all partial waves below $J\leq4$) are involved directly. Further, the predicted off-shell effects because of relativistic Feynman amplitudes for the meson exchanges, which are important in microscopic nuclear structure calculations, are included. It is notable that the first version of the CD-Bonn potential presented in \cite{Bonn3} involved more with the off-shell analyses than the CD issues. \\
Although CSB in the potentials is mainly due to the difference between the proton and neutron masses (the nucleon mass splitting), in CD-Bonn potential an equivalent contribution is due to TBE (mainly $2\pi$ and $\pi \rho$ exchange) diagrams. On the other hand, CIB is mainly due to the difference between the neutral- and charged-pion masses (the pion mass splitting) from OPE diagram, while in CD-Bonn potential an almost equivalent contribution (about 50$\%$) is due to TBE and $\pi \gamma$ diagrams for $\ell > 0$ (or with the predictions of Full-Bonn model due to $2\pi$ as well as $3\pi$ and $4\pi$ exchange diagrams). To see CIB in the potential, we first note that although the OPE amplitudes in the potential are nonlocal, but in the local/static approximation and after a Fourier transformation, the local OPEP in r-space reads
\begin{equation} \label{GrindEQ__59ee_}
\begin{split}
 V_{OPEP}^{(4)}(m_{pi}) =\frac{g_{pi}^2}{12} \left(\frac{m_{pi}}{2M}\right)&^2 \bigg[ \left( \frac{e^{-\mu r}}{r}-\frac{4\pi}{\mu^2} \delta^3(\vec{r}) \right) (\vec{\sigma}_{1}.\vec{\sigma}_{2}) \\
  &+ \left(1+ \frac{3}{\mu r}+ \frac{3}{(\mu r)^3} \right) \frac{e^{-\mu r}}{r} S_{12} \bigg],
\end{split}
\end{equation}
where $\mu = m_{pi}c/\hbar$. Now, because of the pion mass splitting (as the main CIB factor), we have
\begin{equation} \label{GrindEQ__59ff_}
\begin{split}
 & V_{OPEP}^{pp}=V_{OPEP}^{(4)}(m_{\pi_0}), \\
 & V_{OPEP}^{np}=-V_{OPEP}^{(4)}(m_{\pi_0}) \pm 2 V_{OPEP}^{(4)}(m_{\pi_\pm}),
\end{split}
\end{equation}
where in the second relation, $+$ ($-$) is for $T=1$ ($T=0$). We see that because of the pion mass differences, the \emph{np} OPEP with $T=1$ is weaker than that of \emph{pp}, leading to CIB. It is also notable that the $\Delta$-isobar states and multi-meson exchanges in Full-Bonn (Bonn-A) potential caused the energy dependence which was in turn problematic in applying the potentials to direct nuclear calculations. So, in CD-Bonn potential (also in Bonn-B potential) this problem is avoided by using just OBE contributions.

The three potentials of \emph{pp}, \emph{np}, \emph{nn} are not independent but they are related by CSB and CIB. Each of them is first fitted to the related Nijmegen phase shifts; then by minimizing the earned $\chi^2$ from the Nijmegen error matrix and finally minimizing the exact $\chi^2$, which is in turn obtained from comparing with all related scattering data, the potential parameters are adjusted. For the Coulomb force in \emph{pp} case, a similar $V_{C1}$ as in \cite{Haidenbauer} is used, and the relativistic Coulomb interaction besides nuclear phase shifts are considered as well. The base phase shifts are a sum of the Nijmegen-group ones in \cite{NijmPWA00}, \cite{NijmPWA} up to 1992, used also in Arg94 potential, besides the published data after-1992-date and before-2000-date. So, CD-Bonn potential fitted the world 2932 \emph{pp} data below $T_{lab}=$350 MeV available in 2000 with $\chi ^{2}/N_{data} = 1.01$ and the corresponding 3058 \emph{np} data with $\chi^{2}/N_{data}=1.02$. This reproduction of NN data is more accurate than by any other previous NN potentials, according to its authors of course! For more details, such as its first applications to few-and many-body nuclear calculations, CIB, CSB and off-shell effects, see the original papers \cite{Bonn3, Bonn4}; and also look at \cite{4}.

\subsection{Padua-Group Potential}
The Padua model for NN interaction, as a mixture of meson-exchange theory and phenomenological methods, is a special and important effort. The group has tried to set up an NN potential based on their special model for "Nucleon". They have used a nonlocal potential coming from the Padua nucleon-model with similar operators as in Hamada-Johnston \cite{HamadaJohnston}, Yale-group \cite{Yale} and dTRS Super-Soft-Core \cite{deTourreil} Potentials. \\
Indeed, the various terms with the operators shortened as $c, \sigma, \tau, \sigma \tau, t, t\tau, ls, ls \tau, ll, ll \tau, ls2, ls2\tau$ are included. The general form of the potential, in ($S, T$) space, can be written as
\begin{equation} \label{GrindEQ__60aa_}
V = V_{c}(r) + V_{t}(r) S_{12} + V_{ls}(r) \vec{L}. \vec{S} + V_{ls2}(r) (\vec{L}. \vec{S})^2+V_{ll}(r) L^2,
\end{equation}
where the radial functions have special forms almost different from the other potentials mentioned so far. In fact, various contributions of the pion and other single mesons as well as two-pion combinations are introduced through these functions. The functions are in turn in terms of special combinations of some radial functions and operators with included meson masses, their coupling constants, amplitudes, and other free parameters and constant coefficients. Plainly, both $V_{c}(r)$ and $V_{t}(r)$ include the contributions from the mesons of $\pi, \rho, \omega, \eta, \acute{\eta}$, and are written in terms of $c, \sigma, \tau, \sigma \tau$ and $c, \tau$ operators respectively, together with some functions such as $Y(x), Z(x),...$ ($x=m_{\pi_0} r$) and the nucleon and included masses as well as some other coupling and coefficient constants. Similarly, in $V_{ls}(r), V_{ls2}(r), V_{ll}(r)$, the contributions from the mesons of $\rho, \omega, s$ and the operators of $c, \tau$ are included. It is also notable that one may use the operators of $\vec{L}. \left(\vec{\sigma}_{1}- \vec{\sigma}_{2} \right)$ and/or $\vec{L}. \left(\vec{\sigma}_{1}\times \vec{\sigma}_{2} \right)$ instead of $\vec{L}. \vec{S}$ in the Padua model as they are also consistent. \\
In general, the involved radial functions in the potential are more based on theoretical knowledge by aiding of the nucleon-model rather than merely fitting experimental data. Nevertheless, reproducing deuteron parameters and fitting phase shifts are good compared with the counterpart results of its time potentials such as Arg94 \cite{AV14}, Bonn \cite{Machleidt} and Paris \cite{Cottingham} potentials. Although it is rarely used in nuclear calculations, the Padua NN potential is a serious try to find an even more sensible NN potential. For other interesting theoretical and numerical analyses in their method, see the original paper \cite{Minelli}.

\subsection{Nijmegen-Group Potentials}
The Nijmegen-group potentials are mainly the mixtures of meson exchanges with phenomenological characterizes and are often referred to QCD degrees of freedom for the SR part. The group built various Baryon-Baryon (BB) and Baryon-AntiBaryon ($B\bar{B}$) potentials among which are some high-quality NN and Hyperon-Nucleon (YN) potentials. First, they presented a few potential before 1990's and then performed the partial-wave-analysis (PWA) \cite{NijmPWA}, \cite{NijmPWA2} of the experimental scattering data. The insights gained from the analyses were then employed to set up some improved and better potentials. In their NN potentials, besides the famous OBE parts, many new features and other meson contributions are included. The nucleon- and pion- mass splitting are often considered and, for the potentials after PWA93, charge dependence is used. Because of the short-range parameterization, because of the vertex form functions, the potentials are in contact with QCD. The potentials may be divided into at least four classes; the Hard-Core (HC), Soft-Core (SC), Extended Soft-Core (ESC), and High-Quality (HQ) potentials as well as PWA's. We address in the following subsections some of their NN potentials briefly; look also at \cite{Nijmegen}.

\subsubsection{The First Potentials}
The main aim was to form BB potentials below the pion-production threshold. As we know, the OBEP's describe almost well the LR and MR parts next to including uncorrelated $2\pi$ or scalar meson exchanges. Further, to describe the data better, the fictitious meson of $\sigma$ (as a correlated $2\pi$ exchange) is always required. In these models, the heavier meson of $\epsilon$ is sometimes used as well. The Schrodinger equation in r-space is solved with local potentials and Coulomb force (depended on the case) and, in addition, the SR repulsion is considered through HC potentials. The first potentials of the group, named as NijmA, NijmB, NijmC, NijmD, NijmE and NijmF, were represented from 1972 to 1978.

NijmA potential \cite{NijmA} is composed of some OBEP's and a TPEP. Indeed, it includes the members of the pseudoscalar- and vector- meson nonets as well as the Brueckner-Watson TPEP. The potential was to describe low-energy YN data though it was not so good to describe the high NN partial waves.  NijmB and NijmC potentials \cite{NijmBC} are OBEP's fully and reproduced well their time NN scattering data; the group also showed that one can describe the YN channels with this OBEP approach.\\
It is notable that in the pure OBEP's, the mesons were considered in an SU(3) consistent way. That was mainly because one then could extend the calculations from NN to YN systems as well. For example, in the vector-meson (pseudoscalar-meson) nonet, one should use $\rho, \omega, \phi$ ($\pi, \eta, \acute{\eta}$) and all knowledge about $\phi-\omega$ ($\eta-\acute{\eta}$) mixing and coupling constants from SU(3). The OBEP's were constructed in two classes I and II, where both used the nonets of the pseudoscalar- and vector- mesons but they were different in discussing the scalar mesons. In the class I, just the singlet scalar meson of $\epsilon$ was included while in the class II, an octet of the scalar mesons was included. The first model of the class I was NijmB potential with $m_\epsilon=720$MeV and $\Gamma_\epsilon=400$MeV that gave almost $\chi^2/N_{data}=5.9$ for its time NN scattering data below $T_{lab}=$330 MeV of the Livermore-group \cite{Livermore1969} of 1969.

NijmD potential \cite{NijmD} is belong to the class I OBEP's and is similar to NijmB potential except for including the $\eta-\acute{\eta}$ mixture, $m_\epsilon=760$MeV and $\Gamma_\epsilon=640 $MeV, a different ratio of $F/(F+D)$ for the pseudoscalar octet, the slightly different potential forms for vector- and scalar-mesons, as well as some other coupling and parameter changes. Clearly, the NijmD NN potential includes the nonets of the pseudoscalar mesons of $\pi, \eta, \acute{\eta}$ and the vector mesons of $\rho, \omega, \phi$, each with a singlet-octet mixing angle as well as the unitary singlet scalar-meson of $\epsilon$. For short distances, it uses some strong repulsive phenomenological HC potentials, which in turn should simulate the effects of the absent heavier-meson exchanges, inelastic effects and so on. This HC parameterization is suitable, rather than the vertex form factors, in that it is independent of the meson dynamics and is simple to use with Schrodinger equation. The 13 parameters of the potential, which are 8 meson-nucleon couplings and 3 core radiuses, are determined from data fitting.\\
The general form of NijmD potential can be written in the operator format as (\ref{GrindEQ__23aa_}), where now $n=10$
and $c, \sigma, \tau, \sigma \tau, t, t\tau, ls, ls \tau, q, q\tau$ are the indices for the 10 involved operators. In other words, one may say that the potential includes the central, tensor, spin-orbit and quadratic spin-orbit terms in ($S, T$) space. The potentials of $V_{i}$ are gained from field theory with some approximations such as ignoring their total energy dependence, and writing the energy factors as $E\simeq M+k^2/8M$, 
where the notations are those in (\ref{GrindEQ__22dddd_}). This approximation means that just the terms up to the order of $k^2/M^2$ are kept in the p-space potentials. In addition, there are the recoil effects to the quadratic spin-orbit potentials that cause the total energy dependence. Further, in Fourier transform to r-space, all the terms that include $\vec{\nabla}_r$ are neglected except that in $L^2$ operator. It is also notable that the meson bandwidth was settled with a special propagator instead of the static meson propagator of $1/(k^2+m^2)$; and after the Fourier transform to r-space, a superposition of the Yukawa functions resulted.\\
Resultant potentials in r-space are in terms of the functions of $Y(x)$, $Z(x)$ in (\ref{GrindEQ__19d_}) and the proper operators and coupling constants as well as the nucleon and the pion averaged masses. Indeed, we note that the potential for the pseudoscalar mesons is similar to the Full-Bonn potential \cite{Machleidt}, where both have a similar structure as the OPEP of (\ref{GrindEQ__1_}) or that in the Yale-group \cite{Yale} and Reid68 \cite{Reid68} phenomenological potentials. Anyway, it is determined that all mesons contribute to the central potentials (with the function of $Y_c(x)$), the pseudoscalar and vector mesons contribute to the tensor potentials (with the function of $Y_t(x)$), the scalar and vector mesons contribute to the spin-orbit potentials (with the function of $Y_c(x)$) and to the quadratic spin-orbit potentials (with the function of $Y_t(x)$). Still, for the short distances of $r\lesssim 0.5$, the HC radius $x_c$ has four different values for the four channels of $^1S_0$, ${}^3S_1-{}^3D_1$, $\ell=1$, and $\ell \geq 2$.\\
The \emph{pp}+\emph{np} scattering data of the energy-independent phase-shift analyses of the Livermore-group \cite{Livermore1969} were fitted good with $\chi^2/N_{data}=2.4$ for NijmD potential, next to good describing low-energy scattering parameters and deuteron properties. Then, the YN version of the NijmD potential was shown in \cite{NijmD2}. In fact, there, some $\Lambda N$ and $\Sigma N$ potentials were presented with considering charge symmetry between the $\Lambda p$ and $\Lambda n$ channels. The contributions for a scalar octet in this YN potential were neglected (just $\epsilon$ with an important role in YN scattering was included) to prevent introducing more free parameters in the potential. It was argued that the YN interaction there next to NijmD NN potential describe all studied BB systems well.\\
NijmE potential \cite{NijmF} is almost the same as NijmD potential except for the contributions of the scalars in the nonet; meanwhile the results are almost same. NijmF potential \cite{NijmF} completed the HC potentials to describe all experimental known BB systems. Indeed, the need to settle the scalar-octet coupling constants for YN systems, without increasing the number of parameters, led to a different HC potential. Further, that need led to stronger SU(3) constraints between NN and YN analyses than before. With the changes, such as those of the coupling constants and relations among them, they earned better results than Nijm B potential with NijmF potential \cite{NijmF}.

\subsubsection{Nijm78 Potential}
Nijm78 potential \cite{Nijm78}, published in 1978, is a mixture of OBPE's and one-Reggon-exchange-potentials (OREP's). In fact, it includes the vector mesons of $\rho, \omega, \phi$; the pseudoscalar mesons of $\pi$, $\eta$, $\acute{\eta}$, with the couplings and mixings from their suiting SU(3) relations; the scalar mesons of $\delta, S^*, \epsilon(760)$; the dominant $J=0$ contributions of Pomeron (P) (or multi-gluon exchanges), and $f, \acute{f}, A_2$ tensor Regge-trajectories. So, this nonlocal and SC potential is indeed based on Regge-pole theory for low-energy NN interaction and fits high-energy data by using exponential form factors.

In p-space, the general form of Nijm78 potential, the OBEP's with p-dependent central terms and Pomeron-type potentials, reads
\begin{equation} \label{GrindEQ__24bbb_}
\tilde{V}(\vec{p}_i,\vec{p}_f) = \tilde{V}_0(k^2, q^2) +\tilde{V}_{\sigma}(k^2) \vec{\sigma}_1.\vec{\sigma}_2+ \tilde{V}_{t}(k^2)\tilde{S}_{12}^{(0)} + \tilde{V}_{ls}(k^2) \tilde{LS}_1 + \tilde{V}_{q}(k^2) \tilde{Q}_{12},
\end{equation}
where the symbols are those in (\ref{GrindEQ__22dd_}) except $\tilde{S}_{12}^{(0)}=(\vec{\sigma}_{1}.\vec{k})(\vec{\sigma}_{2}.\vec{k})$. With the last relation, one should note that we have just nonlocality in the central potential that means all the momentum dependence in r-space is in the central part of the potential. Meanwhile, in the Fourier transform into r-space, the energy-factor is approximated by $E\simeq M+k^2/8M+q^2/2M$ and that just the first order terms in $k^2/M^2,\ q^2/M^2$ are kept. Now, by the approximations, one can write the potentials of $\tilde{V}_i$ ($i=c,\sigma, t, ls, q$) for all four sets of the involved mesons. The potentials so are some combinations of $k^2, q^2$, meson and nucleon masses, coupling constants and the exponential form-factor of $\Delta$ as
\begin{equation} \label{GrindEQ__24ccc_}
\Delta= \frac{1}{\vec{k}^{2} + m_{mes}^{2}} e^{-\vec{k}^{2} /\Lambda^{2}}, \quad \Delta_P= \frac{1}{M_{p}^{2}} e^{-\vec{k}^{2} /4m_{p}^{2}},
\end{equation}
where $m_{mes}, m_{P}, M_{p}, \Lambda$ are the meson, Pomeron, proton (a scale mass) masses and the cutoff mass (964.52 MeV here), respectively. \\
The Fourier transforms of the potentials into r-space, for the central, tensor, spin-orbit and quadratic spin-orbit potentials are given in the Nijm87 original paper \cite{Nijm78}. The potentials so are in terms of some functions of $\phi_c^0(r), \phi_c^1(r), \phi_c^2(r), \phi_t^0(r), \phi_t^1(r), \phi_{ls}^0(r), \phi_{ls}^1(r)$, which are in turn in terms of  $m_{mes}$, $m_{P}$ (just for the Pomeron-type potentials) and $\Lambda$. Further, the Fourier transform of the form-factor of $\Delta$ in (\ref{GrindEQ__24ccc_}) becomes
\begin{equation} \label{GrindEQ__24ddd_}
\tilde{\Delta}=\frac{m_{mes}}{4\pi} \left[\frac{1}{4}m_{mes}^2 \phi_c^1(r) -\frac{1}{2}\left(\nabla^2 \phi_c^0(r)+\phi_c^0(r) \nabla^2\right) \right],
\end{equation}
and similar for $\Delta_P$ by setting $\frac{1}{2}\Lambda=m_P, m_{mes}=0, \phi_j^{P_n}(r)=\phi_j^{n+1}(r)$ with $j=c, t, ls$ here--For a study of the Fourier transformation in such cases look, for instance, at \cite{Hoshizaki04}. \\
Now, we can write the r-space potential, in ($S,T$) space, as
\begin{equation} \label{GrindEQ__24eee_}
V = V_{c}(r) + V_{t}(r) S_{12} + V_{ls}(r) \vec{L}. \vec{S}+V_q(r) Q_{12}-\frac{1}{M} \left[\nabla^2 \grave{V}_p+\grave{V}_p\nabla^2\right],
\end{equation}
where $V_{c}(r)$ includes the contributions from all mesons and is written in terms of $c, \sigma, \tau, \sigma \tau$ operators together with $\phi_c^0(r), \phi_c^1(r), \phi_c^2(r)$ and the nucleon and included meson masses and couplings as well as some other constants; and similarly for the other three functions of $V_{t}(r), V_{ls}(r), V_q(r)$. But $V_{t}(r)$ includes the contributions from all involved pseudoscalar and vector mesons together with $\phi_t^0(r), \phi_t^1(r)$ and $t, t\tau$ operators; $V_{ls}(r)$ includes the contributions from all involved scalar, vector and Pomeron-type mesons together with $\phi_{ls}^0(r), \phi_{ls}^1(r)$ and $ls, ls\tau$ operators; $V_{q}(r)$ includes the contributions from all involved scalar, vector and Pomeron-type mesons together with $\phi_t^0(r)$ and $q, q\tau$ operators. In is mentionable that the local part of the Pomeron-type potentials is multiplied by the exponential factor of $e^{-m_{p}^{2} r^2}\equiv \phi_P^0(r)$. We also note that in the nonlocal potential, to which all except the pseudoscalar mesons contribute, as the last part in (\ref{GrindEQ__24eee_}), we have
\begin{equation} \label{GrindEQ__24fff_}
 \grave{V}_p=\sum_{\gamma} C_0\ \frac{g_{\gamma}^2}{4\pi} \frac{m_{\gamma}}{4\acute{M}} \phi_c^0(r) - \frac{g_P^2}{4\pi \sqrt{\pi}} \frac{m_{\gamma}^3}{M \acute{M}^2} \phi_P^0(r),
\end{equation}
where $\gamma$ and $g_{\gamma}$ are the suiting meson indices and couplings, $M, \acute{M}$ are for the proton and/or neutron mass ($M_p$ is chosen often), and $C_0=1$ for scalar mesons and $C_0=3$ for vector mesons--It is also notable that the methods to solve Schrodinger equation with nonlocal potentials (such as $[\nabla^2 \grave{V}_p+\grave{V}_p\nabla^2]$ here) is presented in \cite{Green2}.

Anyway, The 13 free parameters of the potential were fitted to the Livermore-group 1969 data up to 330 MeV \cite{Livermore1969} good with $\chi^2/N_{data}=2.09$ besides good describing low-energy parameters such as ${}^1S_0(pp), {}^3S_1(np)$ scattering lengths as well as deuteron properties. The results were very good among the best potentials of its time--The updated and improved version of Nijm78 potential is Nijm93 potential \cite{Nijm93} framed in 1992, which we describe below.

The YN version of Nijm78 potential was presented in 1989 \cite{Nijm78HV} and applied to $B\bar{B}$ systems as well. The form factors, from the Regge-poles, are Gaussian that guarantee the soft behavior of the potentials near the origin. It gave a good description of YN interactions by using SU(3) and meson-nucleon coupling constants from the NN analyses.

\subsubsection{Nijmegen Partial-Wave-Analysis} \label{4.15.3}
The first Nijmegen-group multi-energy phase-shift analysis was published in 1990 for just \emph{pp} interaction \cite{NijmPWA00}. Next in 1993, they published a combined analysis of \emph{np}+\emph{pp} scattering data \cite{NijmPWA}--For a newer PWA look at \cite{NijmPWA3}. Indeed, the basic aim was to provide a more complete database and then to improve the NN phase-shift analyses. To do so, they surveyed the NN data published from Jan 1655 to Dec 1992 in the energy range of $T_{lab}=$0-350 MeV. As a result, from 2078 \emph{pp} data and 3446 \emph{np} data, those survived with an optimized (not a very-high or a very-low) $\chi^{2}$ were 1787 \emph{pp} and 2514 \emph{np} data. Next, they parameterized a special energy-dependent NN potential for each partial-wave up to almost $J=4$. After that, the radial Schrodinger equation was solved by the adjusted potential to get the phase shifts as functions of the adjusted parameters and energy. Then, from the phase shifts, some predictions for observables, and $\chi^{2}$ to fit the experimental scattering data, were made. So, one may call the Nijmegen analysis as an "optimized potential" analysis from which the phase shifts are bought for various partial waves.

By the way, in the Nijmegen PWA's, the potentials for each partial-wave are actually divided into two main parts: A nuclear (N) part and an electromagnetic (EM) part; or a long-range (LR) part, a medium-range (MR) and a short-range (SR) part. That is
\begin{equation} \label{GrindEQ__24aaaa_}
V= V_{EM} +V_{N}=V_{LR}+V_{MR}+V_{SR},
\end{equation}
where the electromagnetic interaction has almost the same structure as that in Arg94 potential \cite{AV18} while the nuclear part includes an LR OPEP, an MR heavy-boson-exchange (HBE) potential and a phenomenological SR potential. The LR potential $V_{LR}$ is indeed a sum of the EM and OPE potentials; the MR potential $V_{MR}$ is mainly from the HBE contributions of Nijm78 potential \cite{Nijm78}; and the SR potential $V_{SR}$ is described by an energy-dependent boundary condition at $r=b=1.4 fm$, where the energy-dependent square wells are used.

The involved EM potential here, in general, reads
\begin{equation} \label{GrindEQ__24bbbb_}
V_{EM}=V_{C1}(r)+V_{C2}(r)+V_{VP}(r)+V_{MM}(r),
\end{equation}
where, as before, the indices of $C1, C2, VP, MM$ stand for the one-photon, two-photon, vacuum-polarization and magnetic-moment interactions. More details were given in the subsection of (\ref{4.12.2}) with two main differences here with the more improved considerations in Arg94 case, where the effects due to the finite-size of the nucleon and a Darwin-Foldy term ($V_{DF}(r)$) were also included and improved.\\
On the other hand, the LR nuclear interaction because of OPE's and the MR nuclear interaction because of HBE's always read
\begin{equation} \label{GrindEQ__24cccc_}
V_{N}= \frac{M}{E} V_{OPE} + f_{med}^s V_{HBE}.
\end{equation}
Indeed, the energy-dependent factor of $M/E$ (where $M$ is as usual the nucleon-mass, $E=\sqrt{M^{2} + q^{2}}$ is the c.m. energy and $q^{2} =MT_{lab}/2$) is required to get a better fit of the data. Also, adding the HBE's (such as $\rho, \omega, \eta$) from Nijm87 potential for $r > b$ to the OPEP tail, give a better fit of the data but the nuclear part is still incomplete. The $f_{med}^s$ factor in the last relation, for the singlet(s) partial waves, makes further improvement with $f_{med}^s(S= 0) = 1.8$ and $f_{med}^s(S= 1)= 1.0$, where $S$ stands for the total spin of NN systems here.\\
For $V_{OPE}$, we first note that one may face with the four isovector coupling constants of $f_{pp \pi_{0}}$, $f_{nn \pi_{0}}$, $f_{np \pi_{-}}$, $f_{pn \pi_{+}}$ in $NN\pi$ vertexes. So, for three possible NN scattering's, one can write
\begin{equation} \label{GrindEQ__24dddd_}
   f_{pp}^{2}  \equiv  f_{pp \pi_{0}} f_{pp \pi_{0}}, \quad
   f_{0}^{2}   \equiv -f_{nn \pi_{0}} f_{pp \pi_{0}}, \quad
   2f_{c}^{2}  \equiv  f_{np \pi_{-}} f_{pn \pi_{+}},
\end{equation}
where one may then take $f_{pp}^{2}=f_{0}^{2}$ when CS and $f_{pp}^{2}=f_{0}^{2}=f_{c}^{2}$ when CI is assumed. Now, we can use the same expression in (\ref{GrindEQ__55a_}) for $V_{OPE}$ in (\ref{GrindEQ__24cccc_}) with a note that again the second term on its RHS is used just for  \emph{np} case and $f_{pp} =-f_{nn}\equiv f$ \cite{Stoks4} with the CI value of $f^{2} = 0.075$; and also we should replace $V_{OPEP}^{(3)}$ there with $V_{OPEP}^{(4)}$ here as
\begin{equation} \label{GrindEQ__24eeee_}
V_{OPEP}^{(4)}(m_{pi})=\frac{1}{3} \left(\frac{m_{pi}}{m_{\pi_\pm}} \right)^{2} \frac{e^{-\mu r}}{r} \left[ \left(\vec{\sigma}_{1}.\vec{\sigma}_{2} \right) + \left(1 + \frac{3}{(\mu r)} + \frac{3}{(\mu r)^{2}} \right) S_{12}  \right],
\end{equation}
where $\mu = m_{pi}c/\hbar$ as before. \\
For the SR potential $V_{SR}$, used for $r < b$ or lower partial waves, in the \emph{pp} PWA's (Nijm90\emph{pp}) \cite{NijmPWA00}, the coordinate-independent energy-dependent square wells were used up to $J=4$ (see, Fig. 2 and 3 of \cite{NijmPWA}). Further, for the isoscalar ($T=0$) \emph{np} partial waves up to $J=4$, and ${}^1S_0$ partial-wave, the same parameterization as the \emph{pp} case was used; whereas for the isovector ($T=1$) \emph{np} phases shifts (except for ${}^1S_0$ phase-shift), the suiting \emph{pp} results by including the pion Coulomb corrections were used. For the middle partial waves of $5\leq J \leq8$, the evaluated phase shifts of the OPE+HBE of Nijm78 potential \cite{Nijm78} were used. And finally, the higher partial waves were obtained from the OPE phase shifts by including the electromagnetic effects depended on the need. \\
It is also notable that the energy dependence of the square-well depth is parameterized through three parameters for each partial-wave. From the total 49 such parameters for the states of $J \le 4$, 21 parameters are for the \emph{pp} case and 18 parameters are for the \emph{np} case besides the pion-nucleon coupling constants ($f_{\pi_{\pm}}$, $f_{0}$) and $f_{med}^s$ determined by fitting the data. By the way, in the combined \emph{pp}+\emph{np} Nijmegen PWA's \cite{NijmPWA}, with 1787 \emph{pp} data (with 1613 degrees of freedom) and 2514 \emph{np} data (with 2332 degrees of freedom) below 360 MeV, published from nearly 1955 to 1992, the "perfect" result of $\chi^{2}/N_{data}\approx 1$ from the data fitting was achieved.\\
Later, in 2004, the Nijmegen-group made a new PWA of \emph{pp} and \emph{np} data up to 500 MeV \cite{NijmPWA3}. There, the NN database was enlarged to almost 5000 \emph{pp} and the same \emph{np} data below that energy. Inelastic effects could be included, and one could gain both the $T=0,1$ phase shifts from the \emph{np} data in contrast to PWA93 \cite{NijmPWA}, where $T=1$ phases shifts were gained from the suiting \emph{pp} ones with some corrections. In the analysis, a chiral TPE potential was added to the LR OPEP used in PWA93, with an improvement of the data fitting--For a new PWA of NN scattering data, by another group, look at \cite{1304.0895}.

\subsubsection{Nijm93, NijmI and NijmII Potentials}  \label{4.15.4}
The Nijmegen high-quality (HQ) potentials, which are Nijm93, NijmI, NijmII and Reid93 NN potentials \cite{Nijm93}, all give almost the perfect value of $\chi^{2}/N_{data} \approx 1$. Nijm93 potential is indeed an updated version of Nijm78 potential in that it is fitted to its time Nijmegen \emph{np}+\emph{pp} database \cite{NijmPWA} (with $\chi^{2}/N_{data}=1.87$) and includes new OPEP's with the pion mass splitting. Both NijmI and NijmII potentials are also built on Nijm78 potential \cite{Nijm78} with some differences and improvements of course. In NijmI potential, in each partial-wave, a few parameters of the potential are adjusted. It includes, like Nijm78 and Nijm93 potentials, the momentum-dependent terms that result in the nonlocal structure of the potential in r-space. However, NijmII potential is completely local that means all momentum-dependent terms in p-space are deliberately removed. These three potentials are regularized at the origin by exponential form factors, are fitted to the same database and have the same number of fitting parameters (15 free parameters) as in PWA93 \cite{NijmPWA}. The results of data fitting signal that NijmI and NijmII potentials have almost the same quality, and that all three potentials reproduce a $\chi^{2}$ close to the suiting value for PWA93.

The general form of the NijmI and NijmII potentials in p-space are as in (\ref{GrindEQ__24bbb_}) except for some differences. The first difference is adding the new operator of $\tilde{LA}=\frac{i}{2} (\vec{\sigma}_{1} -\vec{\sigma}_{2}).\vec{n}=i \vec{A}.\vec{n}$ (which is the Fourier transform of the charge-symmetry operator $\vec{L}.\vec{A}$ used in Arg94 potential \cite{AV18} as well) and so, the new term of $\tilde{V}_{la}(k^2) \tilde{LA}$ is added to the potential. It is notable that for identical-particle scattering, this operator does not contribute; and when CI is supposed, $\tilde{V}_{la}(k^2)$ vanishes. The second difference is that instead of $\tilde{S}_{12}^{(0)}$ in Nijm78, one now uses the complete $\tilde{S}_{12}$ in (\ref{GrindEQ__22dd_}), which is in turn the Fourier transform of the r-space tensor operator of ${S}_{12}$. The third difference is that because $\tilde{Q}_{12}$ in (\ref{GrindEQ__22dd_}) is not an exact Fourier transform of the quadratic spin-orbit operator ${Q}_{12}$ in (\ref{GrindEQ__18_}), to have equivalent r- and p-space potentials with the same phase shifts and bound-states, $\tilde{Q}_{12} \tilde{V}_{q}(k^2)$ in (\ref{GrindEQ__24bbb_}) must be replaced by
\begin{equation} \label{GrindEQ__29c1_}
\tilde{Q}_{12} \tilde{V}_{q}(k^2)-\tilde{Q}_{12}' \int_{\infty}^{{k}^{2}} d{k}'^{2} \tilde{V}_{q}(k^2),
\end{equation}
where
\begin{equation} \label{GrindEQ__29c2_}
\tilde{Q}_{12}'=\left[(\vec{\sigma}_{1}.\vec{q}) (\vec{\sigma}_{2}.\vec{q})-q^{2} (\vec{\sigma}_{1}.\vec{\sigma}_{2})\right] - \frac{1}{4} \left[(\vec{\sigma}_{1}.\vec{k}) (\vec{\sigma}_{2}.\vec{k})-{k}^{2} (\vec{\sigma}_{1}.\vec{\sigma}_{2}) \right].
\end{equation}
One should note that including $Q_{12}$ was indeed necessary there to describe the phase shifts of ${}^1S_0, {}^1D_2$ simultaneously, and its effect could be simulated by including special nonlocal potentials. By the way, the resultant potential forms $V_{i}$ of $V=\sum_{i} V^{i} O_{i}$, where $i=c, \sigma, t, ls, q, la$, are supposed to be same for all partial waves, as the differences among the potentials arise from the vacuum expectation values of the operators in different partial waves. It is also notable that $V_{i}$ may be a function of $r^{2}$, $q^{2}$ and $L^{2}$ in r-space (or $\tilde{V}_{i}(\vec{k}, \vec{q}, \vec{n},E)$ in p-space); meanwhile a $r^{2}$-dependence is always preserved and the $q^{2}$-dependence is included in $V_{c}$, which in turn signals the nonlocal structure of the potential in r-space.\\
The included mesons and Reggons, OBEP's and OREP's as well as the propagators are the same as those in Nijm78 potential except for a few differences. Indeed, the pion- and nucleon-mass splitting are also considered. Taking the mass difference between the neutral- and charged-pions (and also for the $\rho$ meson here) leads to CIB. The coupling constants for the pseudoscalar and vector mesons are related through SU(3) with their special singlet-octet mixing, whereas for the scalar mesons and the Regge poles, the coupling constants are considered as free parameters. Also, for each exchange, an independent cutoff mass is used and so, with the three cutoffs of $\Lambda_{PS}, \Lambda_{V}, \Lambda_{S}$, there are a total number of 14 free parameters. It is also notable that the broad mesons of $\rho$ and $\epsilon$ could be described by a dispersion integral instead of the static formula of $\Delta(k^2)=1/(k^2+m_{mes}^2)$.
In the OPE part, as in PWA93, the pion mass splitting is considered and so, the isovector \emph{np} phase parameters are smaller than the isovector \emph{pp} phase parameters, which in turn means CIB. The plain OPEP's for \emph{pp} and \emph{np} systems are the same as those in PWA93 (and also Arg94 potential in (\ref{4.12.2})) with $f_{pp}^2=f_c^2=f_0^2=f_{\pi}^2=0.075$ (pointing out CI for the pion-nucleon coupling constants), and
\begin{equation} \label{GrindEQ__24eeee_}
V_{OPEP}^{(5)}(m_{pi})=\left(\frac{m_{pi}}{m_{\pi_{\pm}}} \right)^{2} \frac{1}{3} m_{pi}c^{2} \left[\phi_c^1(m_{pi}, r) \vec{\sigma}_{1}.\vec{\sigma}_{2} + 3\phi_t^0(m_{pi}, r) S_{12} \right],
\end{equation}
instead of $V_{OPEP}^{(4)}(m_{pi})$ in (\ref{GrindEQ__24eeee_}).

Describing the data, in the energy range of 0-350 MeV, with the potentials are satisfactory. In fact, Nijm93 potential fits 1787 \emph{pp} data with $\chi^{2}/N_{data}=1.8$ and 2514 \emph{np} data with  $\chi^{2}/N_{data}=1.9$ and so, the whole data with $\chi^{2}/N_{data}=1.87$. This description is better than that of parameterized Paris potential \cite{Paris2} and Full-Bonn potentials \cite{Machleidt}, \cite{Bonn2}. This result suggests that just with the conventional OBEP's one could not describe the data well. On the other hand, NijmI and NijmII Reid-like potentials describe the whole \emph{pp} and \emph{np} data with $\chi^{2}/N_{data}=1.03$ with 41 and 47 fitting parameters, respectively. The potentials are called Reid-like in that, in each partial-wave, just a few parameters are adjusted that is in turn similar to the Reid method in parameterizing the potentials in each partial-wave separately.

It is good to remind that, in making these HQ potentials, the Schrodinger equation of
\begin{equation} \label{GrindEQ__27a_}
(\nabla^2 + k^{2})\Psi = 2M_{r} V \Psi,
\end{equation}
is used, which is a r-space approximation of the full four-dimensional scattering equation. In this equation, $M_{r}$ is the nucleon reduced mass, and the relations between the c.m. energy ($E$) and the squared c.m. momentum ($k^2$) are as $E= k^{2}/2M_{r}$ and $E=\sqrt{k^{2}+M_{p}^{2}}+\sqrt{k^{2}+ M_{n}^{2}}-(M_{p}+M_{n})$ for the nonrelativistic and relativistic kinematics, respectively.

On the other hand, following the discussion in the previous subsections, we know that to regularize the potential at the origin, the form-factor of $F({k}^{2})$ is always used. For the Nijmegen potentials, and to complete the discussion, we quote the following useful Fourier transform (with the $\lambda$ index for the corresponding meson)
\begin{equation} \label{GrindEQ__30a_}
\begin{split}
 \int \frac{d^{3} k}{(2\pi)^{3}} \frac{e^{i\vec{k}.\vec{r}}}{{k}^{2} + m_{\lambda}^{2}}  ({k}^{2})^{n} F({k}^{2}) & \equiv  \frac{m_{\lambda}}{4\pi} (-m_{\lambda}^{2})^{n} \phi _{c}^{n}(r)\\
  & =\frac{m_{\lambda}}{4\pi} (-{\nabla}^{2})^{n} \phi_{c}^{0}(r),
\end{split}
\end{equation}
according to which, for the well-known form functions, we can write
\begin{equation} \label{GrindEQ__31a_}
F(\vec{k}^{2}) = 1 \Rightarrow \phi_{c}^{0}(r)=\frac{e^{-m_{\lambda}r}}{m_{\lambda}r},
\end{equation}
which is the usual Yukawa potential without the form function (the point-like nucleon);
\begin{equation} \label{GrindEQ__31b_}
F({k}^{2}) = \left(\Lambda^{2}-m_{\lambda}^{2} \right) / \left(\Lambda^{2}+{k}^{2}\right) \Rightarrow \phi _{c}^{0}(r)=\left[e^{-m_{\lambda}r}-e^{-\Lambda r}\right] /m_{\lambda}r,
\end{equation}
as the Monopole form-factor normalized such that at the pole, $F(-m_{\lambda}^{2})= 1$; and
\begin{equation} \label{GrindEQ__31d_}
\begin{split}
& F({k}^{2}) = \left(\Lambda^{2}-m_{\lambda}^{2}\right)^{2} / \left(\Lambda^{2}+{k}^{2}\right)^{2}, \\
\Rightarrow & \phi _{c}^{0}(r) = \left[e^{-m_{\lambda}r}-e^{-\Lambda r} \left(1+\frac{\Lambda^{2}-m_{\lambda}^{2}}{2\Lambda^{2}} \Lambda r\right)\right] / m_{\lambda}r,
\end{split}
\end{equation}
as the Dipole form-factor; and
\begin{equation} \label{GrindEQ__31f_}
\begin{split}
&  F({k}^{2}) = e^{-{k}^{2} / \Lambda^{2}}, \\
\Rightarrow & \phi_{c}^{0}(r)=e^{m^{2} / \Lambda^{2}}  \left[e^{-m_{\lambda}r} erfc \left(\frac{m_{\lambda}}{\Lambda} -\frac{\Lambda r}{2} \right)-e^{m_{\lambda}r} erfc \left(\frac{m_{\lambda}}{\Lambda}-\frac{\Lambda r}{2} \right)\right]/2m_{\lambda}r,
\end{split}
\end{equation}
as the Exponential form-factor with
\begin{equation} \label{GrindEQ__31h_}
erfc(y) = \frac{2}{\sqrt{\pi}} \int_{y}^{\infty} dt e^{-t^{2}},
\end{equation}
as the complementary-error-function. \\
It should be also mentioned that, without the form factors, one should use
\begin{equation} \label{GrindEQ__34bb_}
\phi_{c}^{1}(r) = \phi_{c}^{0}(r)-4\pi \delta^{3}(m_{\lambda} \vec{r})
\end{equation}
instead of $\phi_{c}^{0}(r)$ in the presence of the form factors. Besides, with the help of (\ref{GrindEQ__30a_}), one can get the tensor and spin-orbit potentials in terms of the central function of $\phi_{c}^{0}(r)$ as
\begin{equation} \label{GrindEQ__32a_}
\phi_{t}^{0}(r) = \frac{1}{3 m_{\lambda}^{2}} r \frac{d}{dr} \left(\frac{1}{r} \frac{d}{dr} \right) \phi_{c}^{0}(r),
\end{equation}
\begin{equation} \label{GrindEQ__32b_}
\phi_{ls}^{0}(r) = -\frac{1}{m_{\lambda}^{2}} \frac{1}{r} \frac{d}{dr} \phi_{c}^{0}(r).
\end{equation}
Therefore, one can see that with the dipole form-factor (in Reid93) and the exponential form-factor (in Nijm93, NijmI, NijmII) to regularize the potentials, the tensor function is vanished at the origin as well.\\
It is also good to mention the Fourier transform of the momentum-dependent terms (linear in ${q}^{2}$ in Nijm78, Nijm93, NijmI) in the p-space potentials, which lead to the nonlocal structure in r-pace as
\begin{equation} \label{GrindEQ__33a_}
\int \frac{d^{3}k}{(2\pi)^{3}} \frac{e^{i\vec{k}.\vec{r}}}{{k}^{2} + m_{\lambda}^{2}} \left({q}^{2}+\frac{1}{4}{k}^{2}\right) F({k}^{2})= -\frac{m_{\lambda}}{8\pi} \left[\nabla^2 \phi_{c}^{0}(r) + \phi_{c}^{0}(r) \nabla^2\right],
\end{equation}
(to see how to handle such nonlocal terms, look at \cite{Green2}) whereas the absence of the ${q}^{2}$ terms in NijmII (and also Reid93) potential in p-space leads to a radial local potential in r-space.

\subsubsection{Reid93 Potential}
The so-called regularized-Reid (Reid93) potential \cite{Nijm93} is fitted to the updated Nijmegen database, while the quality of the original Reid68 \cite{Reid68} \emph{np} data were poor. Besides, there was a $1/r$ singularity for all partial waves, which are now removed by including the dipole form factors (with the cutoff of $\Lambda=8m_{pi}$); and so the tensor potentials vanish at the origin. For the OPE part, as in the other Nijmegen high-quality potentials, the neutral-and charged-pion mass differences are considered (with $f_{\pi}^{2}=0.075$ again) and so Reid93 potential is charge-dependent. Meanwhile, in (\ref{GrindEQ__24eeee_}), $\phi_{c}^{1}(r)$ is used just for S-wave while, for other partial waves, $\phi_{c}^{0}(r)$ is used instead of $\phi_{c}^{1}(r)$.\\
Besides the OPEP tail, the potential in each partial-wave is parameterized separately by choosing suitable combinations of the central, tensor and spin-orbit terms with arbitrary masses and cutoff parameters. In Reid93 potential, with the coefficients of $\bar{m}=\left(m_{\pi_{0}} + 2m_{\pi_{\pm}} \right)/3$, $\Lambda= 8\bar{m}$, all potentials are written as linear combinations of the following functions
\begin{equation} \label{GrindEQ__36a_}
 \bar{Y}(p)= p\bar{m} \phi_{c}^{0}  (p\bar{m}, r),\quad \bar{Z}(p)= p\bar{m} \phi_{t}^{0}  (p\bar{m}, r),\quad \bar{W}(p)= p\bar{m} \phi_{ls}^{0} (p\bar{m}, r),
\end{equation}
with 50 coefficients of $A_{jp}$, and $B_{jp}$, which are used for isovector potentials, and isoscalar and \emph{np} ${}^{1}S_{0}$ potentials, respectively. These coefficients are fixed by fitting to the relevant \emph{pp}+\emph{np} scattering data. Here, $p$ is an integer and $j$ labels various partial waves, and that $\phi_{t}^{0}$ and $\phi_{ls}^{0}$ are some special radial functions \cite{Nijm93}.\\
One should note that, as in Reid68 potential, in the non-OPE part, for the singlet- and uncoupled triplet-states, the central potentials are used; and for the coupled triplet-states, the potentials having the central, tensor and spin-orbit terms as (\ref{GrindEQ__22_}) are used. For instance, for the uncoupled states of $(T=1, S = 0, L=J)$, they used
\begin{equation} \label{GrindEQ__37a_}
\begin{split}
& V_{pp}(^{1} S_{0})= A_{12} Y(2) + A_{13} Y(3) + A_{14} Y(4) + A_{15} Y(5) + A_{16} Y(6), \\
& V_{np}(^{1} S_{0})= B_{13} Y(3) + B_{14} Y(4) + B_{15} Y(5) + B_{16} Y(6), \\
& V(^{1} D_{2})= A_{24} Y(4) + A_{25} Y(5) + A_{26} Y(6), \\
& V(^{1} G_{4})= A_{33} Y(3), \quad V(^{1} J_{1})= V_{pp}(^{1} S_{0}),\ J\ge 6,
\end{split}
\end{equation}
where the different \emph{pp} and \emph{np} ${}^{1} S_{0}$ potentials are because of the CIB considered in the potentials. We should also mention that, for the coupled states, the potentials have clear forms up to $J=4$; and for the higher partial waves ($J\geq5$), a similar expansion as done by Day \cite{Reid68-Day} (look at subsection (\ref{4.5.2}) is performed. Clearly, for the triplet isovector (isoscalar) partial waves of $J\geq5$, the central and tensor potentials are those of the corresponding $S=1, T=1$ ($S=1, T=0$) $J<5$ partial waves while the spin-orbit potential is set to zero.

By the way, with 50 fitting parameters, Reid93 potential reproduces the result of $\chi^{2}/N_{data}=1.03$ such as the other Nijmegen HQ Potentials. It is also remarkable that the predicted values of the quantities, such as deuteron parameters and low-energy scattering parameters, by Reid93 potential (and also the other HQ potentials of Nijm93, NijmI, NijmII) has a good agreement with the experimental values \cite{Nijm93}. Nowadays, these HQ potentials are extensively used in nuclear structure calculations with many satisfactory results.

\subsubsection{Extended Soft-Core Potentials}
The already mentioned Nijmegen potentials based on OBE and ORE approaches described well the data but with many phenomenological arguments included. The extended-soft-core (ESC) potentials include extra exchanges and are more on the theoretical grounds. That is because by adding a few more free parameters, while preserving the previous advantages, the new potentials reproduce the data well. In the first ESC model \cite{NijmESC}, next to the whole previous exchanges of Nijm78 potential \cite{Nijm78}, they included some two-meson-exchange (TME) (and also $2\pi$-exchange) and meson-pair-exchange (MPE) contributions. That model described the Nijmegen PWA93 (\emph{pp}+\emph{np} database) in the energy range of 25-320 MeV with $\chi^{2}/N_{data}=1.16$, which was the first promising result in the case.\\
The next completed version was presented in 1995 \cite{NijmESC01}. In fact, besides the previous OBE's and ORE's of Nijm78 potential \cite{Nijm78}, TPE's, TME's and MPE's were included as well. In general, TME contributions are from $\pi \otimes \rho, \pi \otimes \omega, \pi \otimes \eta, \pi \otimes \epsilon, \pi \otimes P, ...$, where the parallel and crossed-box Brueckner-Watson diagrams with Gaussian form factors are computed. The MPE contributions are due to one-pair and two-pair $(\pi \pi, \pi \rho, \pi \omega, \pi \eta, \pi \epsilon,...)$, where a Gaussian form-factor (as $e^{-k^2/2\Lambda^2}$) is attached to each vertex. The interaction Lagrangian's are for the effective relativistic theories, with LS equation. The ESC potentials were compared with Nijm93 potential \cite{Nijm93} in that at least both include 14 free parameters. Next, the ESC potentials gave better results with more theoretical grounds besides using the chiral symmetry of the involved Lagrangian's.\\
The chiral-invariant ESC model for NN interaction with 12 free parameters reproduced the data in the energy range of 0-350 MeV with $\chi^{2}/N_{data}=1.75$ \cite{NijmESC02}. It is notable that the TME contributions improved the ESC potential quality with respect to the previous OBEP's. In addition, one notes that the meson-pair vertexes in the triangle and double TME diagrams (supposed to simulate the heavy mesons and resonance degrees of freedom) are analyzed in principle by chiral-symmetry and so, these contributions do not introduce any new parameter. \\
The YN and YY versions of the ESC potentials were then reported in \cite{NijmESC03} first, where next to discussing the usual boson-exchanges, the interactions were discussed in the framework of QCD, flavor SU(3), and chiral SU(3)$\otimes$SU(3) for the low-energy region as well. Then, in 2000, besides reviewing the Nijmegen SC potentials, a new ESC model (called ESC00) was presented to describe NN, YN and YY systems in an unified manner by using $SU(3)_f$ symmetry \cite{NijmESC04}. In the energy range of 0-350 MeV, it described YN and NN systems with $\chi^{2}/N_{data}=1.15$.\\
After that, they modeled the comprehensive p-space versions of the ESC NN potentials in \cite{NijmESC1}. With 20 free parameters (of the masses and coupling constants) there, they reproduced the NN scattering data in the energy range of 0-350 MeV again with $\chi^{2}/N_{data}=1.15$. Some new improvements were because of including the axial-vector mesons and a zero in the scalar meson form factors--It is also mentionable that the SC meson-baryon interactions were discussed in \cite{NijmSC22} as well. From 2005 onwards, some new generations of the ESC BB potentials (called ESC04) have been presented, where the contributions from OPE's, ORE's, MPE's and two-pseudoscalar-meson-exchange (PS-PS exchange) are also included. There are the NN interaction in \cite{NijmESC2}, the YN interaction in \cite{NijmESC3}, and the BB states with the total strangeness of $S = -2$ in \cite{NijmESC4}. For some recent reviews of the Nijmegen ESC potentials, see \cite{NijmESCall1}, \cite{NijmESCall2}.

\subsubsection{Nijmegen Optical Potentials} \label{GrindEQ__NijmOP_}
\begin{figure}[htp]
\centering
 \includegraphics[height=2.8in, width=5.6in, scale=1]{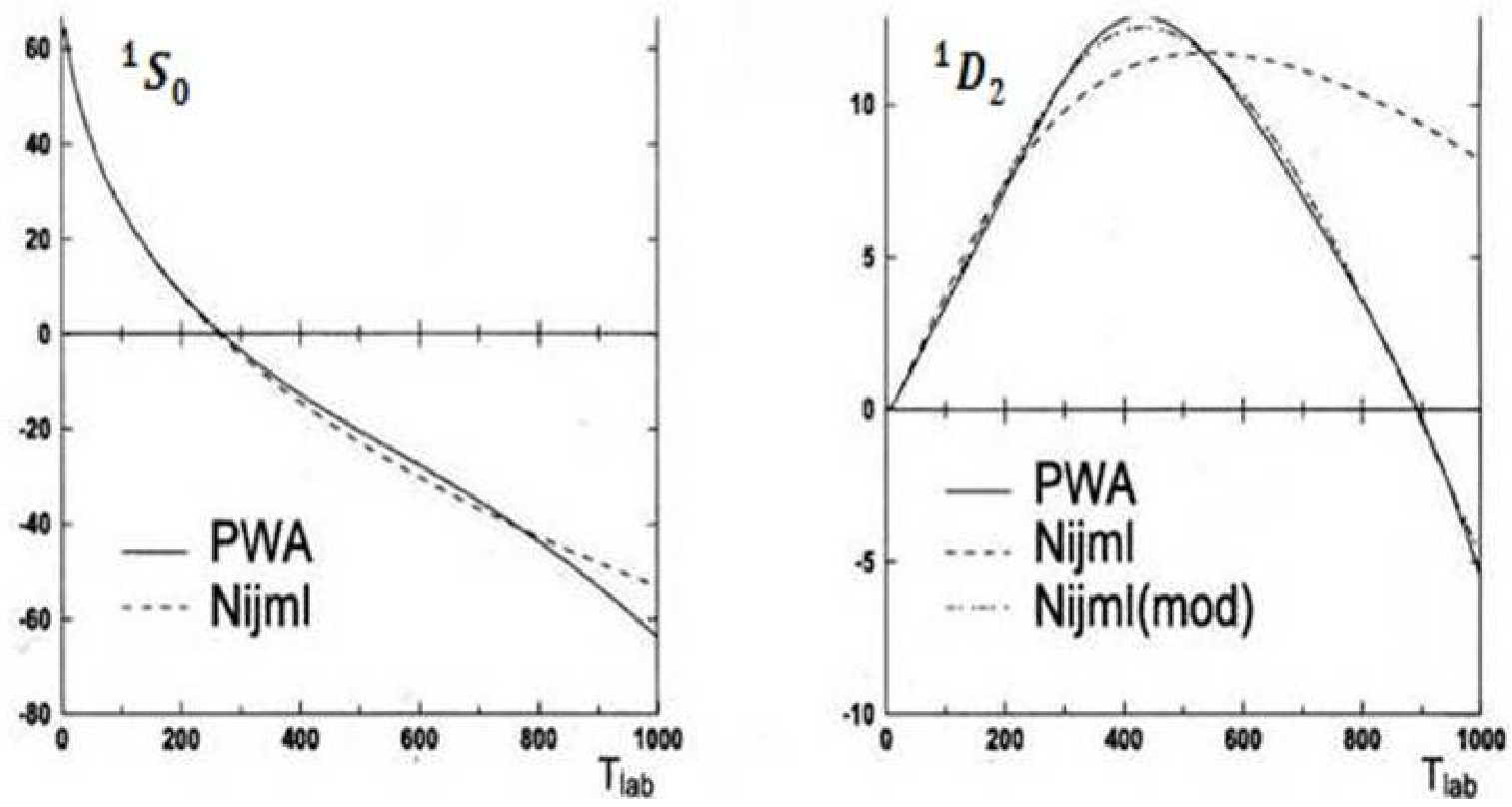} \\
 \caption{\textit{The phase shifts $^{1} S_{0} $(left), and $^{1} D_{2}$(right) for the NijmI optical potential and a modified version of that (quoted from \cite{Nijmegen}).}} \label{Fig3.}
 \end{figure}

NN potentials are often considered to be real below the pion production threshold in about $T_{lab}=$290 MeV. One way to include inelasticity's at the high energies, above the thresholds, is to consider optical potentials. On the other hand, we saw that in the Nijmegen PWA93 for the short distance potential ($V_{SR}$), below $r<b$ (with $b=1.4 fm$; look at subsection of (\ref{4.15.3})), the energy-dependent square wells were used. Now, one may write \cite{NijmOPT1}
\begin{equation} \label{GrindEQ__38a_}
V_{S}=V_{Rel} - iV_{Img},
\end{equation}
where the real SR potential $V_{Rel}$, which is different for each partial-wave, always reads
\begin{equation} \label{GrindEQ__38bab_}
V_{Rel}=\sum_{n=0}^N a_n (k^2)^n,
\end{equation}
and the imaginary SR potential $V_{I}$ is taken as
\begin{equation} \label{GrindEQ__38bab_}
V_{Img}=(k^2-k_{th}^2)V.\theta(E-E_{th}).
\end{equation}
It was established from the Nijmegen PWA's \cite{NijmPWA, NijmPWA2} that the fully real potentials work up to about 500 MeV quite well. Nevertheless, the optical potentials of the Nijmegen-group could be constructed by adding, to the real HQ potentials, the same imaginary part used in the Nijmegen PWA's of the \emph{np} data below 500 MeV, according to the above prescription of course. But, the resultant optical potentials did not give good results for all partial waves in that energy range. Clearly, if one considers all \emph{np} date below 1 GeV, some differences among the results from the preliminary PWA's and the above constructed optical potentials arise in some partial waves. For instance, as it is clear from Figure \ref{Fig3.}, for all \emph{np} data below 1 GeV, the phase shift of ${}^{1}S_{0}$ is well described by both PWA and the optical NijmI potential; whereas for the phase shift of ${}^{1}D_{2}$, the large differences are recognizable clearly. Still, by refitting, the modified NijmI optical potential, NijmI (mod), is obtained that gives a good fitting of the ${}^{1}D_{2}$ phase shift up to 1 GeV. Therefore, it seems that it is not so difficult to model the optical potentials to fit the \emph{np} data up to $T_{lab}=$1 GeV.

\subsection{Hamburg-Group Potentials}
Hamburg potentials are also among meson-exchange models related to QCD substructures. Since, in Bonn-B potential \cite{Bonn2}, the scattering amplitudes were obtained from the meson-baryon Lagrangian in a clear and comprehensive way, that model was used as a base to build one-solitary-boson-exchange potential (OSBEP) by this Hamburg-group. In fact, the Hamburg-group potential somehow refines the common boson exchange picture by seeking for a procedure that reduces markedly the number of free parameters of the conventional boson exchange potentials; whereas the parameters are in turn needed for the quality fittings of the potentials to scattering data. Because of its special mechanism, there is not any cutoff parameter as the only adjustable parameters are the pion self-coupling constant and meson-nucleon coupling constants. The first version \cite{Jaede} was presented in 1996, to fit just elastic \emph{np} scattering data up to $J\leq3$, and was further developed in \cite{Hamburg2} to include both \emph{np}+\emph{pp} data with more improvements.\\
In fact, the features of both QCD-inspired models and common phenomenological boson-exchange potentials are included in OSBEP. We know that below the pion production threshold, chiral symmetry is supposed to be broken. Now, a meson Lagrangian, by including all OBE contributions, with a similar structure as the linear sigma-model, is considered. Because the symmetry conditions are not imposed on the masses and coupling constants, but the latter are used as free parameters in the Lagrangian, the chiral symmetry is so broken. Therefore, the spontaneous chiral symmetry breaking leads to nonlinear terms in the meson part of the Lagrangian. The resultant decoupled nonlinear meson equations are then analyzed, and semiclassical solutions are quantized leading to defining "solitary mesons" from which the propagators come out. Indeed, the nonlinear features of the QCD-inspired models are considered as the nonlinear boson equations result. Because of the nonlinear property of the boson, the form factors are not as those in the Bonn-B potentials; and they are replaced by properly normalized solitary meson fields. As another result, an experimental scaling law arises that relates all the meson parameters and so, reduces the number of fitting parameters. The model also gives a good quantitative description of experimental data and deuteron properties though the quality is not as high as the other mentioned HQ potentials \cite{Hamburg2}.\\
Meanwhile, in \cite{nucl-th/9609054}, NN interaction is discussed from quantum-inversion approach versus meson-exchange picture and especially from OSBEP. In general, we know that by inserting a potential with its special operators into LS equation, one can earn phase shifts and other observables. In full generality, this method involves quark and gluon substructures, and then both on-shell and off-shell data are described well. However, the inversion potentials are in general local and energy-dependent in r-space, whereas BEP's lead to nonlocal potentials in p-space mainly. It is established there that the results from quantum-inversion and boson-exchange potentials are almost same. A main difference is the larger D-state probability for the local potentials, which is in turn related to the different tensor part of the potentials.\\
The next improvement in studying OSBEP's was to extend them for pion-nucleon interactions as well, as in \cite{Hamburg22}, where the OSBEP model was recast into a unique form for NN and $\pi$N interactions. To do so, the $\Delta$-isobar was included besides the chiral-symmetry preserving pseudovector meson-baryon coupling (PS $\pi$NN) instead of the previous pseudoscalar (PS) coupling for $\pi$NN. Describing NN and $\pi$N interactions simultaneously was good as the previous results for just NN interaction.

In 2003, von Geramb et al. proposed another NN potential based on Dirac equations (two coupled Dirac equations with constraints from dynamics) combined with meson-exchange picture (including the $\pi, \eta, \rho, \omega, \sigma$ exchanges) \cite{vonGeramb3}. The resultant potentials, to use in partial-wave Schrodinger-like equations, inspired by meson exchanges, fitted the Arndt et al. \emph{pp}+\emph{np} phase shifts of $T_{lab}=$0-3 GeV \cite{Arndt2000} as well as deuteron properties. The analyses showed a universal core potential coming from relativistic meson-exchange dynamics, and that the high-energy effects such as those of QCD and inelasticity were included. Besides the Dirac meson-exchange potentials, they framed some local and nonlocal optical potentials, which still gave good agreement between theoretical and experimental data.

\subsection{Moscow-Group Potentials}
Moscow-type (M-type) potentials are mainly a hybrid of the quark-model and meson-exchange picture. In general, in short distances, the quark and gluon degrees of freedom are used; whereas for the LR and MR regions, OBEP's+TPEP's are often used. The first major version was presented in 1997 \cite{Kukulin} and latter improved in \cite{Moscow2}. The main features of these potentials are the emphasis on the deep substructures from QCD. One main difference of the M-type potentials from the other NN potentials is that the common SR local repulsive core from the $\omega$ (and also $\rho$) meson is strongly reduced, and is indeed replaced by that of a suitable orthogonality condition plus a deep attractive potential. The orthogonality condition may be interpreted as projecting the compound six-quark states ($\varphi_0$), with the maximal permutational symmetry, into asymptotic NN channels. As a result, one now has a node around $\sim 0.6 fm$ that plays the role of the repulsive core and provides NN phase shifts. Meanwhile, the potential also has strong attraction in intermediate parts commonly assigned to the pseudoscalar-meson exchanges of $\pi, \eta$ and the scalar-meson exchange of $\sigma$. Still, for the SR repulsion of the $\omega$ meson, a repulsive core with a Gaussian form-factor with a positive finite coupling constant is included. In other words, in the short distances ($r\lesssim 1 fm$), the nonlocal and energy-dependent terms, which are in turn coming from the retardation effects and six-quark bags (6q-bags), are replaced by a separable potential.\\
By the way, the main M-type potential in \cite{Kukulin}, with two major parts, reads
\begin{equation} \label{GrindEQ__71aa_}
V=V_{loc.}^{(0)} + V_{sep.},
\end{equation}
where the local potential ($V_{loc.}$) is $\ell$-independent and includes an OPEP and an attractive well, and is depended on the spin and parity of NN system. The separable potential ($V_{sep.}$) is a state-dependent (depended on the $\ell, J$ of NN system) repulsive core with a Gaussian form-factor. They are
\begin{equation} \label{GrindEQ__71bb_}
V_{loc.}= V_{c}(r)+V_{t}(r)S_{12} = V_0 e^{-\eta r^2}+ V_{OPEP}^{(6)}(m_{pi})+\lambda |\varphi><\varphi|,
\end{equation}
in which
\begin{equation} \label{GrindEQ__71cc_}
V_{OPEP}^{(6)}(m_{pi})= -\frac{f_{\pi NN}^2}{4\pi} \frac{\mu^3}{4M^2} \left[f_{tr} Y_c(x) \vec{\sigma}_{1}.\vec{\sigma}_{2} + (f_{tr})^n Y_t(x) S_{12} \right],
\end{equation}
and
\begin{equation} \label{GrindEQ__71dd_}
\varphi=N r^{\ell+1} e^{-\frac{1}{2}\left(\frac{r}{r_0}\right)^2}, \quad \int \varphi^2\ dr =1, \quad f_{tr}=\left(1-e^{-\alpha r}\right),
\end{equation}
where $x=\mu r$ as usual with $\mu$ for the average pion-mass, $n$ as the power of the cutoff factor of $f_{tr}$ is different for the different versions of the potential, $r_0$ is the radius of the repulsive core (different slightly for the different states), $\alpha$ is the cutoff radius of the OPEP, and $\lambda$ is different for different $\ell, J$'s. Because of the freedom to choose the parameters of $\eta, \alpha, V_0$, one can set $\eta=\alpha^2$; and then the width of $\eta$ and the depth of $V_0$ are fitted to scattering data, which are the scattering length and the effective range of the ${}^1S_0$ wave here. It is mentionable that the repulsive core is absent for $\ell \leq 4$ ($T_{lab}< 400$ MeV) because of the second term in (\ref{GrindEQ__71cc_}) or the presence of $Y_t(x)$. Anyway, this M-type potential describes well deuteron properties and NN scattering data up to 500 MeV with 6 free parameters, which are in turn physically meaningful. Also, the off-shell behavior of the potential can be checked in NN bremsstrahlung. It should also be mentioned that, contrary to the other quark-meson hybrid models that use the mixtures of both and so lead to energy-dependent nonlocal potentials, here the quark and meson exchanges are orthogonal besides giving a microscopic description of NN interaction.

In the complete version presented then, the Gaussian form function was replaced by an exponential form function to describe better the phase shifts especially ${}^3S_1-{}^3D_1$ phase-shift. The potential so is written as
\begin{equation} \label{GrindEQ__72aa_}
V=V_{loc.}^{(M)}+V_{OPEP}^{(7)} + V_{sep.},
\end{equation}
in which
\begin{equation} \label{GrindEQ__72bb_}
V_{loc.}^{(M)}= V_0 e^{-\beta r}+ V_0 e^{-\beta r} \vec{L}.\vec{S},
\end{equation}
and
\begin{equation} \label{GrindEQ__72cc_}
V_{OPEP}^{(7)}(\mu)= -\frac{f_{\pi NN}^2}{4\pi} \frac{\mu}{3} (\vec{\tau}_{1}.\vec{\tau}_{2}) \left[ Y_c^{(2)}(x) (\vec{\sigma}_{1}.\vec{\sigma}_{2}) + Y_t^{(2)}(x) \hat{S}_{12} \right],
\end{equation}
where this OPEP is written with a soft dipole form-factor, and now the tensor potential becomes zero at the origin as it must be; $\hat{S}_{12}={S}_{12}/3$ and $\mu=(m_{\pi_0}+2m_{\pi_\pm})/3$; and now, the same as in Arg94 potential \cite{AV18} and Nijmegen HQ potentials \cite{Nijm93}, $f_{\pi NN}^2/{4\pi}=0.075$; $\alpha=\Lambda/\mu$, and
\begin{equation} \label{GrindEQ__72dd_}
\begin{split}
& Y_c^{(2)}(x)=Y_c(x)-\alpha Y_c(\alpha x)- \left(\alpha^2-1\right) \frac{\alpha^2}{2} x\ Y_c(\alpha x), \\
& Y_t^{(2)}(x)=Y_t(x) - \alpha^3 Y_t(\alpha x),
\end{split}
\end{equation}
with the notations in (\ref{GrindEQ__22ff_}). Here just $V_0, \alpha, \beta$ are free parameters in the local part of the potential, which are in turn different for each spin and parity combination. In addition, the parameters of $\lambda$ and $r_0$ are independent for D- and F-waves; whereas for S- and P-waves, $\lambda$ goes to infinity and $r_0$ values are depended on the depth of the attractive local potential. In general, with 32 parameters (a similar number to the so far mentioned HQ potentials) and the $\pi$NN coupling constant, describing deuteron properties and partial waves in the energy range of 0-400 MeV was very good (except for few higher $\ell$'s).

Next, they developed a new mechanism to describe NN interaction in MR and SR parts \cite{Moscow2b}. In fact, instead of the oldest Yukawa formalism for SR interaction, a 6q-bag model, dressed because of the $\pi,\rho, \sigma$ mesons, was used there. That in turn produced an MR attraction that replaced the conventional $\sigma$-meson exchange. On the other hand, the $\rho$ meson, produced in the intermediate six-quark state, caused a nonlocal spin-orbit interaction in the SR part. As a result, the MR attraction and a part of the SR repulsion were described excellently, whereas the SR repulsion was mainly because of the orthogonality of NN- and 6q-channels. \\
In other words, in the common OBE models, there are still many problems. For instance, the cutoff parameters $\Lambda_{m_{\lambda} NN}$ are often larger than the experimental values got by fitting the data; the phenomenological Yukawa functions have also at least the base theoretical problems; and discussing the $\sigma$ meson, as a $2\pi$ resonance in S-state, is also controversial. Further, describing 3N and 4N systems with the settled OBEP's is not addressed satisfactory yet. Therefore, the new M-type potentials try to address some of the existing problems.\\
Here, with the dressed 6q-bag, the $\sigma$- (and even $\rho$-) meson exchange between nucleons is considered because of the transitions between the p-shells of the excited quarks. In other words, each p-shell quark emits a pion and during the transition from the p-shell to s-shell, the pions are absorbed by the di-quark pairs in the intermediate 6q bag-like states (suppose as $qq\rightarrow \sigma+qq$). Further, the $\sigma$ meson, as a scalar-isoscalar excitation of the QCD vacuum, is considered as a quasiparticle inside the hadrons (especially in a multi-quark bag) and not as a real particle in the free space. Therefore, the scalar-isoscalar $\sigma$ meson exists just in a high-density medium and not in the vacuum (contrary to the $\rho, \omega$ mesons).

One can show the main features of the model by a simple phenomenological potential as
\begin{equation} \label{GrindEQ__73aa_}
V=V_{orth.}+ V_{NqN}+V_{OPEP}^{(7)},
\end{equation}
where
\begin{equation} \label{GrindEQ__73bb_}
V_{orth}= \lambda_00 |\varphi_0><\varphi_0|, \quad (\lambda_0\rightarrow \infty),
\end{equation}
as the orthogonality potential, provides the orthogonality condition between the intermediate 6q-bag and the especial NN channel for S- and P-waves. $V_{NqN}$ is the separable potential attributable to the virtual transition of $NN\rightarrow (6q+2\pi)+NN$ as
\begin{equation} \label{GrindEQ__73cc_}
V_{NqN}= \frac{E_0^2}{E^2-E_0^2} \lambda |\varphi><\varphi|,
\end{equation}
for the single channels and
\begin{equation} \label{GrindEQ__73dd_}
V_{NqN}= \frac{E_0^2}{E^2-E_0^2} \left(
                                   \begin{array}{cc}
                                     \lambda_{11} |\varphi_1><\varphi_1| & \lambda_{12} |\varphi_1><\varphi_2|\\
                                     \lambda_{21} |\varphi_2><\varphi_1| & \lambda_{22} |\varphi_2><\varphi_2| \\
                                   \end{array}
                                 \right),
\end{equation}
for the coupled channels. $E_0\approx 600-100$ MeV is a sum of the 6q-bag energy and the $\sigma$-meson mass inside the 6q-bag, and is same for all partial waves with definite parities. The expression for $\varphi_i$ is that in (\ref{GrindEQ__71dd_}) with $\varphi_i, \ell_i$ here instead of $\varphi, \ell$ there. The potential parameters of $\lambda_{jk}(=\lambda_{kj})$, $r_0, E_0$, the phase shifts and the mixing parameters $\varepsilon_1$ are determined by fitting the data with the cutoff parameter of $\Lambda=\Lambda_{dipole}=0.50-0.75$ MeV. The resultant potential describes the partial waves of $\ell <2$, which is in turn equivalent to describing the phase shifts in the energy range of 0-600 MeV and S-waves for an energy about 1200 MeV. Further, the weak contributions because of the vector mesons in the baryon spectra and the strong spin-orbit splitting are explainable by this constituent-quark model. Still, the model leads to some especial 3N and 4N forces because of $2\pi$ and $\rho$ exchanges. \\
In 2005, the group described elastic and inelastic NN scattering's in the energy range of 1-2 GeV by a special EFT \cite{Moscow3}. The previous approach, to describe MR and SR interactions in \cite{Moscow2b}, which made use of six-quark bags and the intermediate mesons of $\pi, \sigma, \rho, \omega$, were employed there and improved as well.

The predictions of these M-type potentials for 3N systems were also analyzed in detail with good results. Large deviations from the conventional NN potentials were established for the momentum-distribution in the high-momentum region. In particular, the coulomb displacement energy for the nuclei of ${}^{3}He-{}^{3}H$ displayed a promising agreement with experiment when the binding energy of $^{3} H$ was extrapolated to the experimental value \cite{Kukulin}, \cite{Moscow2}. Further, in \cite{Moscow2b}, by using a new MR NN interaction model, based on the QCD bag model, an effective energy-dependent NN interaction was constructed. The new potential described experimental data up to 1 GeV and deuteron parameters well. Generalizations of the model to three-nucleon force (3NF) and other related issues were also discussed \cite{Moscow3}. Other mechanisms for the MR and SR parts of NN interaction are addressed by the group members in the references of \cite{Moscow3}.\\
It is remarkable that the M-type potentials, and also the following two potentials/models, do not have necessarily similar structures as the other standard ones and are almost special.

\subsection{Budapest(IS)-Group Potential}
Doleschall et al. set up a set of NN interactions in r-space to get the correct binding energy of triton \cite{IS1}. The potentials are nonrelativistic and almost phenomenological, nonlocal and energy-independent. Nucleons are discussed as point-like objects and effects from their structure is supposed to come from effective NN potentials. In some specific short regions, the potentials are considered to be nonlocal, and in the outside regions as some local Yukawa tails. The first aim was to find a nonlocal potential form to describe triton (${}^3H$ as a 3N bound system) binding energy as well as describe experimental phase shifts and deuteron properties. Later, they used those nonlocal NN potentials to describe some other 3N bound states \cite{IS2}. In fact, they modeled an NN potential respecting the well-known local behaviors in long ranges, whereas it showed a nonlocality at the shorter ranges. The resultant potential provided a satisfactory fit to NN scattering data while including CI and CS. The nonlocality in the NN potential guaranteed that no 3N forces were required to describe 3N bound-states.

\subsection{MIK-Group Potential}
The $J$-matrix inverse scattering approach to make NN potentials was started by Zaitsev et al. in \cite{MIK00}, \cite{nucl-th/0103010}, and was developed by Shirokov et al. in \cite{MIK0}. Indeed, the nonlocal interactions gained in this approach are in the forms of some matrices in oscillator basis in each NN partial-wave separately. In other words, in the approach NN interaction is as a set of potential matrices for various partial waves. However, a main aim to make the potentials was to earn some satisfactory results in nuclear calculations of 3N systems and other light nuclei.\\
In the first serious try \cite{MIK0}, based on \cite{MIK00}, \cite{nucl-th/0103010}, they held the inverse scattering tridiagonal potentials (ISTP), which are tridiagonal (quasi-tridiagonal) in the uncoupled (coupled) partial waves. The dimension of the potential matrix was determined by the maximum value of $N=2n+\ell$ (note that the common nonrelativistic Schrodinger equation is used in the approach), and was refereed as a $N\hbar\omega$ potential. We should note that the resultant interactions are somehow effective and are not related to the usual meson-exchange theories though the main features and results of NN interaction are common. It is also notable that, to describe a wider energy range, the size of the potential matrix, or the oscillator basis parameter $\hbar \omega$, must be increased. The potential \cite{MIK0} was used in nuclear calculations of ${}^3H, {}^3He$ with giving good results also in describing NN scattering data.\\
In \cite{MIK1}, similar to that in \cite{MIK0}, another class of the $J$-matrix inverse scattering potentials (JISP), called JISP6, was constructed. The resultant potentials described well NN scattering data as well as the bound- and resonance-states of the light nuclei up to $A=6$. A remarkable feature of the potentials was that by using the off-shell degrees of freedom, there was not any need to include 3N potentials to describe well the light nuclei. The results to evaluate binding energies of the nuclei ${}^3H, {}^3He, {}^4He, {}^6He, {}^6Li$ were as well as the results of the other HQ potentials such as NijmI, NijmII \cite{Nijm93}, Arg94 \cite{AV18}, and CD-Bonn \cite{Bonn4} potentials. In \cite{MIK1}, the base parameter was $\hbar \omega=40$ MeV, and that the potential described the Nijmegen PWA93 data \cite{NijmPWA} with $\chi^2/N_{data}=1.03$; look also in \cite{Shirokov}.\\
Next, they set up a JISP16 version and then further developed it as a ISP16$_{2010}$ version \cite{MIK2}. The latter potentials were used to evaluate binding energies and spectra of the light nuclei in No-Core-Shell-Model (NCSM) calculations. In a recent study \cite{MIK3}, the progress in developing the JISP NN interactions and other related issues are discussed as well.

\subsection{Imaginary Potentials}
As we know, above the pion production threshold, the inelasticity's and other high-energy effects become important, and then one way to incorporate them is to consider optical or imaginary potentials suitable also to earn high quality descriptions of scattering data in medium and high energies. Among a few imaginary NN potentials, we discussed briefly the Nijmegen ones in subsection of (\ref{GrindEQ__NijmOP_}). In \cite{nucl-th/0105011}, some NN potentials such as Paris, Nijmegen and Argonne potentials, and those traced by quantum inversion, which describe NN interaction for the energies below 300 MeV, are extended to NN optical potentials in r-space. The up-to-date phase-shift analyses, from 300 MeV to 3 GeV, are used to settle the extensions. The imaginary parts of optical potentials account for the flux losing into direct or resonant production processes. The optical potential approach is interesting as it allows one to imagine fusion and resulting fission of nucleus when T-lab energies are above 2 GeV.\\
Discussions about optical potential from quantum inverse-scattering and scattering data as well as modeling an optical potential are also given in \cite{Knyr1}. There is also a relativistic optical NN potential, based on some idea of M-type potentials, in \cite{Knyr2}.

\subsection{QCD-Inspired Potentials}
The QCD-inspired models always use the fundamental quark and gluon degrees of freedom instead of mesons. Indeed, because the SR part of NN interaction is more related to quarks and gluons, so the QCD-inspired potentials are used more to describe this part. Often one uses the hybrids of quarks and mesons to describe the interaction. There, the SR interaction is always attributable to gluon exchanges and the MR and LR parts come from scalar- and pseudoscalar- meson exchanges. One, of course, usually uses OPEP's for the LR part, while the MR part is handled by phenomenological or TPEP's, as we described some potentials briefly. Nevertheless, one should note that by applying the spontaneous chiral symmetry breaking to QCD Lagrangian, one may be able to set up an NN interaction fully based on QCD. Then, the common picture of the nucleon is a quark core surrounded by a pion cloud. So, in large distances, one may use an effective meson-nucleon theory to describe nucleon interactions. In the case, the form factors with free parameters in short distances are often used. For a typical review up to 1988 see \cite{FredMyhrer}, and \cite{nucl-th/9809093} for a study on NN QCD models up to 1998, and \cite{nucl-th/0212044} for a historical and technical review on QCD-inspired models up to 2002. Discussing various aspects of the QCD-inspired NN potentials needs another opportunity and is not in the level and aim of the current note. Nevertheless, we try to address some progresses and more plain potentials.

In fact, a pioneer study of the SR repulsion of NN interaction in the framework of quark model was done in \cite{QCDNN00}. In the first studies, the quark and gluon dynamics (especially one-gluon-exchanges (OGE's) were included), to give quantitative description of SR part, were employed. Then, especially in 1980's, the hybrid quark models were constructed (among the first samples are in \cite{MOkaNN00}), where for the LR and MR parts they always used the potentials from other phenomenological and boson-exchange models; look also at \cite{PRD34-1986}. Describing scattering data and deuteron properties with the earlier quark potential models was not so satisfactory. In \cite{Shimizu00},  nonrelativistic quark-cluster models were used to describe BB, NN and YN interactions especially in the MR's and SR's with some good descriptions. After that, the hybrid models (including quark, gluon and meson (especially pion) exchanges) were developed and more improved. Among the earlier hybrid models, that is an NN potential \cite{VinhMau1} made from Paris potential \cite{Cottingham} for long- and intermediate-distances with the quark-cluster model (QCM) for short distances. There, the effects of the quark degrees of freedom on NN observables were surveyed. But, describing its time \emph{pp} data was not good except if one used some other adjustable potentials in the LR and MR parts. \\
In 1990's, chiral constituent quark models (CCQM) were framed, which were always considered as a result of the spontaneous symmetry breaking of QCD Lagrangian. There, OPE's, OSE's (S for sigma) and OGE's were included besides a phenomenological confining potential. The resultant potentials described NN scattering data and deuteron properties better than any other QCD-inspired potential at that time. They are many of the CCQM potentials that we mention just some studies. In \cite{hep-ph/0211443}, the SR NN interaction is described by a CCQM model as well, where the constituent quarks interact through pseudoscalar meson exchanges. There, projecting the six-quark wave-function into NN channel produces an SR node for S-waves, like M-type potentials \cite{Moscow2}. So, the short distances are described microscopically, whereas the medium and large distances are described through the Yukawa pion and sigma meson between the quarks belong to the nucleons. The CCQM models are further addressed in \cite{nucl-th/0212044}, where it is discussed that they describe well the LR attractive and SR repulsive features in addition that they are universal in describing all baryons on equal footings--Look also at \cite{hep-ph/0205138}.\\
Among other QCD-inspired NN potentials, the hybrid quark-meson models in \cite{nucl-th/9606020} and \cite{JapanQCDNN} are notable, where the former was to apply to finite nuclei calculations and is based on a relativistic quark model. The latter \cite{JapanQCDNN}, which we call it \textbf{Japan-group} potential, is a unified model to describe NN and YN interaction and, with few parameters, gives a good fit of its time scattering data. That is another potential model for NN, $\Lambda$N and $\Sigma$N interactions in \cite{Wu0}, which we call it \textbf{China-group} potential, which is completely based on QCD ingredients without any use of meson exchanges. The China-group potential is based on a quark delocalization color screening model (QDCSM) and describes the SR and MR interactions simultaneously besides a claim that it describes well NB (NN, N$\Lambda$, N$\Sigma$) scattering data.\\
It is also good to mention an extension of the chiral SU(3) quark model (CSQM) to describe BB interactions. Indeed, in CSQM, a nonet of pseudoscalar and a nonet of scalar meson exchanges are used to describe the LR and MR parts of interactions, while the SR part is described by OGEP's and also quark-exchange effects. The model gives a good description for NN and YN systems. In \cite{nucl-th/0404004}, to study whether OGE's or vector-meson exchanges could describe the SR part of the interaction, the CSQM was extended to include vector-meson exchanges as well. In the resultant extended chiral SU(3) quark model (ECSQM), the strength of OGE was largely reduced and the SR repulsion was owned to a combined effect of pseudoscalar and scalar mesons, and was better described with a good fitting of scattering data. \\
The effects of the quark model calculations in the SR part on phenomenological and meson-exchange calculations in the MR and LR parts are studied further in \cite{nucl-th/0010044}. It is settled that the QCD quark models cannot describe the higher partial waves though they could describe the lower partial waves well. Therefore, the hybrid models are indeed essential to describe well scattering data. Then, one should employ LR and MR potentials from the other high-quality meson-exchange or phenomenological potentials next to quark-model potentials to describe the SR part. The potentials so describe experimental NN data and bound states fairly but they are not still as good as the fully phenomenological and meson-exchange HQ potentials.

It should be mentioned that, for the QCD-inspired models, some criteria are more important. Choosing a proper quark model, selecting suitable six-quark ground states, and the methods to evaluate phase shifts are important. It is also notable that the M-type NN potentials \cite{Kukulin} are other clear QCD-inspired potentials, and also the Oxford potential \cite{Oxford1} that we discuss below briefly.

\subsection{The Oxford Potential}
The Oxford potential is among QCD-inspired potentials. The group has applied nonrelativistic constituent-quark models to low-energy NN interaction. They have shown \cite{Oxford1} that the potential reproduces well NN scattering data and deuteron properties as the high-precession potentials such as CD-Bonn \cite{Bonn4}, Nijmegen I \cite{Nijm93}, ArgonneV18 \cite{AV18} and NNNLO \cite{Idaho1} potentials. Indeed, in the Oxford potential, a combination of one-pion-exchange (OPE), one-sigma-exchange (OSE) and one-gluon-exchange (OGE), next to using the charge dependence from CD-Bonn potential, and some other subtleties are involved.

\subsection{The First CHPT NN Potentials}
Although some of the mentioned phenomenological and boson-exchange potentials are related to QCD especially in the SR part, the relation is not systematic and consistent. The works by Weinberg \cite{Weinberg} and considering a Lagrangian, which includes the chiral symmetry of QCD, written in terms of pions and nucleons and their covariant derivatives, was a starting point to build new generations of NN interactions. In the case, the mesons and higher degrees of freedom could be integrated out as their effects might be considered as some undetermined coefficients and higher order terms. From the resultant effective Lagrangian, the potential so is expanded systematically in the powers of $(Q/\Lambda_{QCD})$, where $Q$ is a typical involved momentum. Therefore, the resultant potential is consistent with QCD symmetries and is a logical and systematic way to describe NN interaction and relate it to QCD. Detailed studies in the case need more space and time and are not aim of this concise study. Nevertheless, for more preliminary details and references, look at the subsections of (\ref{3.3}), (\ref{3.4}), (\ref{3.5}).

First, Ordonez, Ray and van Kolck (Texas-group) \cite{Ordonez1}, in 1993, proposed an exact two-nucleon potential based on an effective chiral Lagrangian. For intermediate states, they considered at least one pion ($\pi(140)$) and one isobar ($\Delta(1232)$), and that the resultant NN potential was a sum of the involved irreducible diagrams. NN scattering amplitudes were then evaluated by inserting the potential into LS or modified Schrodinger equation. The lowest order of that perturbative expansion was because of tree graphs which resulted in an LR part OPEP. Still, other diagrams up to the third order of chiral expansion, up to one-loop diagrams, reproduced the other known features of NN interaction such as SR repulsion, MR attraction, spin-orbit force, and many others. The potentials were written in p-space first, and in terms of some operators in (\ref{GrindEQ__22dd_}) and more dependencies of the functions on $k^2, q^2$. The dependence on $k^2$ was common, whereas the $q^2$ dependence was not usual. In addition, the energy dependence in the static OPEP, which was in turn more improved than the previous ones, was because of the recoil effect of pion emission from nucleon. The MR part was because of TPEP with many parameters, while the form factors were used to regularize the potential at the origin. Indeed, for the Fourier transform into r-space, the Gaussian form factors with the cutoffs $\Lambda$ as $e^{k^2\diagup \Lambda^2}$, as in Nijmegen potentials \cite{Nijm93}, were used.\\
The general form of the potential in r-space can be written as (\ref{GrindEQ__23aa_}) with $n=20$,
where the functions $V^{i}$ here are in term of the radial coordinate $r$ and its first and second derivative as well as energy as
\begin{equation} \label{GrindEQ__81bb_}
V_i=V_i^{(0)}(r, E) + V_i^{(1)}(r,E) \frac{\partial}{\partial r} + V_i^{(2)}(r,E) \frac{\partial^2}{\partial r^2},
\end{equation}
and 14 out of the 20 operators are those in (\ref{GrindEQ__23bb_}) of the Urb81 potential and the 6 remaining ones are
 \begin{equation} \label{GrindEQ__81cc_}
O_{i= 15, ..., 20}=S_{12} \big(\vec{L}.\vec{S}\big), S_{12} \big(\vec{L}.\vec{S}\big) \big(\vec{\tau}_{1}.\vec{\tau}_{2}\big), S_{12} L^{2}, S_{12} L^{2} \big(\vec{\tau}_{1}.\vec{\tau}_{2}\big), S_{12} \big(\vec{L}.\vec{S}\big)^{2}, S_{12} \big(\vec{L}.\vec{S}\big)^{2} \big(\vec{\tau}_{1}.\vec{\tau}_{2}\big),
\end{equation}
abbreviated as $t ls, t ls \tau, t ll, t ll \tau, t ls2, t ls2\tau$. The first eight operators exist in almost all potentials with only the radial functions of $V_i$ without derivatives, while the eight functions here are depended on the first and second derivatives of $r$. For the next six operators, $V_i^{(1)}=V_i^{(2)}=0$; and the remaining six operators are among special characterizes here. All extra terms come from the $q^2$ dependence and recoil effects included in the potential.\\
Next, by having the potential, one may solve the Schrodinger equation numerically. The evaluated phase shifts and deuteron properties are depended on the undetermined parameters of the Lagrangian. These parameters are fitted to the Nijmegen PWA93 database \cite{Nijm93} and errors from Arndt et al. \cite{Arndt87}, whereas the cutoff parameter is fixed to $\Lambda=3.90 fm^{-1}$, which is in turn equal to the $\rho$-meson mass. In general, with 26 parameters, the potential is fitted to the \emph{pp}+\emph{np} data up to $T_{lab}=$100 MeV for $J\leq2$. The phase shifts for $J>2$ are determined from OPEP in low-energies, and are not used in the fitting process. The result is a qualitative fitting of deuteron properties and a quantitative fitting of the phase shifts. This means that this new NN potential type can describe well the basic properties from a more fundamental and tight theoretical ground. By including higher orders of the chiral perturbative expansion, one may cover the higher energy ranges as well.

In summary, an advantage of these NN potentials is the systematic expansion of interaction in terms of chiral power counting. Indeed, the Texas-group has earned an NN potential in a certain order of the chiral perturbation expansion in both p- and r-space \cite{Ordonez1}. The group has only used the chiral Lagrangian of QCD in low energies and the resultant potential is free of meson theories. The agreements with deuteron properties and experimental data below 100 MeV are satisfactory. The model has some likenesses with the Paris-group (because of the pion dynamics on the LR part), Bonn-group (likenesses in low energies) and Nijmegen-group potentials (relations to QCD) in some parts of interaction; but, here EFT is used in general. In is mentionable that describing experimental data, by this first CHPT potential, were not good as the phenomenological potentials. It is also notable that the SR nuclear forces from $\chi$EFT were then surveyed by van Kolck in \cite{nucl-th/9808007}, and also in \cite{nucl-th/9902015} as a related general review. Meanwhile, look at \cite{vanKolck1} to study few-nucleon forces with this type potential, where interactions arise in chiral perturbative expansion naturally.

\subsection{Sao Paulo-Group CHPT Potentials}
Robilotta and da Rocha, have tried to estimate tow-pion-exchange contribution to NN interaction based on chiral symmetry with resolving the problem met in the previous TPEP's and including just pions and nucleons \cite{Robilotta1}. In fact, by making use of the similar methods as those in Partovi-Lomon potential \cite{PartoviLomon}, and by employing a chiral model, they framed some $2\pi$-exchange potentials. The model produced the central, spin-spin, spin-orbit and tensor components of the potential with and without isospin dependence. From their view, NN potential reads
\begin{equation} \label{GrindEQ__82aa_}
V=V_{core} + V_S + V_{PS} + V_{OPEP},
\end{equation}
where $V_{core}$ stands for the SR core potential, $V_S$ stands for the contribution from the box and crossed box diagrams, $V_{PS}$ stands for the contribution from  chiral triangle and bubble interactions, and $V_{OPEP}$, as usual, is for OPEP tail. A problem with the first approach was that it could not reproduce experimental data well for the related intermediate region. One might improve the results by including further degrees of freedom such as $\Delta$ resonances as was done in \cite{nucl-th/9611056}, and the results were compared with those from parameterized Paris potential \cite{Paris2}, Arg94 potential \cite{AV14}, dTRS potential \cite{deTourreil} and Bonn87 potential \cite{Machleidt}. \\
Then, a relativistic chiral expansion up to $O(k^4)$ for the TPEP's in p-space and further, its contents and features in r-space, was given in \cite{nucl-th/0304025} by Higa and the former members of this called Sao Paulo-group. One should note that $k<1$ GeV here is for the pion four-momentum and nucleon three-momentum, and is a typical scale for chiral perturbation theory. The resultant potential in r-space reads
\begin{equation} \label{GrindEQ__82bb_}
V=V^+ + V^- (\vec{\tau}_1.\vec{\tau}_2),
\end{equation}
where
 \begin{equation} \label{GrindEQ__82cc_}
V^\pm = V^\pm_{c}(r) +V^\pm_{\sigma}(r) (\vec{\sigma}_1.\vec{\sigma}_2) + V^\pm_{t}(r) S_{12} + V^\pm_{ls}(r) \vec{L}. \vec{S} + V^\pm_{q}(r) Q_{12}.
\end{equation}
These r-space potential functions are in terms of some numerical coefficients (related to pion- and nucleon masses and involved coupling constants) that multiply some dimensionless functions, where the latter are in turn come from the Fourier transforms of the Feynman loop integrals. It is notable that this parameterization is valid to describe NN interaction in the range of about $0.8 fm \leq r \leq 10 fm$. It is also notable that the TPE contribution for 3N force in the same order $O(k^4)$ is presented as well in \cite{0704.0711}; and a review on the subject is given in \cite{0802.2484}.
The differences between the formalism here, to discuss chiral TPEP contributions to NN interactions, and heavy-baryon (HB) formalism in the next subsections, are discussed in \cite{0908.4405}. Indeed, in HB formalism of chiral perturbative expansion, relativistic Lagrangian is expanded in $1/M$ powers, which is in turn a kind of nonrelativistic expansion; for more details look also at \cite{hep-ph/9501384}.

\subsection{Munich-Group CHPT Potentials}
The Munich-group, by using a similar CHPT Lagrangian as \cite{Ordonez1}, and employing a covariant perturbation theory and dimensional regularization, estimated the chiral two-pion-exchange NN potential as well as the usual one-pion-exchange part \cite{Munich1}. The calculations were up to the third order in low external momenta and one-loop order (or NLO). As a result, the phases shifts with $\ell\geq 2$ and the mixing angles with $J\geq 2$ were determined as free parameters, and could be used as input in the next NN phase-shift analyses. By increasing the orbital angular-momentum, a close and better agreement with the usual OPEP became obvious. In other words, the study was to describe NN interaction in terms of OPE's and TPE's for the LR and MR parts in a model independent manner. The potential was composed of the central, spin-spin, tensor, spin-orbit and quadratic spin-orbit terms with and without isospin dependence such as those in Sao Paulo-group potentials--Note that the involved pion-nucleon Lagrangian here, similar to those in the latter group, have the dimension 2 and are based on dimensionally regularized Feynman diagrams; and because the potentials are evaluated perturbatively, the bound-states are not described well! Resultant expressions for the potentials in r-space, coming from irreducible chiral $2\pi$ exchanges, are of the van-der-Waals type with the asymptotic exponential behavior $e^{-2m_{pi}r}\diagup r^n$ valid at least for the range about $1 fm < r < 2 fm$. There is not any pion-nucleon form function in that for $\ell\geq 2, J\geq 2$, the problematic singularities in the Fourier transforms are not so important. Agreement with the phase shifts up to D-wave up to $T_{lab}=150$ MeV are good, and for the higher waves agreements become better and better up to the pion-production threshold in almost 280 MeV. For the lower partial waves, the SR effects become important and so, just TPE is not enough to reproduce the phase shifts. It is also notable that relevant potentials are compared with Paris79 \cite{Paris2} and Full-Bonn (Bonn87) \cite{Machleidt} potentials. \\
Soon later, they also used two-pion exchange diagrams with virtual $\Delta(1232)$-isobar degrees of freedom and correlated $2\pi$ exchange as well as the $\rho, \omega$ vector-meson exchanges in \cite{Munich2}. As a result, they reproduced the experimental data up to 350 MeV for $\ell\geq 3$ and up to 80 MeV for D-waves, without any adjustable parameter. So, this is chiral symmetry that has opened a nice window to NN interaction. It is mentionable that, to describe the lower partial waves, nonperturbative methods and other SR parameterizations are still needed--It is good to mention that the importance of the chiral TPEP's was more confirmed in \cite{nucl-th/9901054} (by some members of the Nijmegen-group and others), when they saw that the chiral TPE loops were important in the LR part of \emph{pp} interaction as they improved the results of just OPEP's. In other words, the group noted that by including both OPE and $\chi$TPE contributions, they could find a good fit of data up to 350 MeV for $r\geq 1.4 fm$. The range below the mentioned one was then parameterized by 23 boundary condition parameters in the energy-dependent partial-wave-analysis.

Further efforts, by Kaiser, have taken to include chiral uncorrelated three-pion exchanges, higher-loop and relativistic corrections to NN interactions \cite{Munich3}. Indeed, it was shown that the uncorrelated $3\pi$ exchanges have negligible effects on NN interactions in $r\geq 0.8 fm$. The local potentials produced by $2\pi$-exchange diagrams in two-loop order of the heavy-baryon chiral perturbation theory, besides including the second order $\pi\pi$N vertexes, and the first relativistic $1/M$ corrections in one-loop $2\pi$-exchange diagrams, were discussed as well. The latter were the components for the chiral NN potential in the next-to-next-to-next-to-leading-order (NNNLO). It should be mentioned that these two-loop diagrams lead to contributions about $O(k^4)$ in chiral expansion and so N$^3$LO. By including $1/M^2$ corrections to $2\pi$-exchange diagrams and their effects on various parts of interaction and various states, the chiral NN potential in this N$^3$LO order is complete. We should remember that the potential structures and operators here are almost the same as those of the Sao Paulo-group; and that in the third reference of \cite{Munich3}, an explicit analytical expression for the potential in r-space from the p-space one is presented. Next, he studied the spin-orbit coupling produced from $2\pi$ exchange in 3N interaction by including the virtual $\Delta$-isobar in \cite{nucl-th/0312058}. \\
It is notable that in \cite{nucl-th/0202039}, by another group, there are also a complete set of $2\pi$-exchange diagrams in the same fourth-order (N$^3$LO) in chiral perturbative expansion. One could see that the fourth-order contribution is less than the third-order one; and this in turn signals the converging of chiral expansion. By employing the analytical expressions in \cite{Munich3}, they applied the methods to NN scattering to calculate scattering amplitudes; and then they compared predictions with experimental phase shifts and those from the usual meson-exchange theories. To make a more sensible comparison, they included OPE and iterated OPE contributions as well, and next showed the phase shifts for $\ell \geq 3$ below the energy of 300 MeV. The agreement between Full-Bonn potential and this N$^3$LO potential was good.

By the way, many other studies are done by the group members. For instance, in \cite{Munich4}, chiral four-nucleon interactions in this framework are studied. A microscopic optical potential from two- and three-body chiral nuclear forces is constructed in \cite{1304.3175}. Some members of the group, next to others, have modeled YN potentials in NLO of chiral effective field theory in \cite{1304.5339}. In the latter, contributions from the one and two pseudoscalar-meson diagrams as well as four-baryon contact terms are included. The SU(3) flavor symmetry was used to set up potentials while its breaking by the physical masses of the pseudoscalar mesons ($\pi, K, \eta$) was considered as well. Excellent results, compared with the counterpart HQ phenomenological potentials, were gained. That is also a relativistic chiral SU(3)-invariant Lagrangian up to  $O(q^2)$  order to describe BB interaction in \cite{1305.3427}.

\subsection{Idaho-Group CHPT Potentials}
Along with various efforts after the first CHPT potential by Texas-group in \cite{Ordonez1}, a better NN potential based on chiral EFT appeared by Entem and Machleidt \cite{Idaho1} in 2001. In the potential both meson and quark degrees of freedom are included, while \cite{Ordonez1} is a meson-free potential. Indeed, that is an NN potential, based on HB formalism of chiral perturbative expansion that includes one-pion and two-pion exchanges up to the third order of  chiral expansion. The short-range force in the fourth order of expansion is involved because of good fitting of the D-wave phase shifts. There, a two-pion exchange potential in the fourth-order of chiral expansion is also presented. The potential has almost the same quality as the HQ Nijmegen potentials \cite{Nijm93}, CD-Bonn \cite{Bonn4} and Arg94 \cite{AV18} potentials. The phase shifts below $T_{lab}=$300 MeV, deuteron properties and low-energy $np$ scattering parameters as well as Triton binding-energy are described well with this potential \cite{Idaho1}.\\
Later, the authors modeled, in fact, the first accurate NN potential in N$^3$LO (fourth-order) of chiral perturbative expansion \cite{Idaho2}. The new potential, in reproducing its time \emph{pp} and \emph{np} data below 290 MeV, is comparable with the best high-precession phenomenological potentials. After mentioning main features of the previous HQ phenomenological and meson-exchange potentials, it is also argued in \cite{Idaho3} that EFT approach to nuclear forces is better than all earlier efforts in that it produces a wished precession, gives satisfactory results in nuclear calculations as well as dealing with few-nucleon interactions on an equal footing as NN interaction. There are also some reviews and many other related issues and progress presented in \cite{0704.0807} and \cite{1110.3022}.

\subsection{Bochum-Julich-Group CHPT Potentials}
Bochum-Julich-group potentials are also based on chiral EFT, similar to the other CHPT potentials mentioned above, except that they extracted the Lagrangian's by using a "unitary transformation" method. In fact, they have studied many NN (also 3N and few-nucleon) forces besides various related aspects in LO, NLO, NNLO and NNNLO of CHPT by taking the most general chiral Hamiltonian with pions and nucleon fields as we describe below concisely.

But before that, we note that in the standard method, such as that of Texas-group \cite{Ordonez1}, the most general Lagrangian including all symmetries such as chiral symmetry of QCD was first written with an infinite number of terms including nucleon and pion fields and their derivatives. The breaking of chiral symmetry was clear in smallness of the pion mass, and then the external momenta of the pion and nucleon should not exceed the scale of $Q$. As a result, the expansion parameter was $Q/\Lambda_{QCD}$ and nucleons were treated nonrelativistically, where $\Lambda_{QCD}\approx 1$ GeV that is almost the $\rho$-meson mass. The other degrees of freedom, such as heavy mesons and other baryons which were then less important, were integrated out (except maybe $\Delta$ isobars) as their information was so included in the Lagrangian's parameters. In the process, a finite set of tree and loop diagrams were included. But a problem was that due to the presence of low energy bound states, perturbative theory failed actually; or in other words, infrared divergences with the few included nucleons disturbed the power counting of chiral expansion. A way to solve the problem was to use the old-fashioned time-ordered perturbation theory by Weinberg \cite{Weinberg}, where the expansion parameter was $Q/M$, instead of the covariant method. Still, in the latter method, the effective potential was not Hermitian as it was depended on the incoming-nucleon energies, and that the nucleon wave functions were not orthogonal there. So the unitary transformations here resolve the problems, where the expansion parameter is now the small momenta of external particles. It is also notable that resultant potentials are energy-independent, which makes the applications to few-body and nuclear-structure calculations simpler.

\subsubsection{LO, NLO and NNLO Potentials}
In general, these potentials include contributions from one- and two-pion exchanges to simulate LR and MR interactions besides contact terms to simulate SR interactions. The resultant interactions, from LO, NLO and NNLO of CHPT by considering the most general chiral Hamiltonian in terms of pions and nucleon fields, are given in \cite{nucl-th/9910064}. The LO interaction includes two four-nucleon contact terms and an OPE potential as
\begin{equation} \label{GrindEQ__83aa_}
V_{cont.}^{(0)}=C_s + C_t (\vec{\sigma}_1.\vec{\sigma}_2),
\end{equation}
and
\begin{equation} \label{GrindEQ__83bb_}
V_{1PEP}^{(0)}=-\left(\frac{g_A}{2f_{\pi}}\right)^2 (\vec{\tau}_1.\vec{\tau}_2) \frac{(\vec{\sigma}_{1}.\vec{k})(\vec{\sigma}_{2}.\vec{k})}{k^2+m_{pi}^2},
\end{equation}
where the low-energy constants (LECs) of $C_s, C_t, C_1, D_1, ...$ are to be determined by fitting some data, $g_A$ is the axial-vector coupling, $f_{\pi}$ is the pion decay-constant, and other symbols are the same as used before. In NLO, the potential is a renormalized sum of one- and tow-pion exchanges and contact interactions. This means that next to above contributions, it includes a TPEP contribution ($V_{TPEP}^{(2)}$) and seven four-nucleon contact terms where the latter reads
\begin{equation} \label{GrindEQ__83aa_}
V_{cont.}^{(2)}=C_1 k^2 +C_2 q^2 + (C_3 k^2 +C_4 q^2) (\vec{\sigma}_1.\vec{\sigma}_2) + C_5 \tilde{LS}_1 + C_6 \tilde{S}^{(0)}_{12} + C_7 \tilde{\grave{S}}^{(0)}_{12},
\end{equation}
where $\tilde{\grave{S}}^{(0)}_{12}=(\vec{\sigma}_{1}.\vec{q})(\vec{\sigma}_{2}.\vec{q})$, the nine LECs are determined by fitting to the \emph{np} S and P and ${}^3S_1-{}^3D_1$ phase shifts, and the mixing parameter $\varepsilon_1$ for the laboratory energies below 100 MeV. TPEP in this NLO includes $k, k^2, q^2$ dependence as well as the operators $I, \tilde{S}_{12}$, isospin dependence and some constants \cite{nucl-th/9910064}. On the other hand, in NNLO, another TPEP ($V_{TPEP}^{(3)}$) is also included, which in turn includes some special combinations of $k, k^2, q^2$ with the operators $I, \tilde{S}_{12}, \tilde{LS}_1$ without and with isospin dependence and some constants. It is mentionable that if one includes the contribution from $\Delta(1232)$-isobar, the resultant NNLO-$\Delta$ potential is almost same as NNLO one especially for low momenta.\\
We should also note that the pion-exchange NN potentials could be written generally, in p-space, as
\begin{equation} \label{GrindEQ__82bb_}
\tilde{V}=\tilde{V}^+ + \tilde{V}^- (\vec{\tau}_1.\vec{\tau}_2),
\end{equation}
where
 \begin{equation} \label{GrindEQ__82cc_}
\tilde{V}^\pm = \tilde{V}^\pm_{c} +\tilde{V}^\pm_{\sigma} (\vec{\sigma}_1.\vec{\sigma}_2) + \tilde{V}^\pm_{ls} \tilde{LS}_1 + \tilde{V}^\pm_{q} \tilde{Q}_{12} + \tilde{V}^\pm_{\sigma k} \tilde{S}^{(0)}_{12} +\tilde{V}^\pm_{\sigma q} \tilde{\grave{S}}^{(0)}_{12},
\end{equation}
and to adjust more with the record in (\ref{GrindEQ__22cc_}), we set $SS_0=(\vec{\sigma}_1.\vec{\sigma}_2)$; and that the functions of $\tilde{V}^\pm_{c},...$ are in terms of $\vec{p}_i, \vec{p}_f, z$ with $z=\cos(\vec{p}_i, \vec{p}_f)$, included masses and coupling constants.\\
To regularize or have right behavior for the potentials in large momenta (short distances), the sharp and exponential form factors are used as
\begin{equation} \label{GrindEQ__82dd_}
F({k}^{2})_{sharp} =\theta\left(\Lambda^{2}-{k}^{2}\right), \quad F({k}^{2})_{exp.} = e^{-{k}^{2n} / \Lambda^{2n}},
\end{equation}
where the sharp cutoff is proper here with $\Lambda=500$ MeV for NLO and $\Lambda=875$ MeV for NNLO; and that in exponential form factors, $n=2,3,...$ with often $n=2$ here, where the latter is used especially to evaluate some deuteron properties with good results. In addition, phase shifts and mixing parameters for high energies and angular momentums are described well for the energies below 300 MeV, with a note that the partial waves higher than P are free of adjustable parameters. Also, various properties of nuclei with $A>2$ and especially the binding energies of ${}^3H$ and ${}^4He$ are evaluated by these NLO and NNLO potentials with an almost same quality as the standard high-precision phenomenological and boson-exchange potentials \cite{nucl-th/9910064}, \cite{BochumJulich1}.

\subsubsection{NNNLO Potentials and More}
Next development of the model was to NNNLO of chiral expansion \cite{BochumJulich2}. The new potential includes one-, two- and three-pion exchanges as well as the contact terms with zero, two and four derivatives. Relativistic corrections and isospin-breaking mechanisms are also included. In fact, next to the previous contact terms of (\ref{GrindEQ__83aa_}) and (\ref{GrindEQ__83aa_}), the new included contact terms are
\begin{equation} \label{GrindEQ__84aa_}
\begin{split}
V_{cont.}^{(4)}= & D_1 k^4 +D_2 q^4 + D_3 k^2 q^2 + D_4 n^2 + (D_5 k^4 +D_6 q^4 + D_7 k^2 q^2 + D_8 n^2) (\vec{\sigma}_1.\vec{\sigma}_2) \\
 + & (D_{9} k^2 + D_{10} q^2) \tilde{LS}_1 + (D_{11} k^2 + D_{12} q^2) \tilde{S}^{(0)}_{12} +(D_{13} k^2 + D_{14} q^2) \tilde{\grave{S}}^{(0)}_{12} + D_{15} \tilde{Q}_{12},
\end{split}
\end{equation}
where one could also include another 24 terms which contain the isospin factor of $(\vec{\tau}_1.\vec{\tau}_2)$. Now, all 26 four-nucleon LECs are determined by fitting the \emph{pp}+\emph{np} Nijmegen-group database \cite{NijmPWA} (the relevant S, P, D phase shifts and mixing parameters) and \emph{nn} scattering-length. \\
On the other hand, for pion-exchange parts, a new three-pion exchange contribution ($V_{3PEP}^{(4)}$) is considered though its effect is negligible (note that the $n$-pion-exchange diagrams become important around $Q^{2n-2}$). These pion-exchange contributions can again be written as (\ref{GrindEQ__82bb_}) with (\ref{GrindEQ__82cc_}), where for instance the lowest order of the scalar function of $\tilde{V}^-_{\sigma k}$ is indeed (\ref{GrindEQ__83bb_}) without $(\vec{\tau}_1.\vec{\tau}_2)$ factor. We remember that LS equations and a relativistic form for kinetic energy are employed to iterate the potential here. Reducing to a nonrelativistic form is more useful in real calculations. The exponential form-factor of (\ref{GrindEQ__82dd_}) with $n=3$ is used to regularize LS equations with the cutoffs of $\Lambda=450-600$ MeV.\\
The isospin-breaking of strong interactions because of different masses of up and down quarks, and from electromagnetic interactions because of different charges of up and down quarks are also included. Indeed, the potentials for different NN systems and isospins are different such that, for instance, $V_{1PEP} (pp)\neq V_{1PEP} (np,\ T=1)\neq V_{1PEP} (np,\ T=0)$ and so on. This is finite-range isospin-breaking, while the long-range isospin-breaking is because of different electromagnetic interactions such that $V_{EM}(pp)\neq V_{EM}(np)\neq V_{EM}\neq(nn)$. In other words, the quark mass splitting causes isospin-breaking in short distances, whereas the contact electromagnetic terms cause isospin-breaking in long distances--Look also at the discussion on Arg94 potential, Nijmegen HQ potentials and CD-Bonn potential in subsections of (\ref{4.12.2}), (\ref{4.15.4}) and (\ref{4.13.2}), respectively.

In summary, the group has set up some NN potentials by using the unitary transformation method applied to the most general chiral invariant Hamiltonian in terms of pion and nucleon fields from LO up to N$^3$LO. In the latter, CIB and CSB in leading order, the pion mass differences in OPEP's, kinematic effects because of the nucleon mass splitting, and electromagnetic corrections such as those in Nijmegen PWA's, and many other subtleties are included. Deuteron properties and the low phase shifts of S, P, D are described excellently, whereas the high partial waves of F, G, H,... are parameter free and are well described depended on the doubts in the cutoffs. In general, several improvements with respect to the lower order expansions and also to the previous CHPT potentials are notable.\\
Among many other studies by the group members, improvements to the Weinberg approach to arrive at the effective potential and the renormalization problem there, a new approach based on an effective Lagrangian with exact Lorentz invariance and by using time-ordered perturbation theory, without using HB expansion, were presented and analyzed in \cite{1301.6134}. Indeed, they improved the heavy chiral perturbation theory for NN interaction and analyzed the OPEP iterations. As a result, it was shown that the used renormalization, for one-and two-loop diagrams of OPEP iterations, removes all nucleon-mass dependencies that disturb the power counting--It is good here to mention a pioneer work to resolve inconsistencies in the Weinberg's chiral expansion. Indeed, in \cite{nucl-th/9801034}, Kaplan et al. used a dimensional regularization scheme with a novel subtraction (renormalization-group techniques) to get a consistent chiral expansion and dissolve the failure of the Weinberg's power counting scheme. They applied the method in the order $O(Q^0)$ to ${}^1S_0$ and ${}^3S_1-{}^3D_1$ NN scattering channels, and then compared the results with Nijmegen PWA93 \cite{NijmPWA} with satisfactory agreements. For some other old, and of course related, typical studies in the phase look at \cite{nucl-th/9807054}, \cite{nucl-th/9906056}. \\
By the way, for a recent review on NN, 3N and few-nucleon interactions especially in the framework of $\chi$EFT, advantages and disadvantages of this approach to nuclear forces, look at \cite{1302.3241} by Epelbaum and references therein. To end the discussion in the phase, we cite \cite{1303.4674} as the last constructed optimized potential at NNLO by other people.

\section{Some Other Models and Potentials}
In general, almost all potentials are belong to one of the four main models. These are, the almost full phenomenological model; the model based on field-theoretical methods, inverse-scattering, quantum-dispersion relations and boson-exchange pictures; the model based on QCD and constituent quark methods (the QCD-inspired model); the model based on CHPT and EFT and their various extensions.\\
We have tried to include and study almost all models and potentials to describe two-nucleon interactions with an emphasis on some in more details as samples of well-known and high-precession NN potentials. Technical studies of some potentials need more space and time next to many physical and mathematical backgrounds that is not the aim of this concise pedagogical review. Nevertheless, there are still some other special NN interaction models and potentials, and related topics, to be addressed. We mention some in what follows.

Among the standard and more theoretical potentials is the\textbf{Virginia-group} potential \cite{Virgina1}, which is a special relativistic OBEP based on field-theoretical and dispersion-relation techniques. In fact, they have framed a few potentials by taking various meson exchanges. The \textbf{Bochum-group} potential \cite{Bochum1} is another fundamental NN potential based on field-theoretical and dispersion-relation methods that also uses various meson exchanges in long distances and QCD effects; meanwhile the direct NN interactions coming from the intrinsic structure of nucleon are considered in short distances. By including some two- and three-pion correlations, they have claimed to hold good description of NN scattering data. The \textbf{Seattle-group} studies on NN interaction are also notable. Indeed, they have studied low-energy NN interactions based on EFT, by using some simple models for interactions, up to NNLO in chiral expansion, next to some other related topics during their study period in 1990's \cite{Seattle00}.

There are the potentials based on \textbf{Mean Field Theory} (MFT), which are of particular interest in many-body calculations in nuclear physics especially--Look, for instance, at \cite{Serra} and \cite{1011.5732} for the first NN interaction made of relativistic mean field theory.

\begin{figure}[htp]
\centering
  \includegraphics[height=3in, width=4.8in, scale=1]{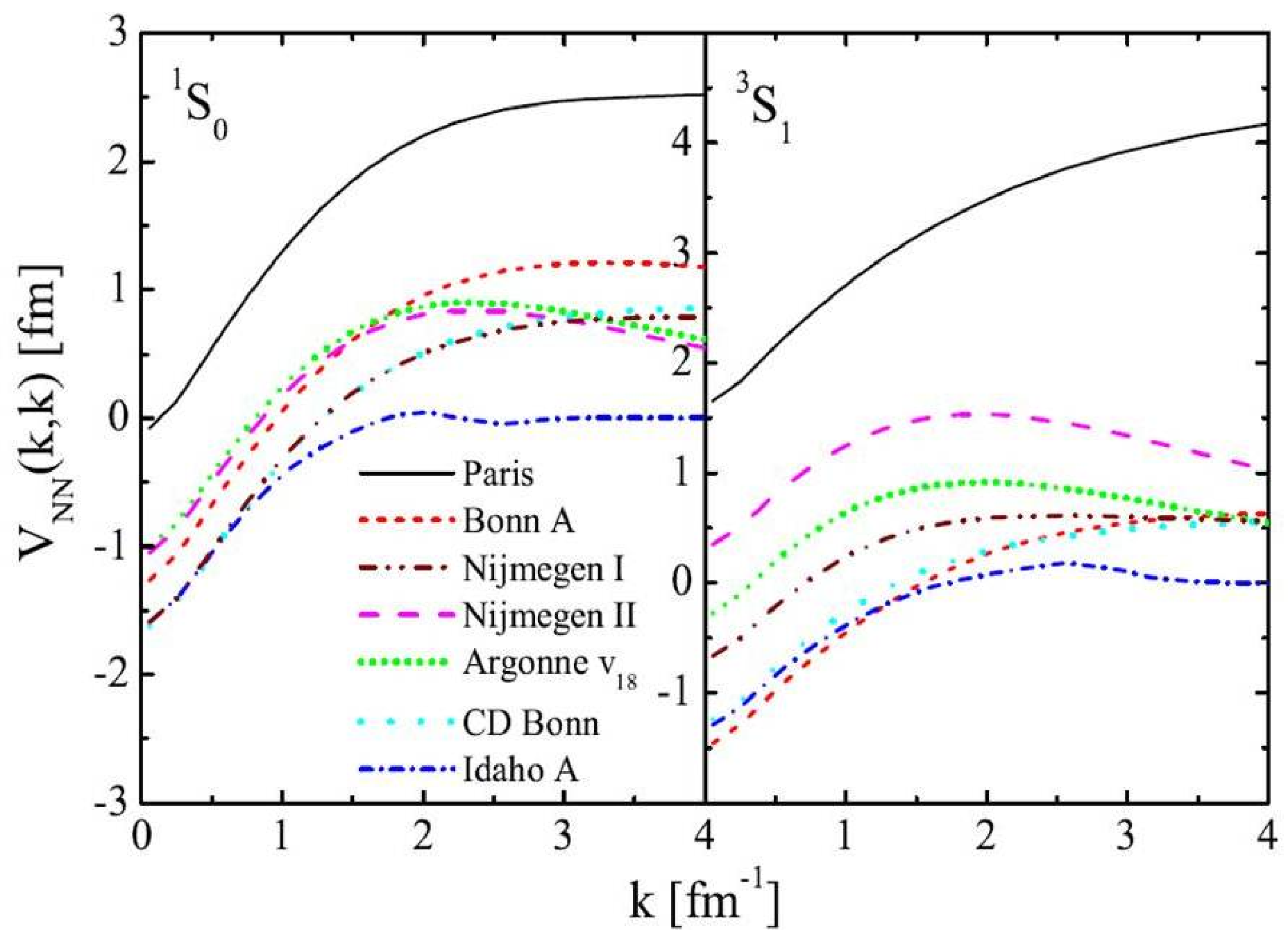} \\
  \caption{\textit{Diagonal matrix elements of some high-quality NN potentials ($V_{NN}$) versus relative-momentum ($k$) for ${}^1 S_0$ and ${}^3S_1$ partial-wave, in momentum-space \cite{nucl-th/0305035}.}} \label{Fig4.}
\end{figure}

\begin{figure}[htp]
\centering
  \includegraphics[height=2.8in, width=4.8in, scale=1]{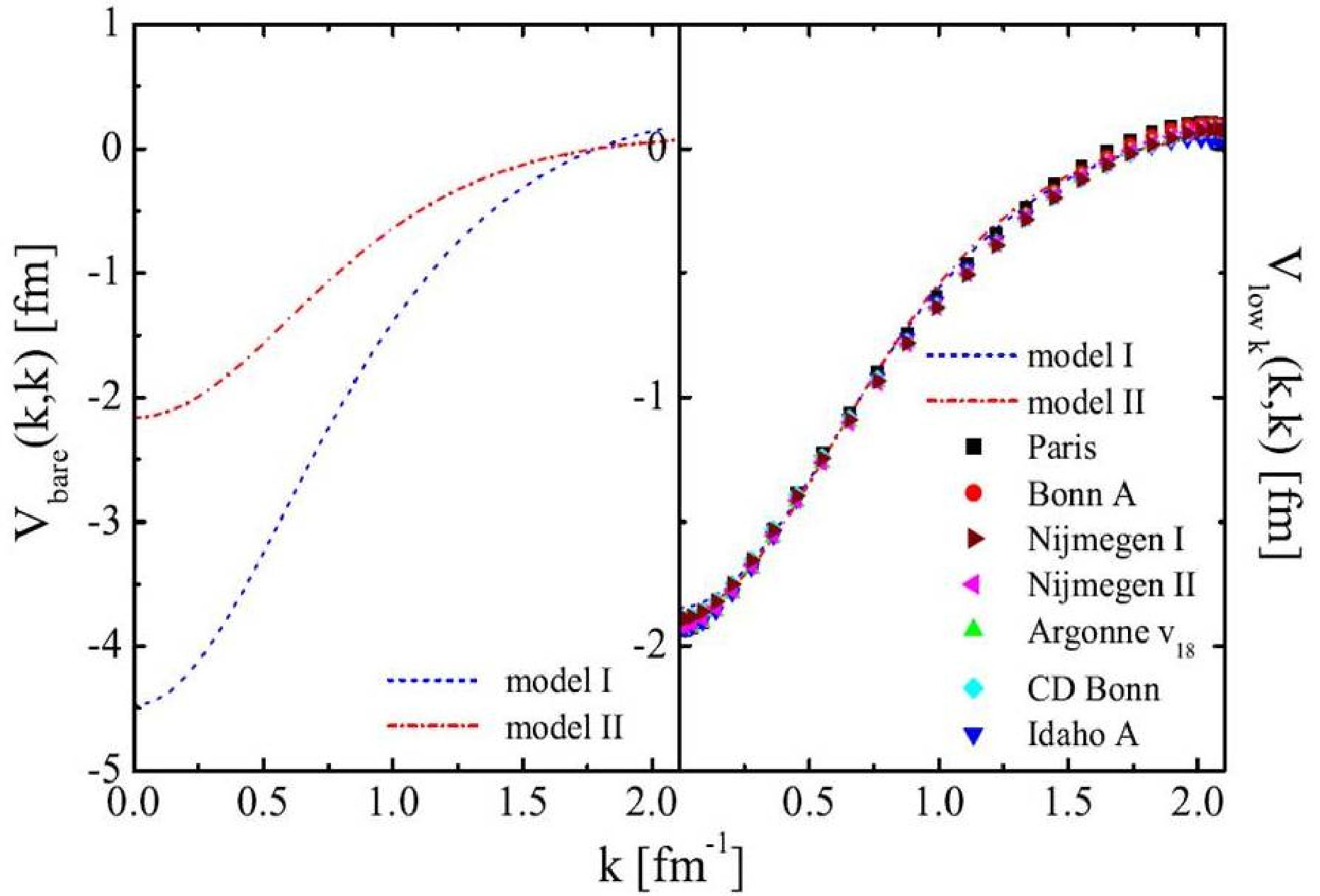} \\
  \caption{\textit{Diagonal matrix elements of $V_{low\ k}$ ($V_{bare}$ in figure) for two simple RG potentials are compared with $V_{low\ k}$ derived from some high-quality NN potentials \cite{nucl-th/0305035}.}} \label{Fig5.}
\end{figure}

\textbf{Renormalization Group} (RG) approaches to NN interaction are other serious efforts. In an RG flow viewpoint, a model-independent low-momentum interaction is obtained by integrating out high-momentum components (cutting out problematic high-momentum modes) of various potential models \cite{nucl-th/0108041}. Indeed, the model independence of resulting potentials shows that the physics of the nucleons interacting at low momenta does not depend on the details of the high-momentum dynamics assumed in conventional potential models. Further developments such as incorporating the method into the Fermi liquid theory are also made \cite{nucl-th/0109059}, \cite{nucl-th/0212034}. In \cite{nucl-th/0305035}, detailed results for the model-independent low-momentum NN potential $V_{low\ k}$ are shown. There, they have applied the approach to some commonly used high-precession NN potentials, and then compared resultant potentials in various ways such as comparing matrix elements of the potentials and various resulting phase shifts in p-space. In Figure \ref{Fig4.} and Figure \ref{Fig5.} are two such sample comparisons of some high-precession NN potentials together and with two simple RG models, respectively.
For a newer "similarity renormalization group" approach, see \cite{nucl-th/0611045} and \cite{1302.3978}, and for a recent review and study of the subject look at \cite{1012.4914}.

\textbf{Lattice QCD} approach to NN interaction is another important way; look, for instance, at \cite{Beane, nucl-th/0611096} and \cite{hep-lat/0601006}. Among some typical studies, see \cite{hep-lat/0609078}, where a spin-dependent potential in lattice QCD is presented; \cite{1103.0619}, and \cite{1109.2889}, where nonlocality of NN potentials, and deuteron and some other two-body bound states in lattice QCD are discussed--Look also at \cite{nucl-th/9807024}, where QCD sum rules are used for NN interactions. Altogether, this phase of study is still improving with giving better quantitative results as the previous good qualitative ones.

\textbf{Tubingen-group} has applied projection techniques on some former NN potentials among the boson-exchange, phenomenological, RG flow and EFT ones to map them over the operator basis of relativistic field theory \cite{nucl-th/0509049}. Indeed, they have presented a model-independent study of NN interaction from its Dirac structure. That is a special way to compare various potentials, where a nice agreement is found as well.\\
They have also built a new energy-independent nonlocal potential above inelastic thresholds in quantum field theories that satisfies a suitable Schrodinger equation at low energies \cite{1212.4896}. The potential is indeed composed of a set of Nambu-Bethe-Salpeter wave functions. By applying the same method, one could set up three-nucleon potentials as well.

By the way, there may be other models and potentials not covered in this note and so, it would be pleasure to hear more about other NN potentials. Meanwhile, there are still many studies on various aspects of NN interaction which need addressing. For examples, nonlocal and local terms and their impact on NN interactions and their roles in some NN potentials are studied, for instance, in \cite{nucl-th/0209058}; and nonlocality of NN potentials in lattice QCD is discussed, for instance, in \cite{1103.0619}. For a study of CIB and CSB of NN interaction, look at \cite{1010.1728} and for parity violation in NN interaction, see, for instance, \cite{0803.2075}.

We should also mention that \textbf{Three- and few-nucleon interactions} are also interesting to which less efforts than two-nucleon interactions are allocated. For there-nucleon force, look at a recent review of \cite{1210.4273}; and for a view to few-nucleon forces, look at \cite{1302.3241, nucl-th/0509032}.

\section{Outlook}
Nowadays the theory of strong nuclear force is well experienced both quantitatively and qualitatively. The best qualitative results are obtained by using phenomenological and boson-exchange potentials based on quantum field theory and dispersion relation techniques, and even new potentials based on chiral perturbation theory. Indeed, more qualitative results are of the QCD-inspired models and the models based on chiral EFT.\\
NN interaction is now under control for the energies below almost $T_{lab}=$500 MeV well. Because of the high-precession experimental NN data, describing the long- and intermediate-range parts of the interaction based on various meson exchanges are quantitatively good and the hybrid models of quark and gluon exchanges for the short-ranges seem to be more suitable.\\
Although we have now many high-precision NN potentials applying to nuclear-structure calculations with satisfactory results, still some questions are remained to be answered. I think the main problem is that we don't have still a unique comprehensive model for including all well-known features of NN interaction. Obviously, chiral EFT methods and models are better in describing nuclear forces in general. They have a standard formalism applicable to few-nucleon systems with including many fundamental physics and mathematics of the problem. But, there are still some problems and limits; look at \cite{1110.3761}, \cite{1210.0992}. Among the issues with EFT potentials, which one may ask, are the proper renormalization of the chiral nuclear potentials and sub-leading chiral few-nucleon forces; few- and multi-nucleon potentials in higher orders of chiral expansion. Meanwhile, lattice QCD models for nuclear forces are still improving and, in some recent studies, a lattice version of chiral EFT is also applied to nuclear forces \cite{1302.3241}.

On the other hand, after the well-conjectured string/gauge, AdS/CFT, duality and thereafter \textbf{Holographic QCD} studies, it seems that the NN interaction issue is faced with another revolution. So, we should be wait for more sophisticated models for two- and many-body nuclear interactions in this language--Look, for instances, at \cite {0901.0012} and \cite {0901.4449}.

Altogether, it seems that the nuclear force issue is still improving. I think that we may someday have a unified scheme for NN interaction and link various known NN models and potentials. Nevertheless, it will also be interesting to compare various NN potentials via some suitable ways and try to understand more nucleon-nucleon interaction subtleties.


\begin{thebibliography}{99}
\bibitem{Bethe1} H. A. Bethe, \textit{"What holds the nucleus together"}, Scientific American 189, 58 (1953).
\bibitem{Yukawa2} H. Yukawa, \textit{"On the interaction of elementary particles"}, Proc. Phys. Math. Soc. Jpn 17, 48 (1935).
\bibitem{0811.1338} E. Epelbaum, H.-W. Hammer and Ulf-G. Meißner, \textit{"Modern theory of nuclear forces"}, Rev. Mod. Phys. 81, 1773 (2009), \href{http://arxiv.org/abs/0811.1338}{[arXiv:0811.1338 [nucl-th]]}.
\bibitem{1110.3761} F. Gross, T. D. Cohen, E. Epelbaum and R. Machleidt, \textit{"Conference discussion of the nuclear force"}, Few-Body Syst. 50, 31 (2011), \href{http://arxiv.org/abs/1110.3761}{[arXiv:1110.3761 [nucl-th]]}.
\bibitem{1210.0992} R. Machleidt, Q. MacPherson, E. Marji, R. Winzer, Ch. Zeoli and D. R. Entem, \textit{"Recent progress in the theory of nuclear forces"}, \href{http://arxiv.org/abs/1210.0992}{[arXiv:1210.0992 [nucl-th]]}.
\bibitem{Tokyo1} M. Taketani, S. Nakamura and M. Sasaki, \textit{"On the method of the theory of nuclear forces"}, Prog. Theor. Phys. 6, 581 (1951).
\bibitem{3} G. E. Brown and A. D. Jackson, \textit{"The Nucleon-Nucleon Interaction"}, North-Holland Publishing (1976).
\bibitem{Garcon} M. Garcon and J. W. Van Orden, \textit{"The deuteron: structure and form factors"}, Adv. Nucl. Phys. 26, 293 (2001), \href{http://arxiv.org/abs/nucl-th/0102049}{[arXiv:nucl-th/0102049]}.
\bibitem{4} R. Machleidt and I. Slaus, \textit{"The nucleon-nucleon interaction"}, J. Phys. G: Nucl. Part. Phys. 27, 69 (2001), \href{http://arxiv.org/abs/nucl-th/0101056}{[arXiv:nucl-th/0101056]}.
\bibitem{Cohen8} B. L. Cohen, \textit{"Concepts of Nuclear Physics"}, McGraw-Hill (1972).
\bibitem{Krane7} K. S. Krane, \textit{"Introductory Nuclear Physics"}, John Wiley \& Sons, (1988).
\bibitem{Wong6} S. S. M. Wong, \textit{"Introductory Nuclear Physics"}, Prentice-Hall, Inc. (1990).
\bibitem{Pal5} M. K. Pal, \textit{"Theory of Nuclear Structure"}, Scientific \& Academic Editions, New York (1983).
\bibitem{Nutshell1} Carlos A. Bertulani, \textit{"Nuclear Physics in a Nutshell"}, Princeton University Press (2007).
\bibitem{Philips10} R. J. N. Philips, \textit{"The two-nucleon interaction"}, Rep. Prog. Phys. 22, 562 (1959).
\bibitem{11} G. Jones, \textit{"Pion production and absorption in nuclei"}, AIP Conf. Proc. 76; Amer. Inst. Phys., New York,  p.150 (1982).
\bibitem{12} H. A. Bethe and P. M. Morrison, \textit{"Elementary Nuclear Theory"}, 2nd Ed., p.388. John Wiley \& Sons, New York (1956).
\bibitem{Goto} J. Goto and S. Machida, \textit{"Nuclear forces in the momentum space"}, Prog. Theor. Phys. 25, 64 (1961).
\bibitem{Hoshizaki} N. Hoshizaki and T. Kadota, \textit{"Nuclear forces in momentum space with One-Boson-Exchange model"}, Prog. Theor. Phys. 50, 1312 (1973).
\bibitem{nucl-th/0103010} S. A. Zaitsev and E. L. Kramar, \textit{"NN potentials from inverse scattering in the J-matrix approach"}, J. Phys. G: Nucl. Part. Phys. 27, 2037 (2001), \href{http://arxiv.org/abs/nucl-th/0103010}{[arXiv:nucl-th/0103010]}.
\bibitem{Shirokov} A. M. Shirokov, J. P. Vary, A. I. Mazur, S. A. Zaytsev and T. A. Weber, \textit{"NN potentials from the J-matrix inverse scattering approach"}, J. Phys. G: Nucl. Part. Phys. 31, 1283 (2005).
\bibitem{Faassen} E. E. van Faassen and J. A. Tjon, \textit{"Relativistic calculations for $NN-N\Delta$ scattering with $\pi$ and $\rho$ exchange"}, Phys. Rev. C 28, 2354 (1983).
\bibitem{Wallace} S. J. Wallacen, \textit{"Relativistic equation for nucleon-nucleus scattering"}, Ann. Rev. Nucl. and Part. Sci. 37, 267 (1987).
\bibitem{HamadaJohnston} T. Hamada and I.D. Johnston, \textit{"A potential model representation of two-nucleon data below 315 MeV"}, Nucl. Phys. 34, 382 (1962).
\bibitem{Yale} K. E. Lassila, M. H. Hull, H. M. Ruppel, F. A. McDonald and G. Breit, \textit{"Note on a nucleon-nucleon potential"}, Phys. Rev. 126, 881 (1962).
\bibitem{Reid68} R. V. Reid, \textit{"Local phenomenological nucleon-nucleon potentials"}, Ann. Phys. (NY) 50, 411 (1968).
\bibitem{Reid68-Day} B. D. Day, \textit{"Three-body correlations in nuclear matter"}, Phys. Rev. C 24, 1203 (1981).
\bibitem{Nijm93} V. G. J. Stoks, R.A. M. Klomp, C. P. F. Terheggen and J.J. de Swart, \textit{"Construction of high-quality nucleon-nucleon potential models"}, Phys. Rev. C 49, 2950 (1994), \href{http://arxiv.org/abs/nucl-th/9406039}{[arXiv:nucl-th/9406039]}.
\bibitem{UV14} I. E. Lagaris and V. R. Pandharipande, \textit{"Phenomenological two-nucleon interaction operator"}, Nucl. Phys. A 359, 331 (1981).
\bibitem{AV14} R.  B. Wiringa, R. A. Smith and T. L. Ainsworth, \textit{"Nucleon-nucleon potentials with and without $\Delta(1232)$ degrees of freedom"}, Phys. Rev. C 29, 1207 (1984).
\bibitem{AV18} R. B. Wiringa, V. G. J. Stoks, R. Schiarilla, \textit{"Accurate nucleon-nucleon potential with charge-independence breaking"}, Phys. Rev. C 51, 38 (1995), \href{http://arxiv.org/abs/nucl-th/9408016}{[arXiv:nucl-th/9408016]}.
\bibitem{Gintautas} G. P. Kamuntavicius and M. Kaminskas, \textit{"Phenomenology of the nucleon-nucleon potential"}, Central Euro. J. Phys. 8, 970 (2010).
\bibitem{Taketani} M. Taketani, S. Machida and S. Ohanuma, \textit{"The meson theory of nuclear forces I: The deuteron ground state and low energy neutron-proton scattering"}, Prog. Theor. Phys. 7, 45 (1952).
\bibitem{BruecknerWatson} K. A. Brueckner and K. M. Watson, \textit{"Nuclear forces in pseudoscalar meson theory"}, Phys. Rev. 92 1023 (1953).
\bibitem{PartoviLomon} M. H. Partovi and E. L. Lomon, \textit{"Field-theoretical nucleon-nucleon potential"}, Phys. Rev. D 2, 1999 (1970).
\bibitem{Erkelenz} K. Erkelenz, \textit{"Current status of the relativistic two-nucleon one boson exchange potential"}, Phys. Rep. 13, 191 (1974).
\bibitem{Bonn2} R. Machleidt, \textit{"The Meson theory of nuclear forces and nuclear structure"}, Adv. Nucl. Phys. 19, 189 (1989).
\bibitem{Gross} F. Gross, J. W. Van Orden and K. Holinde, \textit{"Relativistic one-boson-exchange model for the nucleon-nucleon interaction"}, Phys. Rev. C 45, 2094 (1992).
\bibitem{nucl-th/9609054} L. Jaede, M. Sander and H. V. von Geramb, \textit{"Modeling of nucleon-nucleon potentials, quantum inversion versus meson exchange pictures"}, Springer Lect. Notes in Physics 488, 124 (1997), \href{http://arxiv.org/abs/nucl-th/9609054}{[arXiv:nucl-th/9609054]}.
\bibitem{nucl-th/9611056} M. R. Robilotta and C. A. da Rocha, \textit{"Two-pion Exchange nucleon-nucleon potential: Model independent features"}, Nucl. Phys. A 615, 391 (1997), \href{http://arxiv.org/abs/nucl-th/9611056}{[arXiv:nucl-th/9611056]}.
 \bibitem{Lebed1} Richard F. Lebed, \textit{"NN potentials from inverse scattering in the J-matrix approach"}, J. Phys. G: Nucl. Part. Phys. 27, 2037 (2001), \href{http://arxiv.org/abs/nucl-th/0103010}{[arXiv:nucl-th/0103010]}.
\bibitem{1201.0443} M. Albaladejo and J. A. Oller, \textit{"Nucleon-Nucleon interactions from dispersion relations: coupled partial waves"}, Phys. Rev. C 86, 034005 (2012), \href{http://arxiv.org/abs/1201.0443}{[arXiv:1201.0443 [nucl-th]]}.
\bibitem{Jackson} A. D. Jackson, D. O. Riska and B. Verwest, \textit{"Meson exchange model for the nucleon-nucleon interaction"}, Nucl. Phys. A 249, 397
(1975).
\bibitem{Cottingham} W. N. Cottingham, M. Lacombe, B. Loiseau, J. M. Richard and R. Vinhman, \textit{"Nucleon-Nucleon interaction from pion-nucleon phase-shift analysis"}, Phys. Rev. D 8, 800 (1973).
\bibitem{Machleidt} R. Machleidt, K. Holinde and Ch. Elster, \textit{"The bonn meson-exchange model for the nucleon-nucleon interaction"}, Phys. Rep. 149, 1 (1987).
\bibitem{Minelli} T. A. Minelli, A. Pascolini and C. Villi, \textit{"The Padua model of the nucleon and the nucleon-nucleon potential"}, Il Nuovo Cimento A, 104, 1589 (1991).
\bibitem{Jaede} L. Jaede and H. V. von Geram, \textit{"A nonlinear approach to NN interactions using self-interacting meson fields"}, Phys. Rev. C 55, 57 (1997), \href{http://arxiv.org/abs/nucl-th/9604002}{[arXiv:nucl-th/9604002]}.
\bibitem{FredMyhrer} F. Myhrer and J. Wroldsen, \textit{"The nucleon-nucleon force and the quark degrees of freedom"}, Rev. Mod. Phys. 60, 629 (1988).
\bibitem{nucl-th/9809093} S. A. Zaitsev and E. L. Kramar, \textit{"NN interactions in QCD: Old and new techniques"}, In *Pruhonice 1998, Mesons and light nuclei '98*, 281 (1998), \href{http://arxiv.org/abs/nucl-th/9809093}{[arXiv:nucl-th/9809093]}.
\bibitem{hep-ph/0211443} Fl. Stancu, S. Pepin and L. Ya. Glozman, \textit{"The nucleon-nucleon interaction in a chiral constituent quark model"}, Phys. Rev. C 56, (1997), \href{http://arxiv.org/abs/nucl-th/9705030}{[arXiv:nucl-th/9705030]}; Fl. Stancu, \textit{"The nucleon-nucleon problem in quark models"}, Few Body Syst. Suppl. 14, 83 (2003), \href{http://arxiv.org/abs/hep-ph/0211443}{[arXiv:hep-ph/0211443]}.
\bibitem{nucl-th/0212044} A. Valcarce, F. Fernandez and P. Gonzalez, \textit{"NN interaction in chiral constituent quark models"}, Few Body Syst. Suppl. 15, 25 (2003), \href{http://arxiv.org/abs/nucl-th/0212044}{[arXiv:nucl-th/0212044]}.
\bibitem{hep-ph/0205138} D. Bartz, \textit{"The nucleon-nucleon interaction in a chiral constituent quark model"}, Nucl. Phys. A 699, 316 (2002), \href{http://arxiv.org/abs/hep-ph/0205138}{[arXiv:hep-ph/0205138]}.
\bibitem{Wu0} G. H. Wu, J. L. Ping, L. J. Teng, F. Wang and T. Goldman, \textit{"Quark delocalization, color screening model and nucleon–baryon scattering"}, Nucl. Phys. A 673, 279 (2000), \href{http://arxiv.org/abs/nucl-th/9812079}{[arXiv:nucl-th/9812079]}.
\bibitem{nucl-th/0404004} L. R. Dai, Z. Y. Zhang, Y. W. Yu and P. Wang, \textit{"N-N interactions in the extended chiral $SU(3)$ quark model"}, Nucl. Phys. A 727, 321 (2003), \href{http://arxiv.org/abs/nucl-th/0404004}{[arXiv:nucl-th/0404004]}.
\bibitem{1110.3022} R. Machleidt and D. R. Entem, \textit{"Chiral symmetry and the nucleon-nucleon interaction"}, \href{http://arxiv.org/abs/1110.3022}{[arXiv:1110.3022 [nucl-th]]}.
\bibitem{nucl-th/0007051} I. P. Cavalcante and M. R. Robilotta, \textit{"Nucleon-nucleon interaction in the Skyrme model"}, Phys. Rev. C 63, 044008 (2001), \href{http://arxiv.org/abs/nucl-th/0007051}{[arXiv:nucl-th/0007051]}.
\bibitem{Wambach} J. Wambach and T. Waindzoch, \textit{"From skyrmions to the nucleon-nucleon potential"}, CRM Series in Mathematical Physics, Springer New York, 287 (2000).
\bibitem{Rashdan} M. Rashdan, \textit{"NN interaction derived from the Nambu-Jona-Lasinio model"}, Chaos, Solitons \& Fractals 18, 107 (2003).
\bibitem{Kukulin} V. I. Kukulin, V. N. Pomerantsev, A. Faessler, A. J. Buchmann and E. M. Tursunov, \textit{"Moscow-type NN-potentials and three-nucleon bound states"}, Phys. Rev. C 57, 535 (1998), \href{http://arxiv.org/abs/nucl-th/9711043}{[arXiv:nucl-th/9711043]}.
\bibitem{Oxford1} C. Downum, J. Stone, T. Barnes, E. Swanson and I. Vidana, \textit{"Nucleon-Nucleon interactions from the quark model"}, AIP Conf. Proc. 1257, 538 (2010), \href{http://arxiv.org/abs/1001.3320}{[arXiv:1001.3320 [nucl-th]]}.
\bibitem{0802.2484} M. R. Robilotta, \textit{"Nuclear interactions: The chiral picture"}, Mod. Phys. Lett. A 23, 2273 (2008), \href{http://arxiv.org/abs/0802.2484}{[arXiv:0802.2484 [nucl-th]]}.
\bibitem{0803.4190} D. Shukla, D. R. Phillips and E. Mortenson, \textit{"Chiral potentials, perturbation theory, and the ${}^1 S_0$ channel of NN scattering"}, J. Phys. G: Nucl. Part. Phys. 35, 115009 (2008), \href{http://arxiv.org/abs/0803.4190}{[arXiv:0803.4190 [nucl-th]]}.
\bibitem{Beane} S. R. Beane and M. J. Savage, \textit{"Nucleon-Nucleon Interactions on the Lattice"}, Phys. Lett. B 535, 177 (2002),  \href{http://arxiv.org/abs/hep-lat/0202013}{[arXiv:hep-lat/0202013]}; S. R. Beane, P. F. Bedaque, A. Parreno and M. J. Savage, \textit{"Two nucleons on a lattice"}, Phys. Lett. B 585, 106 (2004), \href{http://arxiv.org/abs/hep-lat/0312004}{[arXiv:hep-lat/0312004]}; S. R. Beane, P. F. Bedaque, K. Orginos and M. J. Savage, \textit{"Nucleon-Nucleon scattering from fully-dynamical lattice QCD"}, Phys. Rev. Lett. 97 012001 (2006), \href{http://arxiv.org/abs/hep-lat/0602010}{[arXiv:hep-lat/0602010]}; S. R. Beane, \textit{"Nuclear forces on the lattice"}, PoS CD 09, 076 (2009),
    \href{http://arxiv.org/abs/0912.5404}{[arXiv:hep-lat/0912.5404 [hep-lat]]}.
\bibitem{nucl-th/0611096} N. Ishii, S. Aoki and T. Hatsuda, \textit{"Nuclear force from lattice QCD"}, Phys. Rev. Lett. 99, 022001 (2007), \href{http://arxiv.org/abs/nucl-th/0611096}{[arXiv:nucl-th/0611096]}.
\bibitem{hep-lat/0601006} T. T. Takahashi, T. Doi and H. Suganuma, \textit{"Nuclear force in lattice QCD"}, AIP Conf. Proc. 842, 249 (2006),  \href{http://arxiv.org/abs/hep-lat/0601006}{[arXiv:hep-lat/0601006]}.
\bibitem{1005.1908} M. I. Buchoff, \textit{"Topics in lattice QCD and effective field theory"}, Ph.D. Thesis, \href{http://arxiv.org/abs/1005.1908}{[arXiv:1005.1908 [hep-lat]]}.
\bibitem{nucl-th/9807024} Y. Kondo and O. Morimatsu, \textit{"QCD sum rules for nucleon-nucleon interactions"}, Prog. Theor. Phys. 100, 1 (1998), \href{http://arxiv.org/abs/nucl-th/9807024}{[arXiv:nucl-th/9807024]}.
\bibitem{Weinberg} S. Weinberg, \textit{"Phenomenological Lagrangians"}, Physica A 96, 327 (1979); S. Weinberg, \textit{"Nuclear forces from chiral lagrangians"},  Phys. Lett. B 251, 288 (1990).
\bibitem{nucl-th/9801034} D. B. Kaplan, M. J. Savage and M. B. Wise, \textit{"A new expansion for nucleon-nucleon interactions"}, Phys. Lett. B 424, 390 (1998), \href{http://arxiv.org/abs/nucl-th/9801034}{[arXiv:nucl-th/9801034]}; D. B. Kaplan, M. J. Savage and M. B. Wise, \textit{"Two-nucleon systems from effective field theory"}, Nucl. Phys. B 534, 329 (1998), \href{http://arxiv.org/abs/nnucl-th/9802075}{[arXiv:nucl-th/9802075]}.
\bibitem{nucl-th/9902015} U. van Kolck, \textit{"Effective field theory of nuclear forces"}, Prog. Part. Nucl. Phys. 43, 337 (1999), \href{http://arxiv.org/abs/nucl-th/9902015}{[arXiv:nucl-th/9902015]}.
\bibitem{nucl-th/9909011} Ulf-G. Meißner, \textit{"Effective field theory for the two-nucleon system"}, PiN Newslett. 1, 65 (1999), \href{http://arxiv.org/abs/nucl-th/9909011}{[arXiv:nucl-th/9909011]}.
\bibitem{0704.0807} R. Machleidt, \textit{"Nuclear forces from chiral effective field theory"}, \href{http://arxiv.org/abs/0704.0807}{[arXiv:0704.0807 [nucl-th]]}.
\bibitem{1303.4674} A. Ekström, G. Baardsen, C. Forssén, G. Hagen, M. Hjorth-Jensen, G. R. Jansen, R. Machleidt, W. Nazarewicz, T. Papenbrock, J. Sarich and S. M. Wild, \textit{"An optimized chiral nucleon-nucleon interaction at next-to-next-to-leading order"}, Phys. Rev. Lett. 110, 192502 (2013), \href{http://arxiv.org/abs/1303.4674}{[arXiv:1303.4674 [nucl-th]]}.
\bibitem{nucl-th/9910064} E. Epelbaoum, W. Glöckle and Ulf-G. Meißner, \textit{"Nuclear forces from chiral Lagrangians using the method of unitary transformation I: Formalism"}, Nucl. Phys. A 637,107 (1998), \href{http://arxiv.org/abs/nucl-th/9801064}{[arXiv:nucl-th/9801064]}; E. Epelbaum, W. Glöckle and Ulf-G. Meißner, \textit{"Nuclear forces from chiral Lagrangians using the method of unitary transformation II: The two-nucleon system"}, Nucl. Phys. A 671, 295 (2000), \href{http://arxiv.org/abs/nucl-th/9910064}{[arXiv:nucl-th/9910064]}.
\bibitem{nucl-th/0608068} R. Machleidt and D. R. Entem, \textit{"Recent advances in the theory of nuclear forces"}, Phys. Conf. Ser. 20, 77 (2005), \href{http://arxiv.org/abs/nucl-th/0608068}{[arXiv:nucl-th/0608068]}.
\bibitem{1302.3241} E. Epelbaum, \textit{"Nuclear physics with chiral effective field theory: state of the art and open challenges"}, J. Phys. G: Nucl. Part. Phys. 35, 115009 (2008), \href{http://arxiv.org/abs/1302.3241}{[arXiv:1302.3241 [nucl-th]]}.
\bibitem{Ordonez1} C. Ordonez, L. Ray and U. van Kolck, \textit{"Nucleon-nucleon potential from an effective chiral Lagrangian"}, Phys. Rev. Lett. 72, 1982 (1994); C. Ordonez, L. Ray and U. van Kolck, \textit{"The Two nucleon potential from chiral Lagrangians"}, Phys. Rev. C 53, 2086 (1996),  \href{http://arxiv.org/abs/hep-ph/9511380}{[arXiv:hep-ph/9511380]}.
\bibitem{Robilotta1} C. A. da Rocha and M. R. Robilotta, \textit{"Two pion exchange nucleon-nucleon potential: The minimal chiral model"}, Phys. Rev. C 49, 1818 (1994); M. R. Robilotta, \textit{"Pion-nucleon scattering and the tail of the two-pion exchange nucleon-nucleon potential"}, Nucl. Phys. A 595, 171 (1995); C. A. da Rocha, M. R. Robilotta and J. L. Ballot, \textit{"The role of chiral symmetry in two-pion exchange nuclear potential"}, \href{http://arxiv.org/abs/nucl-th/9608036}{[arXiv:nucl-th/9608036]}.
\bibitem{Munich1} N. Kaiser, R. Brockmann and W. Weise, \textit{"Peripheral nucleon-nucleon phase shifts and chiral symmetry"}, Nucl. Phys. A 625, 758 (1997), \href{http://arxiv.org/abs/nucl-th/9706045}{[arXiv:nucl-th/9706045]}.
\bibitem{Idaho1} D. R. Entem and R. Machleidt, \textit{"Accurate nucleon-nucleon potential based upon chiral perturbation theory"}, Phys. Lett. B 524, 93,(2002), \href{http://arxiv.org/abs/nucl-th/0108057}{[arXiv:nucl-th/0108057]}.
\bibitem{BochumJulich2} E. Epelbaum, A. Nogga, W. Gloeckle, H. Kamada, U.-G. Meissner and H. Witala, \textit{"The two-nucleon system at next-to-next-to-next-to-leading order"}, Nucl. Phys. A 747, 362 (2005), \href{http://arxiv.org/abs/nucl-th/0405048}{[arXiv:nucl-th/0405048]}.
\bibitem{Gammel1} J. L. Gammel, R. S .Christian and R. M. Thaler, \textit{"Calculation of phenomenological nucleon-nucleon potentials"}, Phys. Rev. 105, 311 (1957).
\bibitem{Gammel2} J. L. Gammel and R. M. Thaler, \textit{"Spin-Orbit coupling in the proton-proton interaction"}, Phys. Rev. 107, 291 (1957).
\bibitem{SingellMarshak} P. S. Signell and R. E. Marshak, \textit{"Phenomenological two-nucleon potential up to 150 Mev"}, Phys. Rev. 106, 832 (1957).
\bibitem{Gartenhaus} S. Gartenhaus, \textit{"Two-nucleon potential from the cut-off Yukawa theory"}, Phys. Rev. 100, 900 (1955).
\bibitem{OkuboMarshak} S. Okubo and R. E. Marchak, \textit{"Velocity dependence of the two-nucleon interaction"}, Ann. Phys. 4, 166 (1958).
\bibitem{Sugawara1} M. Sugawara and S. Okubo, \textit{"Two-nucleon potential from pion field theory with pseudoscalar coupling"}, Phys. Rev. 117, 605 (1960).
\bibitem{Sugawara2} H. Sugawara and F. Von Hippel, \textit{"Zero-parameter model of the N-N potential"}, Phys. Rev. 172, 1764 (1968).
\bibitem{Yale2} G. Breit, M. H. Hull, K. E. Lassila and K. D. Pyatt,\textit{"Phase-parameter representation of proton-proton scattering from 9.7 to 345 Mev"}, Phys. Rev. 120, 2227 (1960); G. Breit, M. H. Hull, K. E. Lassila, K. D. Pyatt and H. M. Ruppel, \textit{"Phase-parameter representation of proton-proton scattering from 9.7 to 345 MeV. II"}, Phys. Rev. 128, 826 (1962).
\bibitem{Livermore} R. A. Arndt and M. H. Macgregor, \textit{"Determination of the nucleon-nucleon elastic-scattering matrix. IV. Comparison of energy-dependent and energy-independent phase-shift analyses"}, Phys. Rev. 141, 873 (1966).
\bibitem{0706.2195} R. A. Arndt, W. J. Briscoe, I. I. Strakovsky and R.L. Workman, \textit{"Updated analysis of NN elastic scattering to 3 GeV"}, Phys. Rev. C. 76, 025209 (2008), \href{http://arxiv.org/abs/0706.2195}{[arXiv:0706.2195 [nucl-th]]}.
\bibitem{Hoshizaki00} N. Hoshizaki, I. Lin and Sh. Machida, \textit{"Nonstatic one-boson-exchange potentials"}, Prog. Theor. Phys. 26, 680 (1961).
\bibitem{Wong00} D. Y. Wong, \textit{"Meson resonances and nucleon-nucleon potentials"}, Nucl. Phys. 55, 212 (1964).
\bibitem{Bryan} R. A. Bryan and B. L. Scott, \textit{"Nucleon-Nucleon scattering from one-boson-exchange potentials"}, Phys. Rev. 135, 434 (1964); R. A. Bryan and B. L. Scott, \textit{"Nucleon-Nucleon scattering from one-boson-exchange potentials. 2. Inclusion of all momentum-dependent terms through order $p^2$"}, Phys. Rev. 164, 1215 (1967); R. A. Bryan and B. L. Scott, \textit{"Nucleon-nucleon scattering from one-boson-exchange potentials. iii. s waves included"}, Phys. Rev. 177, 1435 (1969).
\bibitem{Nijm78} M. M. Nagels, T. A. Rijken and J. J. de Swart, \textit{"Low-energy nucleon-nucleon potential from Regge-pole theory"}, Phys. Rev. D 17, 768 (1978).
\bibitem{Virgina1} F. Gross, J. W. Van Orden and K. Holinde, \textit{"Relativistic one-boson-exchange model for the nucleon-nucleon interaction"}, Phys. Rev. C 45, 2094 (1992).
\bibitem{Bochum1} D. Pluemper, J. Flender and M. F. Gari, \textit{"Nucleon-nucleon interaction from meson exchange and nucleonic structure"}, Phys. Rev. C 49, 2370 (1994).
\bibitem{Paris2} M. Lacombe, B. L. Seau, J. M. Richard, R. Vinhman, J. Côté, P. Pirés and R. de Tourreil, \textit{"Parametrization of the Paris N-N potential"}, Phys. Rev. C 21, 861 (1980).
\bibitem{Bonn4} R. Machleidt, \textit{"The high-precision, charge-dependent Bonn nucleon-nucleon potential (CD-Bonn)"}, Phys. Rev. C 63, 024001 (2001),  \href{http://arxiv.org/abs/nucl-th/0006014}{[arXiv:nucl-th/0006014]}.
\bibitem{VinhMau1} R. Vinh Mau, C. Semay, B. Loiseau and M. Lacombe, \textit{"Nuclear forces and quark degrees of freedom"}, Phys. Rev. Lett. 67, 1392 (1991).
\bibitem{nucl-th/0108041} S. K. Bogner, T. T. S. Kuo, A. Schwenk, D. R. Entem and R. Machleidt, \textit{"Towards a model independent low momentum nucleon nucleon interaction"}, Phys. Lett. B 576, 265 (2003), \href{http://arxiv.org/abs/nucl-th/0108041}{[arXiv:nucl-th/0108041]}.
\bibitem{HamadaJohnstonM} T. Hamada, K. Nakamura and R. Tamgaki, \textit{"Modification of Hamada-Johnston potential"}, Prog. Theor. Phys. 33, 769 (1965).
\bibitem{Bressel} C. N. Bressel, A. K. Kerman and B. Rouben, \textit{"Soft-core nucleon-nucleon potential"}, Nucl. Phys. A 124, 624 (1969).
\bibitem{Livermore1969} M. M. MacGregor, R. A. Arndt and R. M. Wright, \textit{"Determination of the nucleon-nucleon scattering matrix. X. (p,p) and (n,p) analysis from 1 to 450 MeV"}, Phys. Rev. 182, 1714 (1969).
\bibitem{Paris3A} J. Haidenbaur and W. Plessas, \textit{"Separable representation of the Paris nucleon-nucleon potential"}, Phys. Rev. C 30, 1822 (1984).
\bibitem{Paris3} J. Haidenbaur and W. Plessas, \textit{"Modified separable representation of the Paris nucleon-nucleon potential in the $^{1} S_{0}$ and $^{3} P_{0}$ states"}, Phys. Rev. C 32, 1424 (1985).
\bibitem{deTourreil0} R. de Torreil and D. W. L .Sprang, \textit{"Construction of a nucleon-nucleon soft-core potential"}, Nucl. Phys. A 201, 193 (1973).
\bibitem{deTourreil} R. de Tourreil, B. Rouben and D. W. L. Sprung, \textit{"Super-soft-core nucleon-nucleon interaction with $\pi$-, $\rho$- and $\omega$-exchange contributions"}, Nucl. Phys. A 242, 445 (1975).
\bibitem{deTourreil2} J. Cote, B. Rouben, R. de Tourreil and D. W. L. Sprang, \textit{"Third-order perturbation calculations in nuclear matter for realistic potentials"}, Nucl. Phys. A 273, 269 (1976).
\bibitem{Petris9} L. Petris, \textit{"A phenomenological nucleon-nucleon interaction"}, J. Phys. G: Nucl. Phys. 7, 309 (1981).
\bibitem{Funabashi00} T. Obinata and M. Wada, \textit{"Nonstatic one-boson-exchange potential with retardation"}, Prog. Theor. Phys. 53, 732 (1975).
\bibitem{Funabashi01} T. Obinata and M. Wada, \textit{"Nonstatic one-boson-exchange potential with retardation and nuclear matter- Including velocity-dependent tensor potential-"}, Prog. Theor. Phys. 57, 1984 (1977).
\bibitem{Funabashi02} T. Obinata, \textit{"Refinements of the R-space funabashi potentials"}, Prog. Theor. Phys. 73, 1270 (1985).
\bibitem{Funabashi2} I. Arisaka, K. Nakagawa, T. Obinata and M. Wada, \textit{"Realistic nucleon-nucleon potentials expressed in terms of the gaussian basis"}, Prog. Theor. Phys. 92, 281 (1994).
\bibitem{Green} A. M. Green and P. Haapakoski, \textit{"The effect of the $\Delta(1236)$ in the two-nucleon problem and in neutron matter"}, Nucl. Phys. A 221, 429 (1974).
\bibitem{Green2} A. M. Green,\textit{"Velocity dependent nuclear forces and their effect in nuclear matter"}, Nucl. Phys. 33, 218 (1962).
\bibitem{Arndt3} R. A. Arndt, R. H. Hackman and L. D. Roper, \textit{"Nucleon-nucleon scattering analyses. II. Neutron-proton scattering from 0 to 425 MeV and proton-proton scattering from 1 to 500 MeV"}, Phys, Rev. C 15, 1002 (1977); R. A. Arndt, R. H. Hackman and L. D. Roper, \textit{"Nucleon-nucleon scattering analyses. III. np phase-shift analyses: Complex structure and multiple solutions at 50 and 325 MeV"}, Phys, Rev. C 15, 1021 (1977) .
\bibitem{Bugg} D. V. Bugg, J .A. Edgington, C. Amster, R. C. Brown, C. J. Oram and K. Shakarchi, \textit{"Proton-proton elastic scattering from 150 to 515 MeV"}, J. Phys. G: Nucl. Part. Phys. 4, 1025 (1978).
\bibitem{Henley} E. M. Henley and G. A. Miller, \textit{"Mesons in Nuclei"}, P. 406 , North-Holland (1979).
\bibitem{Stoks4} V. G. J. Stoks, R. Timmermans and J. J. de Swart, \textit{"Pion-nucleon coupling constant"}, Phys. Rev. C 47, 512 (1993), \href{http://arxiv.org/abs/nucl-th/9211007}{[arXiv:nucl-th/9211007]}.
\bibitem{NijmPWA} V. G. J. Stoks, R. A. M. Klomp, M. C. M. Rentmeester and J .J. Swart, \textit{"Partial-wave analysis of all nucleon-nucleon scattering data below 350 MeV"}, Phys. Rev. C 48, 792 (1993).
\bibitem{NijmPWA00} J. R. Bergervoet, P. C. van Campen, R. A. M. Klomp, J.-L. de Kok, T. A. Rijken, V. G. J. Stoks and J. J. de Swart, \textit{"Phase shift analysis of all proton-proton scattering data below $T_{lab}$=350 MeV"}, Phys. Rev. C 41, 1435 (1990).
\bibitem{AV18pq} R. B. Wiringa, A. Arriaga and V. R. Pandharipande, \textit{"Quadratic momentum dependence in the nucleon-nucleon interaction"}, Phys. Rev. C. 68, 054006 (2003), \href{http://arxiv.org/abs/nucl-th/0306018}{[arXiv:nucl-th/0306018]}.
\bibitem{1106.1934} S. Veerasamy and W. N. Polyzou, \textit{"A momentum-space Argonne V18 interaction"}, Phys. Rev. C. 84, 034003 (2011), \href{http://arxiv.org/abs/1106.1934}{[arXiv:1106.1934 [nucl-th]]}.
\bibitem{nucl-th/9211013} V. Stoks and J. J. de Swart, \textit{"Comparison of potential models with the pp scattering data below 350 MeV"}, Phys. Rev. C 47, 761 (1993), \href{http://arxiv.org/abs/nucl-th/9211013}{[arXiv:nucl-th/9211013]}.
\bibitem{Haidenbauer} J. Haidenbauer and K. Holinde, \textit{"Application of the Bonn potential to proton-proton scattering"}, Phys. Rev. C 40, 2465 (1989).
\bibitem{nucl-th/9301019} R. Machleidt and G. Q. Li, \textit{"Nucleon-Nucleon potentials in comparison: Physics or polemics?"}, Phys. Rept. 242, 5 (1994), \href{http://arxiv.org/abs/nucl-th/9301019}{[arXiv:nucl-th/9301019]}.
\bibitem{Bonn3} R. Machleidt, F. Sammaruca, and Y. Song, \textit{"Nonlocal nature of the nuclear force and its impact on nuclear structure"}, Phys. Rev. C 53, 1483 (1996), \href{http://arxiv.org/abs/nucl-th/9510023}{[arXiv:nucl-th/9510023]}.
\bibitem{NijmPWA2} R. Timmermans, Th. A. Rijken and J. J. de Swart, \textit{"Antiproton-proton partial-wave analysis below 925 MeV/c"}, Phys. Rev. C 50, 48 (1994), \href{http://arxiv.org/abs/nucl-th/9403011}{[arXiv:nucl-th/9403011]}.
\bibitem{Nijmegen} J. J. de Swart, R. A. M. M. Klomp, M. C. M. Rentmeester and Th. A. Rijken, \textit{"The Nijmegen potentials"}, Few Body Syst. Suppl. 8, 438 (1995), \href{http://arxiv.org/abs/nucl-th/9509024}{[arXiv:nucl-th/9509024]}.
\bibitem{NijmA} M. M. Nagels, T. A. Rijken and J. J. de Swart, \textit{"A potential model for hyperon-nucleon scattering"}, Annals Phys. 79, 338 (1973).
\bibitem{NijmBC} M. M. Nagels, T. A. Rijken and J. J. de Swart, \textit{"Determination of the mixing angle, $f/(f+d)$ ratio, and coupling constants of the scalar-meson nonet"}, Phys. Rev. Lett. 31, 569 (1973).
\bibitem{NijmD} M. M. Nagels, T. A. Rijken and J. J. de Swart, \textit{"Baryon-baryon scattering in a one-boson-exchange-potential approach. I. Nucleon-nucleon scattering"}, Phys. Rev. D 12, 744 (1975).
\bibitem{NijmD2} M. M. Nagels, T. A. Rijken and J. J. de Swart, \textit{"Baryon-baryon scattering in a one-boson-exchange-potential approach. II. Hyperon-nucleon scattering"}, Phys. Rev. D 15, 2547 (1977).
\bibitem{NijmF} M. M. Nagels, T. A. Rijken and J. J. de Swart, \textit{"Baryon-baryon scattering in a one-boson-exchange-potential approach. III. A nucleon-nucleon and hyperon-nucleon analysis including contributions of a nonet of scalar mesons"}, Phys. Rev. D 20 1633 (1979).
\bibitem{Hoshizaki04} N. Hoshizaki and Sh. Machida, \textit{"Two nucleon potential with full recoil. I---General formalism and one-pion-exchange potential---"}, Prog. Theor. Phys. 24, 1325 (1960).
\bibitem{Nijm78HV} P. M. M. Maessen, Th. A. Rijken and J.J. de Swart, \textit{"Soft-core baryon-baryon one-boson-exchange models. II. Hyperon-nucleon potential"}, Phys. Rev. C 40, 2226 (1989).
\bibitem{NijmPWA3} M. C. M. Rentmeester, R. G. E. Timmermans and J. J. de Swart \textit{"Partial-wave analyses of all proton-proton and neutron-proton data below 500 MeV"}, AIP Conf. Proc. 768, 59 (2005), \href{http://arxiv.org/abs/nucl-th/0410042v}{[arXiv:nucl-th/0410042]}.
\bibitem{1304.0895} R. Navarro Perez, J. E. Amaro and E. Ruiz Arriola, \textit{"Partial wave analysis of nucleon-nucleon scattering below pion production threshold"}, Phys. Rev. C 88, 024002 (2013), \href{http://arxiv.org/abs/1304.0895}{[arXiv:1304.0895 [nucl-th]]}.
\bibitem{NijmESC} Th. A. Rijken, \textit{"Baryon-Baryon interactions"}, Few Body Syst. Suppl. 7, 1 (1994), \href{http://arxiv.org/abs/nucl-th/9401004}{[arXiv:nucl-th/9401004]}.
\bibitem{NijmESC01} Th. A. Rijken and V. G. J. Stoks, \textit{"Soft two-meson-exchange nucleon-nucleon potentials. I. Planar and crossed-box diagrams"}, Phys. Rev. C 54, 2851 (1996), \href{http://arxiv.org/abs/nucl-th/9509029}{[arXiv:nucl-th/9509029]}; Th. A. Rijken and V. G. J. Stoks, \textit{"Soft two-meson-exchange nucleon-nucleon potentials. II. One-pair and two-pair diagrams"}, Phys. Rev. C 54, 2869 (1996), \href{http://arxiv.org/abs/nucl-th/9509031}{[arXiv:nucl-th/9509031]}.
\bibitem{NijmESC02} Th. A. Rijken and V. G. J. Stoks, \textit{"Meson-baryon coupling constants from a chiral-invariant SU(3) Lagrangian and application to NN scattering"},  Nucl. Phys. A 613, 311 (1997), \href{http://arxiv.org/abs/nucl-th/9611002}{[arXiv:nucl-th/9611002]}.
\bibitem{NijmESC03} Th. A. Rijken, \textit{"The Nijmegen hyperon-nucleon and hyperon-hyperon interactions"}, Nucl. Phys. A 639, 29c (1998).
\bibitem{NijmESC04} Th. A. Rijken, \textit{"Recent Nijmegen soft-core hyperon nucleon and hyperon hyperon interactions"}, Nucl. Phys. A 691, 322 (2001).
\bibitem{NijmESC1} Th. A. Rijken, H. Polinder and J. Nagata, \textit{"ESC NN-potentials in momentum space. I. PS-PS exchange potentials"}, Phys. Rev. C 66, 044008 (2002), \href{http://arxiv.org/abs/nucl-th/0201018}{[arXiv:nucl-th/0201018]}; Th. A. Rijken, H. Polinder and J. Nagata, \textit{"ESC NN-potentials in momentum space. II. Meson-Pair exchange potentials"}, Phys. Rev. C 66, 044009 (2002), \href{http://arxiv.org/abs/nucl-th/0201020}{[arXiv:nucl-th/0201020]}.
\bibitem{NijmSC22} H. Polinder and Th. A. Rijken, \textit{"Soft-core meson-baryon interactions. I. One-hadron-exchange potentials"}, Phys. Rev. C 72, 065210 (2005), \href{http://arxiv.org/abs/nucl-th/0505082}{[arXiv:nucl-th/0505082]}; H. Polinder and Th. A. Rijken, \textit{"Soft-core meson-baryon interactions. II. $\pi N$ and $K^+ N$ scattering"}, Phys. Rev. C 72, 065211 (2005), \href{http://arxiv.org/abs/nucl-th/0505083}{[arXiv:nucl-th/0505083]}.
\bibitem{NijmESC2} Th. A. Rijken, \textit{"Extended-soft-core baryon-baryon model I. nucleon-nucleon scattering (ESC04)"}, Phys. Rev. C 73, 044007 (2006), \href{http://arxiv.org/abs/nucl-th/0603041}{[arXiv:nucl-th/0603041]}.
\bibitem{NijmESC3} Th. A. Rijken and Y. Yamamoto, \textit{"Extended-soft-core baryon-baryon model. II. Hyperon-nucleon interaction"}, Phys. Rev. C 73, 044008 (2006), \href{http://arxiv.org/abs/nucl-th/0603042}{[arXiv:nucl-th/0603042]}.
\bibitem{NijmESC4} Th. A. Rijken and Y. Yamamoto, \textit{"Extended-soft-core baryon-baryon model III. $S=-2$ hyperon-hyperon/nucleon interaction"}, \href{http://arxiv.org/abs/nucl-th/0608074}{[arXiv:nucl-th/0608074]}.
\bibitem{NijmESCall1} Th. A. Rijken and Y. Yamamoto, \textit{"Recent soft-core baryon–baryon interactions"}, Nucl. Phys. A 804, 51 (2008).
\bibitem{NijmESCall2} Th. A. Rijken, M. M. Nagels and Y. Yamamoto, \textit{"Baryon-Baryon Interactions---Nijmegen Extended-Soft-Core Models---"}, Prog. Theor. Phys. Suppl. 185, 14 (2010).
\bibitem{NijmOPT1} R.A. M. M. Klomp, J. -L. de Kok, M. C. M. Rentmeester, Th. A. Rijken and J. J. de Swart, \textit{"Partial Wave Analyses of the pp data alone and of the np data alone"}, AIP Conf. Proc. 334, 367 (1995), \href{http://arxiv.org/abs/nucl-th/9406022}{[arXiv:nucl-th/9406022]}.
\bibitem{Hamburg2} L. Jaede and H. V. VonGeramb, \textit{"Nucleon-nucleon scattering observables from the solitary boson exchange potential"}, Phys. Rev. C 57, 496 (1998), \href{http://arxiv.org/abs/nucl-th/9707023}{[arXiv:nucl-th/9707023]}.
\bibitem{Hamburg22} L. Jaede, \textit{"Consistent description of NN and pi-N interactions using the solitary boson exchange potential"}, Phys. Rev. C 58, 96 (1998), \href{http://arxiv.org/abs/nucl-th/9802039}{[arXiv:nucl-th/9802039]}.
\bibitem{vonGeramb3}  H. V. von Geramb, B. Davaadorj and St. Wirsching, \textit{"Relativistic nucleon-nucleon potentials using Dirac's constraint instant form dynamics"},, \href{http://arxiv.org/abs/nucl-th/0308004}{[arXiv:nucl-th/0308004]}.
\bibitem{Arndt2000} R. A. Arndt, I. I. Strakovsky and R. L. Workman, \textit{"Nucleon-nucleon elastic scattering to 3 GeV"}, Phys. Rev. C 62, 034005 (2000), \href{http://arxiv.org/abs/nucl-th/0004039}{[arXiv:nucl-th/0004039]}.
\bibitem{Moscow2} V. I. Kukulin, V. N. Pomerantsev and A. Faessler, \textit{"The complete version of Moscow NN potential"}, \href{http://arxiv.org/abs/nucl-th/9903056}{[arXiv:nucl-th/9903056]}.
\bibitem{Moscow2b}  A. Faessler, V. I. Kukulin, I. T. Obukhovsky and V. N. Pomerantsev, \textit{"New mechanism for intermediate- and short-range nucleon-nucleon interaction"}, J. Phys. G: Nucl. Part. Phys. 27, 1851 (2001), \href{http://arxiv.org/abs/nucl-th/9912074}{[arXiv:nucl-th/9912074]}; V. I. Kukulin, I. T. Obukhovsky, V. N. Pomerantsev and A. Faessler, \textit{"Two-component dressed-bag model for NN interaction: Deuteron structure and phase shifts up to 1-GeV"}, Int. J. Mod. Phys. E 11, 1 (2002).
\bibitem{Moscow3} A. Faessler, V. I. Kukulin, I. T. Obukhovsky and V. N. Pomerantsev \textit{"Description of intermediate- and short-range NN nuclear force within a covariant effective field theory"}, Ann. Phys. 320, 71 (2005), \href{http://arxiv.org/abs/nucl-th/0505026}{[arXiv:nucl-th/0505026]}.
\bibitem{IS1} P. Doleschall, \textit{"Phenomenological non-local N-N interactions and the triton binding energy"}, Nucl. Phys. A 602, 60 (1996); P. Doleschall and I. Borbely, \textit{"Properties of the nonlocal NN interactions required for the correct triton binding energy"}, Phys. Rev. C 62, 054004 (2000).
\bibitem{IS2} P. Doleschall, I. Borbely, Z. Papp and W. Plessas, \textit{"Nonlocality in the nucleon nucleon interaction and three nucleon bound states"}, Phys. Rev. C 67, 064005 (2003).
\bibitem{MIK00} S. A. Zaitsev, \textit{"Tridiagonal parametrization of interaction in the discrete approach to the scattering problem"}, Theor. Math. Phys. 115, 575 (1998).   .
\bibitem{MIK0} A. M. Shirokov, A. I. Mazur, S. A. Zaytsev, J. P. Vary and T. A. Weber, \textit{"Nucleon-nucleon interaction in the J-matrix inverse scattering approach and few-nucleon systems"}, Phys. Rev. C 70, 044005 (2004), \href{http://arxiv.org/abs/nucl-th/0312029}{[arXiv:nucl-th/0312029]}.
\bibitem{MIK1} A. M. Shirokov, J. P. Vary, A. I. Mazur, S. A. Zaytsev and T. A. Weber, \textit{"Novel NN interaction and the spectroscopy of light nuclei"}, Phys. Lett. B 621, 96 (2005), \href{http://arxiv.org/abs/nucl-th/0407018}{[arXiv:nucl-th/0407018]}.
\bibitem{MIK2} A. M. Shirokov, J. P. Vary, A. I. Mazur and T. A. Weber, \textit{"Many-body nuclear Hamiltonian: Ab exitu approach"}, Phys. Lett. B 644, 33 (2007), \href{http://arxiv.org/abs/nucl-th/0512105}{[arXiv:nucl-th/0512105]}; A. M. Shirokov, V. A. Kulikov, P. Maris, A. I. Mazur, E. A. Mazur and J. P. Vary, \textit{"NN interaction JISP16: Current status and prospect"}, EPJ Web Conf. 3, 05015 (2010), \href{http://arxiv.org/abs/0912.2967}{[arXiv:0912.2967 [nucl-th]]}.
\bibitem{MIK3} A. M. Shirokov, V. A. Kulikov, P. Maris, A. I. Mazur, E. A. Mazur and J. P. Vary, \textit{"New development of realistic J-matrix inverse scattering NN interaction and ab initio description of light nuclei"}, Proc. 3rd Int. Conf. Current Problems in Nucl. Phys. and Atomic Energy, Kyiv, Part I, 321 (2011), \href{http://arxiv.org/abs/1009.2993}{[arXiv:1009.2993 [nucl-th]]}.
\bibitem{nucl-th/0105011} A. Funk, H. V. von Geramb and K. A. Amos, \textit{"Nucleon-Nucleon optical model for energies to 3 GeV"}, Phys. Rev. C 64, 054003 (2001), \href{http://arxiv.org/abs/nucl-th/0105011}{[arXiv:nucl-th/0105011]};  H. V. von Geramb, A. Funk and A. Faltenbacher, \textit{" Nucleon-Nucleon pptical potentials and fusion of $p$N, KN, $pp$ and NN Systems"}, Few Body Syst. Suppl. 13, 274 (2000), \href{http://arxiv.org/abs/nucl-th/0010057}{[arXiv:nucl-th/0010057]}.
\bibitem{Knyr1} N. A. Khokhlov and V. A. Knyr \textit{"Reconstruction of the optical potential in the inverse quantum scattering. Application to the relativistic inelastic NN scattering"}, \href{http://arxiv.org/abs/nucl-th/0410092}{[arXiv:nucl-th/0410092]}; N. A. Khokhlov and V. A. Knyr \textit{"Reconstruction of the optical potential from scattering data"}, Phys. Rev. C 73, 024004 (2006), \href{http://arxiv.org/abs/quant-ph/0506014}{[arXiv:quant-ph/0506014]}.
\bibitem{Knyr2} V. A. Knyr, V. G. Neudachin and N. A. Khokhlov, \textit{"Relativistic optical model on the basis of the Moscow potential and lower phase shifts for nucleon nucleon scattering at laboratory energies of up to 3-GeV"}, Phys. Atom. Nucl. 69, 2034 (2006); Yad. Fiz. 69, 2079 (2006).
\bibitem{QCDNN00} V. G. Neudatchin, Yu. F. Smirnov and R. Tamagaki, \textit{"An explanation of N-N "repulsive core" in terms of forbidden states based on the quark model"}, Prog. Theor. Phys. 58, 1072 (1977).
 \bibitem{MOkaNN00} M. Oka and K. Yazaki, \textit{"Nuclear force in a quark model"}, Phys. Lett. B 90, 41 (1980); M. Oka and K. Yazaki, \textit{"Short range part Of nuclear force and deuteron in a quark model"}, Nucl. Phys. A 402, 477 (1983).
\bibitem{PRD34-1986} P. LaFrance and E. L. Lomon , \textit{"Six-quark resonance structures in nucleon-nucleon scattering"}, Phys. Rev. D 34, 1341 (1986).
\bibitem{Shimizu00} K. Shimizu, \textit{"Study of baryon baryon interactions and nuclear properties in the quark cluster model"}, Prog. Theor. Phys. 52, 56 (1989); K. Shimizu, S. Takeuchi and A. J. Buchmann, \textit{"Study of nucleon nucleon and hyperon nucleon interaction"}, Prog. Theor. Phys. Suppl. 137, 43 (2000).
\bibitem{nucl-th/9606020} K. Saito, K. Tsushima and A. W. Thomas, \textit{"Self-consistent description of finite nuclei based on a relativistic quark model"}, Nucl.Phys. A 609, 339 (1996), \href{http://arxiv.org/abs/nucl-th/9606020}{[arXiv:nucl-th/9606020]}.
\bibitem{JapanQCDNN} Y. Fujiwara, C. Nakamoto and Y. Suzuki, \textit{"Unified description of NN and YN interactions in a quark model with Effective meson-Exchange potentials"}, Phys. Rev. Lett. 76, 2242 (1996).
\bibitem{nucl-th/0010044} D. Hadjimichef, J. Haidenbauer and G. Krein, \textit{"Long- and medium-range components of the nuclear force in quark-model based calculations"}, Phys. Rev. C 63, 035204 (2001), \href{http://arxiv.org/abs/nucl-th/0010044}{[arXiv:nucl-th/0010044]}.
\bibitem{Arndt87} R. A. Arndt, J. S. Hyslop, III and L. D. Roper, \textit{"Nucleon-nucleon partial-wave analysis to 1100 MeV"}, Phys. Rev. D 35, 128 (1987).
\bibitem{nucl-th/9808007} U. van Kolck, \textit{"Effective field theory for short-range forces"}, Nucl. Phys. A 645,273 (1999), \href{http://arxiv.org/abs/nucl-th/9808007}{[arXiv:nucl-th/9808007]}.
\bibitem{vanKolck1} U. van Kolck, \textit{"Few nucleon forces from chiral Lagrangians"}, Phys. Rev. C 49, 2932 (1994).
\bibitem{nucl-th/0304025} R. Higa and M. R. Robilotta, \textit{"Two-pion exchange nucleon-nucleon potential:$O(q^4)$ relativistic chiral expansion"}, Phys. Rev. C 68 024004 (2003), \href{http://arxiv.org/abs/nucl-th/0304025}{[arXiv:nucl-th/0304025]}; R. Higa, M. R. Robilotta and C. A. da Rocha, \textit{"Relativistic $O(q^4)$ two-pion exchange nucleon-nucleon potential: configuration space"}, Phys. Rev. C 69, 034009 (2004), \href{http://arxiv.org/abs/nucl-th/0310011}{[arXiv:nucl-th/0310011]}; R. Higa, M. R. Robilotta and C. A. da Rocha, \textit{"Relativistic $O(q^4)$ two-pion exchange nucleon-nucleon potential: parametrized version"}, \href{http://arxiv.org/abs/nucl-th/0501076}{[arXiv:nucl-th/0501076]}.
\bibitem{0704.0711} R. Higa and M. R. Robilotta, \textit{"Two-pion exchange three-nucleon potential: $O(q^4)$ chiral expansion"}, Phys. Rev. C 76, 014006 (2007), \href{http://arxiv.org/abs/0704.0711}{[arXiv:0704.0711 [nucl-th]]}.
\bibitem{0908.4405} R. Higa, \textit{"NN potentials from IR chiral EFT"}, PoS CD 09, 025 (2009), \href{http://arxiv.org/abs/0908.4405}{[arXiv:0908.4405 [nucl-th]]}.
\bibitem{hep-ph/9501384} V. Bernard, N. Kaiser and U.-G. Meißner, \textit{"Chiral dynamics in nucleons and nuclei"}, Int. J. Mod. Phys. E 4, 193 (1995), \href{http://arxiv.org/abs/hep-ph/9501384}{[arXiv:hep-ph/9501384]}.
\bibitem{Munich2} N. Kaiser, S. Gerstendoerfer and W. Weise, \textit{"Peripheral NN-scattering: Role of delta-excitation, correlated two-pion and vector meson exchange"}, Nucl. Phys. A 637, 395 (1998), \href{http://arxiv.org/abs/nucl-th/9802071}{[arXiv:nucl-th/9802071]}.
\bibitem{nucl-th/9901054} M. C. M. Rentmeester, R. G. E. Timmermans, J. L. Friar and J. J. de Swart, \textit{"Chiral Two-Pion Exchange and Proton-Proton Partial-Wave Analysis"}, Phys. Rev. Lett. 82, 4992 (1999), \href{http://arxiv.org/abs/nucl-th/9901054}{[arXiv:nucl-th/9901054]}.
\bibitem{Munich3} N. Kaiser, \textit{"Chiral 3 pi exchange N N potentials: Results for representation invariant classes of diagrams"}, Phys. Rev. C 61, 014003 (2000), \href{http://arxiv.org/abs/nucl-th/9910044}{[arXiv:nucl-th/9910044]}; N. Kaiser, \textit{"Chiral 2 pi exchange N N potentials: Two loop contributions"}, Phys. Rev. C 64, 057001 (2001), \href{http://arxiv.org/abs/nucl-th/0107064}{[arXiv:nucl-th/0107064]}; N. Kaiser, \textit{"Chiral 2 pi exchange NN potentials: Relativistic $1/M2$ corrections"}, Phys. Rev. C 65, 017001 (2002), \href{http://arxiv.org/abs/nucl-th/0109071}{[arXiv:nucl-th/0109071]}.
\bibitem{nucl-th/0312058} N. Kaiser, \textit{"Three-body spin-orbit forces from chiral two-pion exchange"}, Phys. Rev. C 68, 054001 (2003), \href{http://arxiv.org/abs/nucl-th/0312058}{[arXiv:nucl-th/0312058]}.
\bibitem{nucl-th/0202039} D. R. Entem and R. Machleidt, \textit{"Chiral $2\pi$ exchange at fourth order and peripheral NN scattering"}, Phys. Rev. C 66, 014002 (2002), \href{http://arxiv.org/abs/nucl-th/0202039}{[arXiv:nucl-th/0202039]}.
\bibitem{Munich4} N. Kaiser, \textit{"Chiral four-body interactions in nuclear matter"}, Eur. Phys. J. A 48, 135 (2012), \href{http://arxiv.org/abs/1209.4556}{[arXiv:1209.4556 [nucl-th]]}.
\bibitem{1304.3175} J. W. Holt, N. Kaiser, G. A. Miller and W. Weise, \textit{"Microscopic optical potential from chiral nuclear forces"}, Phys.Rev. C 88, 024614 (2013), \href{http://arxiv.org/abs/1304.3175}{[arXiv:1304.3175 [nucl-th]]}.
\bibitem{1304.5339} J. Haidenbauer, S. Petschauer, N. Kaiser, U.-G. Meißner, A. Nogga and W. Weise, \textit{"Hyperon-nucleon interaction at next-to-leading order in chiral effective field theory"}, Nucl. Phys. A 915, 24 (2013), \href{http://arxiv.org/abs/1304.5339}{[arXiv:1304.5339 [nucl-th]]}.
\bibitem{1305.3427} S. Petschauer and N. Kaiser, \textit{"Relativistic SU(3) chiral baryon-baryon Lagrangian up to order $q^2$"}, Nucl. Phys. A 916, 1 (2013), \href{http://arxiv.org/abs/1305.3427}{[arXiv:1305.3427 [nucl-th]]}.
\bibitem{Idaho2} D. R. Entem and R. Machleidt, \textit{"Accurate charge-dependent nucleon-nucleon potential at fourth order of chiral perturbation theory"}, Phys. Rev. C 68, 041001 (2003), \href{http://arxiv.org/abs/nucl-th/0304018}{[arXiv:nucl-th/0304018]}.
\bibitem{Idaho3} R. Machleidt and D. R. Entem, \textit{"Towards a consistent approach to nuclear structure: EFT of two- and many-body forces"}, J. Phys. G: Nucl. Part. Phys. 31, 1235 (2005), \href{http://arxiv.org/abs/nucl-th/0503025}{[arXiv:nucl-th/0503025]}.
\bibitem{BochumJulich1} E. Epelbaum, A. Nogga, W. Gloeckle, H. Kamada, U.-G. Meissner and H. Witala, \textit{"Few-nucleon systems with two-nucleon forces from chiral effective field theory"}, Eur. Phys. J. A 15, 543 (2002), \href{http://arxiv.org/abs/nucl-th/0201064}{[arXiv:nucl-th/0201064]}.
\bibitem{1301.6134} E. Epelbaum and J. Gegelia, \textit{"Weinberg's approach to nucleon-nucleon scattering revisited "}, Phys. Lett. B 716, 338 (2012), \href{http://arxiv.org/abs/1207.2420}{[arXiv:1207.2420 [nucl-th]]}; E. Epelbaum and J. Gegelia, \textit{"The two-nucleon problem in EFT reformulated: Pion and nucleon masses as soft and hard scales"}, PoS CD12, 090 (2013), \href{http://arxiv.org/abs/1301.6134}{[arXiv:1301.6134 [nucl-th]]}.
\bibitem{nucl-th/9807054} T.-S. Park, K. Kubodera, D.-P. Min and M. Rho, \textit{"The power of effective field theories in nuclei: The deuteron, NN scattering and electroweak processes"}, Nucl. Phys. A 646, 83 (1999), \href{http://arxiv.org/abs/nucl-th/9807054}{[arXiv:nucl-th/9807054]}.
\bibitem{nucl-th/9906056} S. Fleming, T. Mehen and I. W. Stewart, \textit{"NN scattering ${}^3S_1-{}^3D_1$ mixing angle at next-to-next-to-leading order"}, Phys. Rev. C 61, 044005 (2000), \href{http://arxiv.org/abs/nucl-th/9906056}{[arXiv:nucl-th/9906056]}.
\bibitem{Seattle00} J.-W. Chen, G. Rupak and M. J. Savage, \textit{"Nucleon-nucleon effective field theory without pions"}, Nucl. Phys. A 653, 386 (1999), \href{http://arxiv.org/abs/nucl-th/9902056}{[arXiv:nucl-th/9902056]}; G. Rupak and N. Shoresh, \textit{"NNLO calculation of two nucleon scattering in EFT for a two Yukawa toy model"}, Phys. Rev. C 60, 054004 (1999), \href{http://arxiv.org/abs/nucl-th/9902077}{[arXiv:nucl-th/9902077]}; G. Rupak and N. Shoresh, \textit{"Nucleon-nucleon scattering in effective field theory"}, In *Seattle 1999, Nuclear physics with effective field theory* 77-99, \href{http://arxiv.org/abs/nucl-th/9906077}{[arXiv:nucl-th/9906077]}.
\bibitem{Serra} M. Serra, T. Otsuka, Y. Akaishi, P. Ring and Sh. Hirose, \textit{"Relativistic mean field models and nucleon-nucleon interactions"}, Prog. Theor. Phys. 113, 1009 (2005).
\bibitem{1011.5732} B. B. Singh, M. Bhuyan, S. K. Patra and R. K. Gupta \textit{"A new microscopic nucleon-nucleon interaction derived from relativistic mean field theory"}, J. Phys. G: Nucl. Part. Phys. 39, 025101 (2012), \href{http://arxiv.org/abs/1011.5732}{[arXiv:1011.5732 [nucl-th]]}.
\bibitem{nucl-th/0109059} A. Schwenk, G. E. Brown and B. Friman, \textit{"Low momentum nucleon-nucleon interaction and Fermi liquid theory"}, Nucl. Phys. A 703, 745 (2002), \href{http://arxiv.org/abs/nucl-th/0109059}{[arXiv:nucl-th/0109059]}; S. K. Bogner, A. Schwenk, T. T. S. Kuo and G. E. Brown, \textit{"Renormalization group equation for low momentum effective nuclear interactions"}, \href{http://arxiv.org/abs/nucl-th/0111042}{[arXiv:nucl-th/0111042]}; A. Schwenk, \textit{"Nuclear interactions from the renormalization group"}, Int. J. Mod. Phys. B 20, 2724 (2006), \href{http://arxiv.org/abs/nucl-th/0411070}{[arXiv:nucl-th/0411070]}.
\bibitem{nucl-th/0212034} L. Coraggio, A. Covello, A. Gargano, N. Itaco and T. T. S. Kuo, \textit{"Low-momentum nucleon-nucleon potential and Hartree-Fock calculations"}, \href{http://arxiv.org/abs/nucl-th/0212034}{[arXiv:nucl-th/0212034]}.
\bibitem{nucl-th/0305035} S. K. Bogner, T. T. S. Kuo and A. Schwenk, \textit{"Model independent low momentum nucleon interaction from phase shift equivalence"}, Phys. Rept. 386, 1 (2003), \href{http://arxiv.org/abs/nucl-th/0305035}{[arXiv:nucl-th/0305035]}.
\bibitem{nucl-th/0611045} S. K. Bogner, R. J. Furnstahl and R. J. Perry, \textit{"Similarity renormalization group for nucleon-nucleon interactions"}, Phys. Rev. C 75, 061001 (2007), \href{http://arxiv.org/abs/nucl-th/0611045}{[arXiv:nucl-th/0611045]}.
\bibitem{1302.3978} E. Ruiz Arriola, V. S. Timoteo and S. Szpigel, \textit{"Nuclear symmetries of the similarity renormalization group for nuclear forces"}, Phys. Rev. C 75, 061001 (2007), \href{http://arxiv.org/abs/1302.3978}{[arXiv:1302.3978 [nucl-th]]}.
\bibitem{1012.4914} Michael C. Birse, \textit{"The Renormalisation group and nuclear forces"}, Phil. Trans. Roy. Soc. Lond. A 369, 2662 (2011), \href{http://arxiv.org/abs/1012.4914}{[arXiv:1012.4914 [nucl-th]]}.
\bibitem{hep-lat/0609078} Y. Koma and M. Koma, \textit{"Spin-dependent potentials from lattice QCD"}, Nucl. Phys. B 769, 79 (2007),  \href{http://arxiv.org/abs/hep-lat/0609078}{[arXiv:hep-lat/0609078]}.
\bibitem{1103.0619} K. Murano, N. Ishii, S. Aoki and T. Hatsuda, \textit{"Nucleon-Nucleon potential and its non-locality in lattice QCD"}, Prog. Theor. Phys. 125, 1225 (2011), \href{http://arxiv.org/abs/1103.0619}{[arXiv:1103.0619 [nucl-th]]}.
\bibitem{1109.2889} S. R. Beane, E. Chang, W. Detmold, H. W. Lin, T. C. Luu, K. Orginos, A. Parreno, M. J. Savage, A. Torok and A. Walker-Loud, \textit{"The deuteron and exotic two-body bound states from lattice QCD"}, Phys. Rev. D 85, 054511 (2012), \href{http://arxiv.org/abs/1109.2889}{[arXiv:1109.2889 [nucl-th]]}.
\bibitem{nucl-th/0509049} O. Plohl, C. Fuchs and E. N. E. van Dalen, \textit{"Model independent study of the Dirac structure of the nucleon-nucleon interaction"}, Phys. Rev. C 73, 014003 (2006), \href{http://arxiv.org/abs/nucl-th/0509049}{[arXiv:nucl-th/0509049]}.
\bibitem{1212.4896} O. Plohl, C. Fuchs and E. N. E. van Dalen, \textit{"Construction of energy-independent potentials above inelastic thresholds in quantum field theories"}, Phys. Rev. D 87, 034512 (2013), \href{http://arxiv.org/abs/1212.4896}{[arXiv:1212.4896 [nucl-th]]}.
\bibitem{nucl-th/0209058} A. Amghar and B. Desplanques, \textit{"More about the comparison of local and non-local NN interaction models"}, Nucl. Phys. A 714, 502 (2003), \href{http://arxiv.org/abs/nucl-th/0209058}{[arXiv:nucl-th/0209058]}; B. Desplanques and A. Amghar, \textit{"Nucleon-nucleon interaction models and non-locality"}, Few Body Syst. Suppl. 14, 59 (2003), \href{http://arxiv.org/abs/nucl-th/0210028}{[arXiv:nucl-th/0210028]}.
\bibitem{1010.1728} A. C. Cordon, M. P. Valderrama and E. R. Arriola, \textit{"Nucleon-Nucleon interaction, charge symmetry breaking and renormalization"}, Phys. Rev. C 85, 024002 (2012), \href{http://arxiv.org/abs/1010.1728}{[arXiv:1010.1728 [nucl-th]]}.
\bibitem{0803.2075} B. Desplanques, C.H. Hyun, S. Ando and C.-P. Liu, \textit{"Parity-violating nucleon-nucleon interaction from different approaches"}, Phys. Rev. C 77, 064002 (2008), \href{http://arxiv.org/abs/0803.2075}{[arXiv:0803.2075 [nucl-th]]}.
\bibitem{1210.4273} H.-W. Hammer, A. Nogga and A. Schwenk, \textit{"Three-body forces: From cold atoms to nuclei"}, Rev. Mod. Phys. 85, 197 (2013), \href{http://arxiv.org/abs/1210.4273}{[arXiv:1210.4273 [nucl-th]]}.
\bibitem{nucl-th/0509032} Evgeny Epelbaum, \textit{"Few-nucleon forces and systems in chiral effective field theory"}, Prog. Part. Nucl. Phys. 57, 654 (2006), \href{http://arxiv.org/abs/nucl-th/0509032}{[arXiv:nucl-th/0509032]}.
\bibitem{0901.0012} K.-Y. Kim and I. Zahed, \textit{"Nucleon-Nucleon potential from holography"}, JHEP 0903, 131 (2009), \href{http://arxiv.org/abs/0901.0012}{[arXiv:0901.0012 [nucl-th]]}.
\bibitem{0901.4449} K. Hashimoto, T. Sakai and S. Sugimoto, \textit{"Nuclear force from string theory"}, Prog. Theor. Phys. 122, 427 (2009), \href{http://arxiv.org/abs/0901.4449}{[arXiv:0901.4449 [nucl-th]]}.








\end{thebibliography}
\end{document}